%% file: g1paper.tex
\documentclass[letterpaper,twocolumn,preprintnumbers,superscriptaddress,amssymb,showpacs,amsmath,floatfix,showkeys,nofootinbib]{revtex4}

\usepackage{bm}       
\usepackage{dcolumn}  
\usepackage{graphicx} 
\usepackage{mathrsfs}
\usepackage{multirow}
\usepackage{longtable}       

\def\PPP{\csname PPP\endcsname}
\expandafter\def\csname PPP\endcsname{\thetable}

\newcommand{\tx}{\textrm}
\newcommand{\be}{\begin{equation}}
\newcommand{\ee}{\end{equation}}
\newcommand{\beqa}{\begin{eqnarray}}
\newcommand{\eeqa}{\end{eqnarray}}
\newcommand{\nn}{\nonumber}

\newcommand{\D}{\ensuremath{\textrm{d}}}

\newcommand{\Qs}{\ensuremath{Q^{2}}}

\newcommand{\Apar}{\ensuremath{A_{||} } }

\newcommand{\Ap}{\ensuremath{A_{1}^{\textrm{p}}}}
\newcommand{\Ad}{\ensuremath{A_{1}^{\D}}}

\newcommand{\spar}{{\ensuremath\stackrel{\rightarrow}{\Rightarrow}}}
\newcommand{\sant}{{\ensuremath\stackrel{\rightarrow}{\Leftarrow}}}

\newcommand{\sparq}{{\ensuremath\stackrel{\rightharpoonup}{\Rightarrow}}}
\newcommand{\santq}{{\ensuremath\stackrel{\rightharpoonup}{\Leftarrow}}}

\newcommand{\dt}{\!\!{\rm\,d}t\,}
\newcommand{\dx}{\!\!\!{\rm\,d}x\,}
\newcommand{\de}{{\rm\,d}}
\newcommand{\detwo}{{\rm\,d^2\!}}

\newcommand{\MSeq}{\ensuremath{\stackrel{\scriptstyle \overline{MS}}{=}}}

\graphicspath{{./figs/}{./xfigs/}}

\begin{document}

\preprint{DESY/06-142}

\title{
 Precise  determination of the spin structure function $\mathbf{g_1}$ \\
 of the proton, deuteron and neutron }

\include{rec-g1long}

\date{\today}


\begin{abstract}
Precise measurements of the spin structure functions of the proton 
$g_1^p(x,Q^2)$ and deuteron $g_1^d(x,Q^2)$ are presented over the kinematic 
range $0.0041 \leq  x \leq 0.9$ and 
$0.18 $ GeV$^2$ $\leq Q^2 \leq 20$ GeV$^2$. The data
were collected at the HERMES experiment at DESY, in
deep-inelastic scattering of 27.6~GeV  longitudinally polarized positrons
off longitudinally polarized hydrogen and deuterium gas targets 
internal to the HERA storage ring. The neutron spin structure function
$g_1^n$ is extracted by combining proton and deuteron data. 
The integrals of $g_1^{p,d}$ at $Q^2=5$~GeV$^2$  are evaluated over the 
measured $x$ range. 
Neglecting any possible contribution to the $g_1^d$ integral from the region 
$x \leq 0.021$,  a value of 
$0.330 \pm 0.011\mathrm{(theo.)}\pm0.025\mathrm{(exp.)}\pm 0.028$(evol.)
is obtained for the flavor-singlet axial charge $a_0$
in a leading-twist NNLO analysis. 
\end{abstract}


\pacs{13.60.-r, 13.60.Hb, 13.88.+e, 14.20.Dh, 14.65.-q, 67.65.+z}


\maketitle


\section{\label{sec:intro}Introduction}

A major goal in the study of Quantum Chromo-Dynamics (QCD) in recent years 
has been the detailed investigation of the spin structure of the nucleon 
 and the determination of the partonic composition of its spin projection: 
\begin{equation}\label{spinsum}
S_z=\frac{1}{2} = \frac{1}{2} \Delta \Sigma(\mu^2) + \Delta g(\mu^2) 
+ L_z^q(\mu^2) + L_z^g(\mu^2)\,.
\end{equation}
Here $\frac{1}{2}\Delta \Sigma$  ($\Delta g$) describes the net integrated 
contribution of quark and anti-quark (gluon) helicities to the nucleon 
helicity and $L_z^q$  ($L_z^g$) is the $z$ component of  the orbital 
angular momentum among all quarks (gluons). 
The individual terms in the sum 
are dependent on scale $\mu$ and factorization scheme, 
and the decomposition of the gluon total angular
momentum $J^g_z(\mu^2) =\Delta g(\mu^2) +  L_z^g(\mu^2)$ is 
not gauge invariant.

Detailed information about $\Delta \Sigma$ and its flavor decomposition into 
the contributions from quarks and antiquarks 
 can be obtained from various sources. In the context of this paper,
double-spin asymmetries of cross sections 
in inclusive deep-inelastic scattering (DIS) $\ell + N \to \ell + X$ of 
longitudinally polarized charged leptons off longitudinally polarized 
nucleons are considered, where only the scattered charged lepton is 
observed but not 
the hadronic final state $X$. 
Inclusive scattering is sensitive to the square of the quark charges,
and therefore cannot distinguish quarks from anti-quarks. This
distinction can be made in semi-inclusive deep-inelastic 
scattering $\ell + N \to \ell + h +X$, where in addition to the scattered 
lepton one or more hadrons, produced in the reaction, are 
recorded~\cite{hermesdeltaq,smc-deltaq}.
  
The theoretical and experimental status
on the spin structure of the nucleon  has been discussed in great detail in several 
recent reviews 
(see, e.g., Refs.~\cite{anselmino-1995,reya-lampe,hughes-voss,filippone-ji,leader-book} and 
references therein). Here  only the essential ingredients will be summarized. 

In lowest order perturbation theory, the deep-inelastic reaction 
$\ell + N \to \ell + X$ proceeds via the exchange of a neutral virtual boson 
($\gamma^*, Z^0$). At HERMES center-of-mass energies contributions 
from $Z^0$-exchange to the cross section can be 
safely neglected. Therefore only the electromagnetic interaction in the 
approximation of one-photon exchange is taken into account here. 
In this approximation, the cross section
of polarized inclusive DIS is parameterized by two spin 
structure functions $g_1$ and $g_2$. These functions cannot
presently be calculated from the QCD Lagrangian.

In the QCD-improved quark parton model (QPM), i.e., at leading twist,
and to leading logarithmic order in the running strong coupling 
constant $\alpha_s(Q^2)$ of Quantum-Chromodynamics (LO QCD), 
the deep-inelastic scattering off the nucleon can be 
interpreted as the incoherent superposition of virtual-photon interactions with quarks of 
any flavour $q$. By angular momentum conservation, a spin-$\frac{1}{2}$
 parton can absorb a hard photon only when their 
spin orientations are opposite. 
Polarized photons with the same (opposite) helicity as the 
polarized target nucleon consequently probe 
the quark number density $q_\sparq(x,Q^2)$ ($q_\santq(x,Q^2)$) for quarks with 
the  same (opposite) helicity as that of the parent nucleon.
The spin structure function $g_1$  has then a 
probabilistic interpretation, which for the proton and the 
neutron reads:
\begin{eqnarray}\label{g1-lo}
g_1^{p,n}(x,Q^2)=\frac{1}{2} \sum_{q} e^2_q \left[
\Delta q^{p,n}(x,Q^2) + \Delta \bar{q}^{p,n}(x,Q^2)\right]
\nn\\
= \frac{1}{2}\langle e^2\rangle \left[\Delta q_S (x,Q^2)
+\Delta q^{p,n}_{NS} (x,Q^2)\right]. 
\end{eqnarray}
Here, the quantity $-Q^2$ is the squared four-momentum transferred by
the virtual photon, $x$ is the fraction of the nucleon's light-cone
momentum carried by the struck quark, $e_q$ is the charge, in units of
the elementary charge $\vert e \vert$, of quarks of flavor $q$,
$\langle e^2\rangle=\sum_q e^2_q/N_q$ is the average squared charge of
the $N_q$ active quark flavors, 
and $\Delta{q}(x,Q^2) = q_\sparq(x,Q^2) -
q_\santq(x,Q^2)$ is the quark helicity distribution for
massless\footnote{ Among the power corrections are terms of order
$m_q^2/Q^2$.  } quarks of flavor $q$ in a longitudinally polarized
nucleon in the `infinite-momentum frame'.
Correspondingly, $\Delta{\bar{q}}(x,Q^2), \bar{q}_\sparq(x,Q^2)$ and
$\bar{q}_\santq(x,Q^2)$ are anti-quark distributions.  The flavor
singlet and flavor non-singlet quark helicity distributions are
defined as
\begin{eqnarray}\label{deltaq-s}
\Delta q_S(x,Q^2)&=&\sum_{q}\left[\Delta q^{p,n}(x,Q^2) + \Delta 
\bar{q}^{p,n}(x,Q^2)\right]\nn\\
 &\equiv& \Delta {\Sigma}(x,Q^2),
\end{eqnarray}
and 
\begin{eqnarray}\label{deltaq-ns}
\Delta q_{NS}^{p,n}(x,Q^2)\!\!&=&\!\!\frac{1}{\langle 
e^2\rangle}\!\sum_qe^2_q\left[\Delta q^{p,n}(x,Q^2)\! +\!\Delta
\bar{q}^{p,n}(x,Q^2)\right]\nn\\
&~&-\Delta q_S(x,Q^2). 
\end{eqnarray}
The functions $g_1^p$ and $g_1^n$ differ only in their 
non-singlet components, which through isospin symmetry are obtained from 
each other by exchanging $u$ and $d$ quarks: 
$\Delta u^p =\Delta d^n \equiv \Delta u,\Delta d^p = \Delta u^n \equiv \Delta d$.
For the analysis presented in this paper,
only the three lightest quark flavors, $q=u,d,s$, are taken into 
account  and the number of active quark flavors $N_q$ is 
equal to three.

Quantities of particular importance are the first moments of the quark 
helicity distributions:
\begin{equation}
\Delta q(Q^2) = \int_0^1\dx \Delta q(x,Q^2)\,.
\end{equation}
The quantity ($\Delta q(Q^2) + \Delta{\bar{q}}(Q^2))$ is the net number of quarks 
plus antiquarks of flavor $q$ with positive helicity inside a nucleon 
with positive helicity and thus 
$\frac{1}{2} \Delta \Sigma (Q^2)$ is the net contribution to the 
nucleon's helicity that can be attributed to the helicities of the quarks. 

In LO QCD the gluon distribution does not contribute explicitly in Eq.~(\ref{g1-lo}) to the
structure function $g_1(x,Q^2)$, but it does appear in the QCD
evolution equations~\cite{altarelli77} for $g_1$. Beyond leading order QCD,  
also gluons have to be taken explicitly into 
account in the expression for $g_1$, with $\Delta g(x,Q^2)$ being the gluon helicity 
distribution. The first moment $\Delta g(Q^2)= \int_0^1\dx \Delta g(x,Q^2)$
represents the total gluon helicity contribution to the helicity of the 
nucleon.

If the electromagnetic currents in the nucleon are treated as fields
of free quarks of only the lightest three flavors,
the first moment of the structure function $g_1$ can be decomposed into 
contributions from the axial charges $a_3, a_8$, and $a_0$, which are 
related to the hadronic matrix elements of the octet plus singlet
quark SU(3) axial vector currents.
In the scaling (Bjorken) limit,
\begin{equation}\label{gamma1-ai}
\Gamma_1^{p,n}(Q^2) = \int_0^1\dx g_1^{p,n}(x,Q^2) = \frac {1}{36} 
\left( a_8\pm 3a_3 +4 a_0\right).
\end{equation}
Here the +($-$) sign  of the $a_3$ term holds for the proton
(neutron).  In the naive quark model,
the axial charges are related to the first moments of quark helicity
distributions by 
\begin{equation}\label{a3-deltaq}
a_3 = (\Delta u + \Delta \bar{u}) - (\Delta d + \Delta \bar{d}),
\end{equation}
\begin{equation}\label{a8-deltaq}
a_8 =( \Delta u + \Delta \bar{u})  + (\Delta d + \Delta \bar{d}) - 
2(\Delta s + \Delta \bar{s}),
\end{equation}
\begin{equation}\label{a0-deltaq}
a_0 = (\Delta u + \Delta \bar{u}) +(\Delta d + \Delta \bar{d})  +
(\Delta s + \Delta \bar{s})\equiv\Delta \Sigma.
\end{equation}
The quantities on either side of Eqs.~(\ref{a3-deltaq}) and (\ref{a8-deltaq})
are flavor non-singlet quantities and 
independent of $Q^2$ to any order in $\alpha_s(Q^2)$.
Beyond LO, $a_0$ becomes dependent on $Q^2$, and
$\Delta \Sigma$ may or may not be the same and may or may not depend
on $Q^2$, depending on the factorization scheme chosen~\cite{Adler:69,Bell:69}.

In the approximation of SU(3) flavor symmetry and of 
identical masses of up-, down-, and strange-quarks, 
the fundamental non-singlet quantities $a_3$ and $a_8$ can be related to 
the two decay constants $F$ and $D$ which govern  the Gamow-Teller part of 
the flavor-changing weak decays in the spin-$\frac{1}{2}$ baryon octet~\cite{anselmino-1995}:
$a_3 = F + D $ and $a_8 = 3F - D$. 
Here the values $F = 0.464\pm
0.008 $ and $D = 0.806 \pm 0.008$ are used as obtained from a fit to recent hyperon
decay data~\cite{pdg}, leading to 
$a_3  = F + D = g_A/g_V   = 1.269 \pm 0.003$ and 
$a_8 = 3F - D = 0.586 \pm 0.031$, 
with negligible correlation between $a_3$ and $a_8$.
Measurements of one of the 
spin structure functions $g_1^{p,n,d}$ and its first moment 
provide via Eqs.~(\ref{gamma1-ai}) to
(\ref{a8-deltaq}) the third necessary input  for the 
determination of the  flavor-singlet axial charge $a_0$,
and thereby also of $\Delta{\Sigma}$ and the 
moments of the helicity distributions 
of the three quark flavors ($\Delta u + \Delta \bar{u}$), ($\Delta d + \Delta \bar{d}$) and
 ($\Delta s + \Delta \bar{s}$).

At any order in $\alpha_s(Q^2)$ and in a leading-twist approximation,
the structure function $g_1$ is  a  
convolution of  quark, anti-quark and gluon helicity 
distributions~\cite{abfr1} 
with Wilson coefficient
functions $\Delta C(x,\alpha_s(Q^2))$~\cite{wilson}: 
\begin{multline} 
g_1^{p,n}(x,Q^2) = \\
\displaystyle\frac{1}{2} \langle e^2\rangle \int_x^1
\frac{\mathrm{d}x'}{x'} \left[ 
\Delta C_S \left(\frac{x}{x'},\alpha_s(Q^2)\right)\Delta q_S (x',Q^2) 
\right.\\
+\left. 
2 N_q \Delta C_g \left(\frac{x}{x'},\alpha_s(Q^2)\right)\Delta g(x',Q^2) 
\right.\\
+\left.
\Delta C_{NS}^{p,n} \left(\frac{x}{x'},\alpha_s(Q^2)\right)
\Delta q_{NS}^{p,n}(x',Q^2)\right].
\label{g1_nlo}
\end{multline}
In LO Eq.~(\ref{g1_nlo}) reduces to Eq.~(\ref{g1-lo}) since then 
$\Delta C_{S}$ and $\Delta
C_{NS}$ become $\delta$ functions and $\Delta C_g$ vanishes.  
The factorization between the helicity distributions and the 
coefficient functions involves some arbitrary choice, and hence the
distributions and their moments depend on the factorization scheme. 
The structure function $g_1(x,Q^2)$, as a 
physical observable, is scheme independent.
There are straightforward transformations that relate different schemes 
and their results to each other. In the `modified minimal subtraction' 
($\overline{MS}$) scheme~\cite{msbar}, 
the factorization scheme commonly used in most of 
the present  NLO analyses of unpolarized deep-inelastic and hard 
processes, the first moment of the gluon coefficient function vanishes
and $\Delta{g}(Q^2)$ does not contribute to the first moment of $g_1$. 
Therefore $\Gamma_1$ can be directly related to 
$a_0(Q^2) 
\MSeq
\Delta{\Sigma}(Q^2)$.

In the $\overline{MS}$ scheme, Eq.~(\ref{gamma1-ai}) becomes
\begin{eqnarray}\label{gamma1-ai-nlo} 
\Gamma_1^{p,n}(Q^2)& =& \frac {1}{36} \left[(a_8\pm 3 a_3) 
\Delta C_{NS}^{\overline{MS}}(\alpha_s(Q^2))\right.\nn\\
&~&\left.+ 4a_0 \Delta C_{S}^{\overline{MS}}(\alpha_s(Q^2))\right],
\end{eqnarray}
where $\Delta C_{NS}^{\overline{MS}}(\alpha_s(Q^2))$ and
 $\Delta C_{S}^{\overline{MS}}(\alpha_s(Q^2))$ are the first
moments of the non-singlet and singlet Wilson coefficient functions, respectively.

The difference of the $g_1$ moments for proton and neutron leads to the 
Bjorken Sum Rule~\cite{bjsr,Bjorken:70}, which in leading twist reads:
\begin{equation}\label{bjsr} 
\Gamma_1^{p}(Q^2) - \Gamma_1^{n}(Q^2) = \frac {1}{6} a_3 \Delta C_{NS}^{\overline{MS}}(\alpha_s(Q^2)),
\end{equation}
while their sum is given by:
\begin{eqnarray}\label{gamma1d-a8} 
\Gamma_1^{p}(Q^2) + \Gamma_1^{n}(Q^2) &=&  \frac {1}{18} 
\left[a_8 \Delta C_{NS}^{\overline{MS}}(\alpha_s(Q^2))\right.
\nn\\
&~&\left.+ 4a_0 \Delta C_{S}^{\overline{MS}}(\alpha_s(Q^2))\right].
\end{eqnarray}
This sum equals twice the deuteron moment apart from a small 
correction due to the
D-wave admixture to the deuteron wave function (see Eq.~(\ref{g1dfromg1pn})).
The measurement of $\Gamma_1^d$  hence allows for a straightforward
determination of $a_0$ using only $a_8$ as additional input.

In the $\overline{MS}$ scheme,
the non-singlet (singlet) coefficient  has been 
calculated up to third (second) order in the strong coupling constant~\cite{larin}:
\begin{eqnarray}
\Delta C_{NS}^{\overline{MS}}(\alpha_s(Q^2))  \!&=&\!1-\!\frac{\alpha_s}{\pi}\! -\!
3.583 \left(\frac{\alpha_s}{\pi}\right)^2\!\!-\! 20.215\left(\frac{\alpha_s}{\pi}\right)^3\nn\\
&~&\\
\Delta C_S^{\overline{MS}}(\alpha_s(Q^2))  &=& 1 - \left(\frac{\alpha_s}{\pi}\right)  - 
1.096 \left(\frac{\alpha_s}{\pi}\right)^2,
\end{eqnarray}
 for $N_q = 3$~\cite{lrv}. 
Estimates exist for the fourth (third)  order non-singlet (singlet) term~\cite{kataev}.

The first determination of $\Delta \Sigma$  was a moment analysis of 
the EMC proton data~\cite{emc-spin-crisis},
using Eq.~(\ref{gamma1-ai-nlo}) and  the moments of the Wilson coefficients in 
 $\mathcal{O}(\alpha_s^1)$.
It resulted in $\Delta \Sigma=0.120 \pm
0.094\mathrm{(stat)}\pm 0.138$(sys), much smaller than the expectation
($\Delta \Sigma\approx 0.6$)~\cite{jaffe90,schreiber99} from the relativistic constituent quark model.
This result caused
enormous activity in both experiment and theory.
A series of   high-precision 
scattering experiments with polarized beams and targets 
were completed 
at CERN~\cite{SMC:1999pdlox,SMC:1997p,Adeva:1998vv}, 
SLAC~\cite{Abe:1998wq,Anthony:1999rm,Anthony:2000fn}, 
DESY~\cite{g1n_hermes} and continue at CERN~\cite{compass} and JLAB~\cite{jlab:g1n}. 
Such measurements are always restricted to certain $x$ and $Q^2$ ranges due to
the experimental conditions.
However, any determination of $\Delta \Sigma$ 
requires an `evolution' to a fixed value of $Q^2$ and an
extrapolation of $g_1$ data to the full $x$ range 
and substantial uncertainties might arise from 
the necessary extrapolations  $x \rightarrow 0$ and $x\rightarrow 1$. 
This limitation applies also to recent determinations of $\Delta \Sigma$ based on NLO
fits~\cite{e154qcd,smcqcd,aac,grsv,bb} of the $x$ and $Q^2$
dependence of $g_1$ for proton, deuteron, and neutron,
using Eq.~(\ref{g1_nlo}) and the corresponding evolution equations.

This paper reports final results obtained by the HERMES experiment on
the structure function $g_1$  for the proton, deuteron, and neutron.
The results include an analysis of the proton data collected in 1996, 
a re-analysis of 1997 proton data 
previously published~\cite{HERMES:g1p}, as well as the analysis 
of the deuteron data collected in the year 2000. 
While the accuracy of the HERMES proton data is comparable to
that of earlier measurements, the HERMES deuteron data
are more precise than all published data.
By combining HERMES proton and deuteron data, 
precise  results on the neutron spin structure
function $g_1^n$ are obtained.

For this analysis, the kinematic range has been extended with respect
to the previous proton 
analysis, to include the region at low $x$ ($0.0041\leq x \leq 0.0212$) with low
$Q^2$. 
In this region the information available on $g_1$ was sparse.
As will be discussed in Sect.~\ref{sec:finalresults}, the first moment $\Gamma_1^d$
determined from HERMES data appears to saturate for $x < 0.04$. 
This observation allows for a determination of 
$a_0$ with small uncertainties and for a test of the 
Bjorken Sum Rule,  as well as scheme-dependent estimates of $\Delta \Sigma$
and the first moments of the flavor 
separated quark helicity distributions, $\Delta u + \Delta \bar{u}$,  
$\Delta d + \Delta \bar{d}$ and  $\Delta s + \Delta \bar{s}$.

The paper is organized as follows: the formalism leading to the
extraction of the structure function $g_1$  will be briefly reviewed in
Sect.~\ref{sec:formalism}, Sect.~\ref{sec:hermes} deals with  the
HERMES experimental
arrangement and the data analysis is described in
Sect.~\ref{sec:extraction}. Final
results are presented  in Sect.~\ref{sec:results} and discussed in 
Sect.~\ref{sec:finalresults}.


\section{\label{sec:formalism}Formalism}

\begin{figure}[t!]
\center{
\includegraphics[width=\columnwidth]{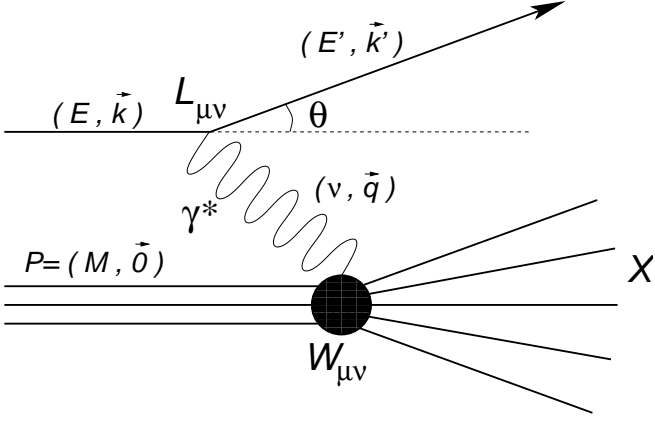}
   \caption{ Schematic picture of Deep-Inelastic Scattering 
for one photon exchange. The kinematic variables are defined in Tab.~\ref{table:kinevar}.}
   \label{fig:dis_DIAGRAM}
}
\end{figure}

In the one-photon-exchange approximation, the
 differential cross section for inclusive deep-inelastic scattering of
polarized charged leptons off polarized nuclear targets can be written~\cite{lea82}
as:
\begin{equation}
\displaystyle\frac{\mathrm{d}^2\sigma(s,S) }{\mathrm{ d} x~ \mathrm{d}Q^2}= \frac{2\pi  \alpha^2 y^2}{Q^6}
{{\mathbf{L_{\mu\nu}}(s) \mathbf {W^{\mu\nu}}(S)}}\,,
\end{equation}
where $\alpha$ is the fine-structure constant. 
As depicted
in  Fig.~\ref{fig:dis_DIAGRAM}  the leptonic tensor $\mathbf{L_{\mu\nu}}$ describes the
emission of a virtual photon at the lepton vertex, and the hadronic
tensor $\mathbf{W^{\mu\nu}}$ describes the hadron vertex.
The main kinematic variables used for the description of 
deep-inelastic scattering are defined in Tab.~\ref{table:kinevar}. 
The tensor $\mathbf{L_{\mu\nu}}$ can be calculated  precisely in
Quantum  Electro-Dynamics (QED)~\cite{Bjorken:70}:
\begin{eqnarray}\label{lmn}
\mathbf{L_{\mu\nu}}(s)&=&2(k_\mu k'_\nu+k_\nu k'_\mu-g_{\mu\nu}(k\cdot
k'-m^2_l))\nn\\
&+&2i\epsilon_{\mu\nu\alpha\beta}(k-k')^\alpha s^\beta\,.
\end{eqnarray}
Here the spinor normalization $s^2=-m_l^2$ is used.
In the following the lepton mass $m_l$ is neglected.
For a  spin-1/2 target the representation of $\mathbf{W^{\mu\nu}}$ requires 
four \textit{structure functions}  to describe the nucleon's internal structure. 
It can be written as~\cite{Bjorken:70,close,first-spin-review}:
\begin{eqnarray}
\!\!\!\!\!\!\!{ \mathbf{W^{\mu\nu}_\frac{1}{2}}}(S)\!\!\!&=&\!\!\left(-g^{\mu
\nu}-\frac{q^\mu q^\nu}{Q^2}\right)F_1\nn\\
&+&\!\!\left(P^\mu + \frac{P\cdot q }{Q^2}q^\mu\right)
\left(P^\nu + \frac{P\cdot q }{Q^2}q^\nu\right)
\frac{F_2}{P\cdot q}\nn\\
&-&\!\!\!i\epsilon^{\mu\nu\lambda\sigma}\frac{q_\lambda}{P\cdot q}
\left[S_\sigma g_1+\!\left(
S_\sigma\! -\frac{S\cdot q }{P\cdot q} P_\sigma\right)\!g_2\right]\!\!
.
\end{eqnarray}
\noindent 
Here, $F_1$ and $F_2$ are polarization-averaged structure functions
(in the following also called 'unpolarized'),
while $g_1$ and $g_2$ are spin structure functions, all
depending on both $x$ and $Q^2$, which have been suppressed here for simplicity.
The sensitivity of
the cross section   to $g_1$ and $g_2$ arises from the product of the
anti-symmetric parts of the $\mathbf{L_{\mu\nu}}$ and
$\mathbf{W^{\mu\nu}}$ tensors,  which is 
non-zero only when both target and beam are polarized. 
\renewcommand{\baselinestretch}{1.}
\begin{table}[t!]
{\renewcommand{\baselinestretch}{1.}\caption{\label{table:kinevar}Kinematic variables used in the description of deep-inelastic scattering.}}
\begin{ruledtabular}
 \begin{tabular}{ll}
\parbox{0.40\columnwidth}{\raggedright $ m_l$ }                             & \parbox{0.55\columnwidth}{\raggedright Mass of incoming lepton (considered as negligible)  }\\[4mm]  
\parbox{0.40\columnwidth}{\raggedright $ M$ }                               & \parbox{0.55\columnwidth}{\raggedright Mass of target nucleon                              }\\[4mm]  
\parbox{0.40\columnwidth}{\raggedright $k=(E,\vec{k})$, $k'=(E',\vec{k'})$} & \parbox{0.55\columnwidth}{\raggedright 4--momenta of the initial and final state leptons   }\\[4mm]
\parbox{0.40\columnwidth}{\raggedright $ s$, $S$ }              & \parbox{0.55\columnwidth}{\raggedright Lepton's and target's spin 4-vectors                }\\[4mm]  
\parbox{0.40\columnwidth}{\raggedright $\theta,\; \phi$}                    & \parbox{0.55\columnwidth}{\raggedright Polar and azimuthal angle of the scattered lepton   }\\[4mm]
\parbox{0.40\columnwidth}{\raggedright $P\stackrel{\mathrm{lab}}{=}(M,0)$}  & \parbox{0.55\columnwidth}{\raggedright 4--momentum of the initial target nucleon           }\\[4mm]
\parbox{0.40\columnwidth}{\raggedright $q=(E-E',\vec{k}-\vec{k'})$ }        & \parbox{0.55\columnwidth}{\raggedright 4--momentum of the virtual photon                   }\\[4mm]
\parbox{0.40\columnwidth}{\raggedright $Q^2=-q^2$\\\hspace*{1em}$\stackrel{\mathrm{lab}}{\approx} 4EE'\sin^2\frac{\theta}{2}$ }& \parbox{0.55\columnwidth}{\raggedright Negative squared 4--momentum transfer   }\\[4mm]
\parbox{0.40\columnwidth}{\raggedright $ \displaystyle{\nu=\frac{P\cdot q}{M}\stackrel{\mathrm{lab}}{=} E-E'}$ }                              & \parbox{0.55\columnwidth}{\raggedright Energy of the virtual photon in the target rest frame }\\[4mm]
\parbox{0.40\columnwidth}{\raggedright $ \displaystyle{x=\frac{Q^2}{2\,P\cdot q}=\frac{Q^2}{2\,M\nu}}$ }             & \parbox{0.55\columnwidth}{\raggedright Bjorken scaling variable }\\[4mm]
\parbox{0.40\columnwidth}{\raggedright $ \displaystyle{y=\frac{P\cdot q}{P\cdot k}=\frac{\nu}{E}}$ }                 & \parbox{0.55\columnwidth}{\raggedright}\\[4mm]
\parbox{0.40\columnwidth}{\raggedright $W^2=(P+q)^2$\\\hspace*{2em} $=M^2+2M\nu-Q^2$}                 & \parbox{0.55\columnwidth}{\raggedright Squared invariant mass of the photon--nucleon system }
\end{tabular}
\end{ruledtabular}
\end{table}
\renewcommand{\baselinestretch}{1.}

For a spin-1 target such as the deuteron, the hadronic tensor
has four additional structure functions arising from
its electric quadrupole structure~\cite{Jaffe, Sather}. 	
Only three appear at leading-twist level: $b_1$, $b_2$ and $\Delta$. 
 In the scaling (Bjorken) limit the structure function 
$b_2$ is related to  $b_1$ by  $b_2(x)=2xb_1(x)$; $\Delta$
describes the double helicity-flip (virtual) photon-deuteron
amplitude~\cite{Jaffe_delta}.
The structure function  $b_1$ appears in a product with  the
tensor polarization of the target,  which can coexist
 with the vector polarization in spin-1 targets.
The influence of the tensor polarization on the
$g_1$ measurement is discussed in Sect.~\ref{subsec:extraction-asy}. In this
analysis the unmeasured  function $\Delta$ is neglected
since its contribution is suppressed for longitudinally
polarized targets.

\bigskip

The structure function $g_1$ is related directly 
to the cross section  difference:
\begin{equation}
\label{eq:formalism-sigLL}
\sigma_{LL} \equiv  \frac{1}{2}(\sigma^\sant - \sigma^\spar) \,,
\end{equation}
where longitudinally ($L$) polarized leptons ($\rightarrow$) scatter on
longitudinally  ($L$) polarized nuclear targets with polarization direction
either parallel  or anti-parallel ($\spar$, $\sant$) to the spin
direction of the beam.
The relationship to spin structure functions is:
\begin{multline} \label{eq:sigLL}
\frac{\detwo\sigma_{LL}(x,Q^2)}{\de x \de Q^2} =
\frac{8\pi\alpha^2y}{Q^4} \\
\times \left[
\left( 1-\frac{y}{2} -\frac{y^2}{4}\gamma^2\right)\, g_1(x,Q^2) 
- \frac{y}{2}\gamma^2 \, g_2(x,Q^2) \right]\,,
\end{multline}
where $\gamma^2=Q^2/\nu^2$.
The spin 
structure function $g_2(x,Q^2)$ does not have any probabilistic 
interpretation in the QPM. It will not be discussed 
further in this paper, but it is taken into account in the extraction 
of $g_1$ by using
a parameterization of the published data. The 
second term is small compared to the first.  
Averaged  over all $(x,Q^2)$ bins of this analysis it is of order 
0.54$\%$ for the proton and 1.9$\%$ for the deuteron.
Therefore the existing precision for $g_2$ has only a
marginal effect on an extraction of $g_1$.

For only the purely technical reason that
absolute cross sections are difficult to measure,
asymmetries are the usual direct experimental observable:
\begin{equation}
\label{eq:formalism-Apardef}
\Apar    \equiv  \frac{\sigma_{LL}}{\sigma_{UU}} \,,
\end{equation}
where $\sigma_{UU}$ is the polarization-averaged cross 
section\footnote{ In presenting the extraction of spin observables of interest 
from the experimental data, we depart from the traditional 
formalism with the purpose of making a clearer distinction 
between technical issues and spin physics.  In particular, 
we avoid entanglement in a so-called `depolarization factor'
of the crucial spin dependence of the leptonic tensor with 
one of several parameters used to represent previous data 
for $\sigma_{UU}$ that are needed to convert $A_{||}$ into 
the needed $\sigma_{LL}$. }, wherein
the subscript $UU$ indicates that both beam and target are unpolarized.
Parity conservation implies 
$\sigma_{UU} = \sigma_{LU} \equiv \frac{1}{2}(\sigma^\sant + \sigma^\spar)$.
Values of absolute polarization-averaged DIS cross sections 
$\sigma_{UU}$ are introduced from previous experiments that 
were designed to measure them precisely.  These cross sections
are all that is needed to extract the
spin structure function $g_1$ at the {\em measured} combinations of
$x$ and $Q^2$.  They are also needed in the process of correcting 
the measured asymmetry $A_{||}^m$ for higher order QED (radiative) 
effects to obtain the asymmetry $A_{||}$ at Born level.  This is
because the polarization-averaged yields must be normalized to known
values of $\sigma_{UU}$ in order to subtract QED radiative background 
calculated as absolute cross sections.

Substituting $\sigma_{LL}=\sigma_{UU} A_{||}$ into Eq.~(\ref{eq:sigLL}),
and solving for $g_1$, we obtain
\begin{multline} \label{eq:g1fromApar}
g_1(x,Q^2) = 
\frac{1}{1-\frac{y}{2}-\frac{y^2}{4}\gamma^2} \\
\times \left[\frac{Q^4}{8\pi\alpha^2y}\frac{\detwo\sigma_{UU}(x,Q^2)}
{\de x \de Q^2} \Apar(x,Q^2)
+ \frac{y}{2} \gamma^2 \, g_2(x,Q^2) \right]\,.
\end{multline}

The  structure functions $g_1^p$ and $g_1^n$ on proton and
neutron targets are related to that of the deuteron by the
relation:
\begin{equation}\label{g1dfromg1pn}
{ {{g_1^d}}}~=~\frac{1}{2}({ {{g_1^p}}}
+{ {{g_1^n}}})\left(1-\frac{3}{2}\omega_D \right)\,,
\end{equation}
where $\omega_D$ takes into account 
the D-state admixture to the deuteron wave function. A value of 
$\omega_D=0.05\pm 0.01$ is used, which covers most of the available estimates~\cite{omegad:lacombe:1981,omegad:Buck:1979,omegad:Zuilhof:1980,omegad:Machleidt:1987,omegad:Umnikov:1994}.
In this paper the neutron structure function $g_1^n$ is
evaluated according to Eq.~(\ref{g1dfromg1pn}).

Alternatively, $g_1^n$ can be obtained, e.g., from  measurements on a
polarized $^3$He target: 
\begin{equation}
g_1^{^3\textrm{He}} \simeq P_n g_1^n +2P_p g_1^p\,,
\end{equation} 
where the effective
polarizations of neutron and proton are $P_n=0.86\pm 0.02$ and
$P_p=-0.028\pm 0.004$~\cite{neutronpol}; i.e., the polarized $^3$He acts effectively as a
polarized neutron target. 
The HERMES measurement of $g_1^n$ off $^3$He has been previously
published in \cite{g1n_hermes}.


\section{\label{sec:hermes}The Experiment}

The HERA facility at DESY comprises a proton and a lepton 
storage ring. 
HERMES is a fixed-target experiment using exclusively the lepton ring, 
which can be filled with either electrons or positrons (only positron
data are used for this analysis), while the proton beam 
passes through the non-instrumented horizontal mid-plane of the spectrometer. 
Internal to the lepton ring, an open-ended storage cell is installed that can be 
fed with either polarized or unpolarized target gas. The three major components 
to the HERMES experiment (beam, target, spectrometer) are briefly described in 
the following, while detailed descriptions can be found elsewhere
(\cite{Ackerstaff:1998av,Bernreuther:1998qm,Avakian:1998bz,Andreev:2001kr,Brack:2001qy,Benisch:2001rr,Akopov:2000qi,Airapetian:2004yr,Barber:1995ew,Beckmann:2002}).

\subsection{\label{subsec:hermes-beam}Polarized {HERA} Beam}

Spin rotators and polarimeters
are essential components of the HERA lepton beam.
They are described in detail in 
Refs.~\cite{Buon:1986,Barber:1993,Barber:1994,Beckmann:2002}.
The initially unpolarized beam becomes transversely polarized by an
asymmetry  in the emission of
synchrotron radiation  associated with a spin  
flip (Sokolov-Ternov mechanism~\cite{sokolov-ternov}). 
The beam polarization grows  
and approaches asymptotically an equilibrium value, 
with a time constant depending on the characteristics
of the ring, typically over 30-40 minutes. 
The transverse guide field generates transverse
beam polarization 
in the ring, while spin rotators in front of  and behind  the experiment provide longitudinal 
polarization at the interaction point and at one of the two beam polarimeters.

The two HERA beam polarimeters are based on Compton back-scattering of
circularly polarized laser light.
The transverse  beam polarization at the opposite side of the ring 
causes an up-down asymmetry
in the direction  of the back-scattered photons. The resulting
position  asymmetry is  measured by the top and bottom halves of the
lead-scintillator sampling calorimeter of the transverse polarimeter~\cite{Barber:1993},
with  fractional systematic uncertainties of 3.5\% (1996-97 data). 
The longitudinal beam polarization in the region of the experiment 
leads to an asymmetry in the energy
of the back-scattered Compton photons measured  in the NaBi(WO$_4$)$_2$ crystal 
calorimeter of the longitudinal polarimeter~\cite{Beckmann:2002},
with  fractional systematic uncertainties of 1.6\% (2000
data). The average beam polarization was typically larger than 0.5.

In order to account for  the time dependence of the beam polarization,
its continuously monitored values are used in the
analysis. The average values of the beam polarizations for each year
covered in this paper are shown in Tab.~\ref{tab:beampol}.
The numbers are weighted by the
luminosity so that the value near the beginning of each fill
dominates. 
{
\begin{table}[b]
{\renewcommand{\baselinestretch}{1.}
\caption{\label{tab:beampol}
Average value and uncertainty of the beam 
  polarization for each year of measurements.}}
\begin{ruledtabular}
\begin{tabular}{c|ccc}
$~~~~~~~~~~$Year$~~~~~~~~~~$&\multicolumn{3}{c}{Average Polarization} \\
\hline
1996   & ~~ &   $0.528  \pm 0.018$  & ~~\\
1997   & ~~ &   $0.531  \pm 0.018$ & ~~ \\
2000   & ~~ &   $0.533  \pm 0.010$ & ~~ \\
\end{tabular}
\end{ruledtabular}
\end{table}

\begin{figure}[b!]
\center{
\includegraphics[width=\columnwidth]{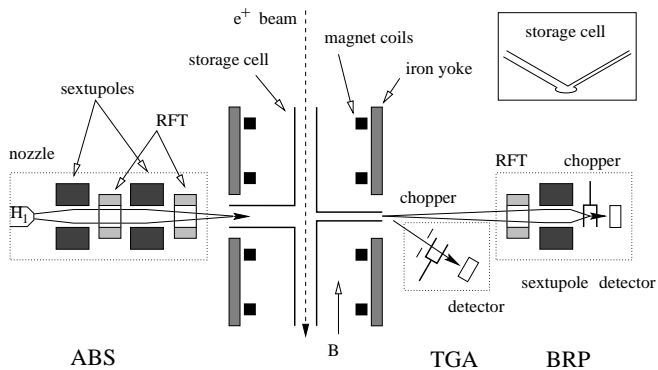}
\caption[Schematic view of the target]{ Schematic view of the
longitudinally polarized target. From left to right: Atomic Beam
Source (ABS) containing Radio-Frequency Transitions (RFT), target
chamber with cell and magnet, diagnostic system composed by Target Gas Analyzer
(TGA), and Breit-Rabi Polarimeter (BRP).}
\label{fig:target}}
\end{figure}

\subsection{\label{subsec:hermes-target}The HERMES Polarized Gas Target }

The combination of a fixed target with the HERA lepton
storage ring requires the employment of a gaseous target internal to
the beam line.
This has the  advantages
of being almost free of dilution (the fraction $f$ of
polarizable nucleons being close to 1), of providing a high degree
of vector polarization ($P_z>0.8$), and of being able to invert
the direction of the spin of the nucleons within milliseconds.

The HERMES longitudinally polarized gas target~\cite{Airapetian:2004yr}, 
schematically shown in Fig.~\ref{fig:target}, consists of an Atomic 
Beam Source (ABS)~\cite{Baumgarten:2003yp} which produces a polarized
jet of  atomic 
hydrogen or deuterium and focuses it into a thin-walled storage cell
along  the beam 
line~\cite{Baumgarten:2003yo}. The atomic gas is produced in a
dissociator and is formed into a beam using a cooled nozzle,
collimators and a series of differential pumping stations. 
A succession of magnetic sextupoles and radio-frequency
fields are used to select (by Stern-Gerlach separation) and exchange
(by radio-frequency transitions) the atomic hyperfine states that have
a given nuclear polarization to be injected into the cell. The storage
cell,  inside  the HERA beam pipe, is a windowless 40~cm long 
elliptical tube, coaxial to the beam, 
with 75~$\rm \mu m$ thick Al walls coated to inhibit surface recombination and
depolarization.
The use of the storage cell technique results in a typical areal density increase
of about two orders of magnitude compared to a free jet target.
A sample of gas (ca.~5$\%$) diffuses from the middle of the cell into a 
Breit-Rabi Polarimeter (BRP)~\cite{Baumgarten:2001ym} which measures
the atomic  polarization, or into a Target Gas Analyzer
(TGA)~\cite{Baumgarten:2003yq} which  
measures the atomic and the molecular content of the sample. 
A magnet surrounding the storage cell provides a holding field 
defining the polarization axis and prevents spin relaxation via spin exchange 
or wall collisions by effectively decoupling the magnetic moments of electrons 
and nucleons.
A gaseous helium cooling system keeps the cell temperature at the
lowest value for which atomic recombination and spin relaxation during
wall collisions are minimal.

The vector polarization $P_z$ is defined as 
$P_z\equiv(n^+ - n^-)/(n^+ + n^-)$ and 
$P_z\equiv(n^+ - n^-)/(n^+ + n^- + n^0)$ for spin-1/2 and spin-1 
targets, respectively. Here $n^+$, $n^-$ and $n^0$ are the atomic 
populations with positive, negative and zero spin projection on the 
beam direction.
 The sign of the target polarization is randomly chosen each 60~s 
for hydrogen and 90~s for deuterium. The target parameters are
measured for each such interval.
The rapid cycling of the target 
polarization  reduces the systematic 
uncertainty in the measured spin asymmetries related to
the stability of the experimental setup.
Due to the very stable performance of the target 
operation, luminosity-average polarization values are  
used in the analysis.

\vspace{0.5cm}	

\noindent{\it Hydrogen data}\\
\noindent During the years 1996-97 a  longitudinally polarized hydrogen target
was employed at a  nominal temperature $T_{cell}=100\,$K. 
The average target polarization for the year 1997
reached the value $P_z^+=P_z^-=0.851\pm0.032$; the average target areal
density was determined to be $7.6\cdot 10^{13}$~nucleons/cm$^{2}$.
The target polarization for the year 1996, which contributes only 
about 27$\%$ to the total statistics, was determined from the
normalization of the 1996 inclusive asymmetry to that of  1997.
For this limited data-set this method provides a smaller uncertainty on the
target polarization w.r.t. the direct measurement.
In 1997, a set of data at higher temperature ($T_{cell}=260$~K) was 
collected in order to measure the polarization of the recombined 
molecules~\cite{Airapetian:2004ys}, thus reducing the systematic
uncertainty on the target polarization measurement by a factor close 
to 2.

\begin{figure*}[!t]
\center{
\includegraphics[width=\textwidth]{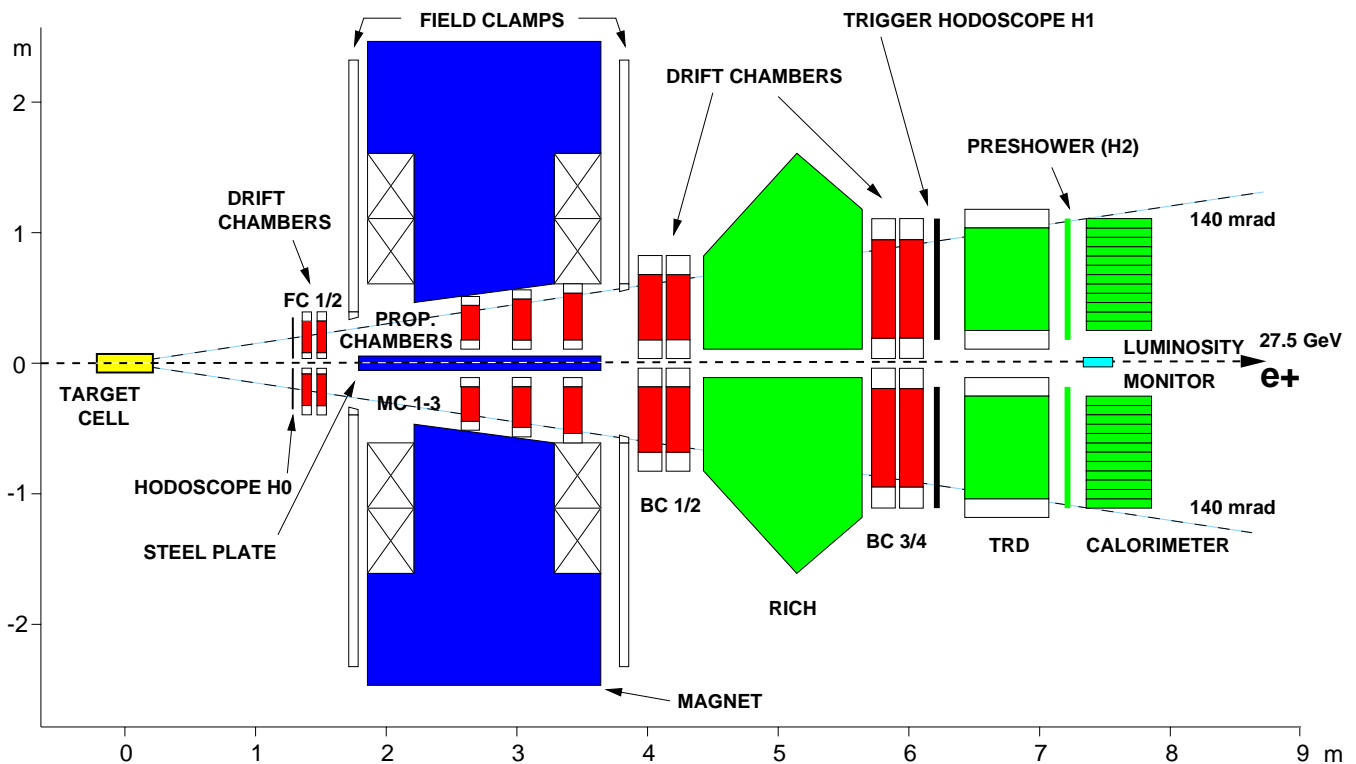}
\caption[The HERMES spectrometer.]{A schematic side view of the HERMES spectrometer.
}\label{fig:spectrometer}
}
\end{figure*}

\vspace{0.5cm}

\noindent{\it Deuterium data}\\
\noindent
From the end of 1998 till the end of  2000 deuterium was used as target material.
The data taken in 1998 and 1999 have been excluded from the present analysis
because of the overwhelming statistics of the  2000 data.
In the year 2000 a  target  cell with smaller cross section 
was installed with reduced nominal temperature $T_{cell}=60\,$K,
thus increasing the target density by a factor of 2.
The average values of the target vector polarization were 
 $P_z^+=0.851\pm0.031$ and $P_z^-=0.840\pm0.028$. The two  polarization 
values  are due to
different injection efficiencies of the ABS.
The average  target areal density was determined to be 
$2.1\cdot 10^{14}$~nucleons/cm$^{2}$.
In a dedicated running period, about 3.5 million deep-inelastic
scattering events were taken with tensor polarization, allowing the
first measurement ever of the tensor structure function $b_1^d$
and the first assessment of the effect on the $g_1^d$ measurement 
by the coexistent  tensor polarization of the deuteron target
(see Sect.~\ref{subsec:extraction-asy}). 

Tab.~\ref{tab:targetpol} shows the average polarizations of the target
and their 
uncertainties for the three data sets used in this analysis. The uncertainties
are  dominated by systematics, which depend on varying running conditions 
and quality of the target cell surface.
\renewcommand{\baselinestretch}{1.3}
\begin{table}[th]
{\renewcommand{\baselinestretch}{1.}\caption{\label{tab:targetpol}
Average magnitude of the target vector 
  polarization $P_z$ for the various data sets.}}
\begin{ruledtabular}
\begin{tabular}{c|c|c|c}
~~~Year~~~&Target gas~~~&Sign of $P_z$~~~~~&$|P_z|$~~ \\
\hline
1996   &  H   & $\pm$ &  $0.759$  $ \pm $   $0.032 $   \\
1997   &  H   & $\pm$ &  $0.851$  $ \pm $   $0.032 $   \\
2000   &  D   & $+$   &  $0.851$  $ \pm $   $0.031 $  \\
2000   &  D   & $-$   &  $0.840$  $ \pm $   $0.028 $   \\
\end{tabular}
\end{ruledtabular}
\end{table}
\renewcommand{\baselinestretch}{1.}

\subsection{\label{subsec:hermes-spectr}The HERMES Spectrometer}

A detailed description of the HERMES spectrometer (see
Fig.~\ref{fig:spectrometer}) is given in  
Ref.~\cite{Ackerstaff:1998av}. It constitutes a forward
spectrometer with multiple tracking stages before and after a 1.5~Tm dipole 
magnet, and good particle identification (PID) capabilities. 
A horizontal iron plate shields the HERA beam lines from the
dipole field, thus dividing the spectrometer in two identical halves,
top and bottom.
The  geometrical acceptance of
$\pm170$~mrad horizontally and $\pm(40-140)$~mrad vertically
results in detected scattering angles ranging from 40 to 220~mrad.

In each spectrometer half, the intersection points of charged particle 
trajectories with the 36  planes  of the Front Chambers (FC 1-2)
 and Back Chambers (BC 1-4)
are used  for track reconstruction in space. 
These  detectors are horizontal-drift chambers
with alternating cathode and anode wires between two cathode foils, all
operated with the gas  
mixture Ar:CO$_2$:CF$_4$ (90:5:5), the average drift velocity  being
7~cm/$\mu s$.  
They are assembled in 
modules of six layers in three coordinate doublets ($XX', UU'$, and
$VV'$), where  
the primed planes are offset by half a cell width to help resolve left-right
ambiguities. 
The cell width is  7~mm for 
FCs~\cite{Brack:2001qy}, while it is 15~mm for  
BCs~\cite{Bernreuther:1998qm} behind  the magnet. 

The proportional wire chambers {\it MC1-3}, also shown in 
Fig.~\ref{fig:spectrometer}, are not included in the 
tracking algorithm~\cite{Ackerstaff:1998av} used for this analysis.

This tracking algorithm determines 
partial tracks before (front track) and after the magnet (back track).
The track projections are found in a fast tree search and then
combined to determine the particle momentum.
The algorithm  uses the intersection point of the back track with the
magnet mid-plane to refine front tracks. From their scattering angles
and positions the event vertex is determined, while the  
back tracks are also used to identify hits in the PID detectors. Monte Carlo 
simulations show that the intrinsic momentum resolution $\Delta p/p$ is 
between 0.015 and 0.025 over the accessible momentum range.
The resolution for hydrogen was better  than for deuterium as the
RICH detector installed before the deuterium data taking period
introduced some additional material. 

The scattered positron is identified through a combination of 
a lead-glass calorimeter, a pre-shower detector, a  
Transition-Radiation Detector (TRD), and the \v{C}erenkov detectors. 
(While the threshold \v{C}erenkov was
used in 1996-97 primarily for pion identification, the
Ring-Imaging \v{C}erenkov (RICH) detector~\cite{Akopov:2000qi,rich2} was used 
thereafter to identify pions, kaons, and protons.)

The Calorimeter~\cite{Avakian:1998bz} is used to suppress hadrons by a factor of 10 
at the trigger level and
 a factor of 100 in the off-line analysis, to measure the
energy of electrons/positrons and also of photons. 
Each half consists of 42x10  
blocks of radiation-resistant F101 lead glass~\cite{calo2}. Each block has a cross
section of 9x9~cm$^2$ and 50~cm depth, and 
is viewed from the back by a photo-multiplier tube. 
The calorimeter's resolution was measured to be 
$\sigma(E)/E [\%] = (5.1 \pm 1.1)/\sqrt{E [GeV]} +  
(2.0\pm 0.5)+(10.0\pm 2.0)/E$~\cite{Avakian:1998bz}. 

The scintillator hodoscope H2, consisting of 42 vertical 9.3x91~cm$^2$ 
`paddles' of 1~cm thickness, forms the pre-shower detector in combination
with two radiation   
lengths of lead preceding it. As pions deposit only about 2~MeV in H2, 
as compared to 20-40~MeV for leptons, the pion contamination 
 can be reduced by more than a factor of 10 with 95\%
efficiency if this detector were used alone. 

The TRD rejects hadrons 
by a factor of more than 100 at 90\% 
electron/positron efficiency, if used alone. A TRD half comprises six proportional
wire chambers to detect the  photons from transition radiation in 
the preceding radiator. All TRD proportional
chambers  use Xe:CH$_4$ (90:10) gas.

The luminosity monitor~\cite{Benisch:2001rr} 
 detects in coincidence
$e^+e^-$ pairs originating from Bhabha scattering of the beam positrons off
electrons from the target atoms, and also $\gamma\gamma$ pairs from $e^+e^-$
 annihilations.
It consists of two small calorimeters made of
highly radiation resistant NaBi(WO$_4$)$_2$ crystals covering a
horizontal acceptance  of 4.6-8.9~mrad. 
They are mounted to the right and left of the beam pipe, 7.2~m downstream of the target. 


\section{\label{sec:extraction}Data Analysis}

The inclusive data sample is selected from the recorded events to
satisfy the following criteria:
\begin{itemize}
\item  there exists a trigger composed of signals in the Calorimeter  and in the 
hodoscopes H0, H1 and Pre-Shower (H2) (see Fig.~\ref{fig:spectrometer}),
\item data quality criteria are met,
\item the particles identified by the Particle Identification scheme
as leptons are selected,
\item the highest momentum lepton in the event originating from the
  target region is selected,
\item geometric and kinematic constraints are applied. 
\end{itemize}


\subsection{The kinematic {\bf range}}

The kinematic range of the events selected for this analysis is shown in
Fig.~\ref{fig:xq2plane}, together with the requirements imposed  on the
kinematic variables. 
\begin{figure}[!b]
\includegraphics[width=\columnwidth]{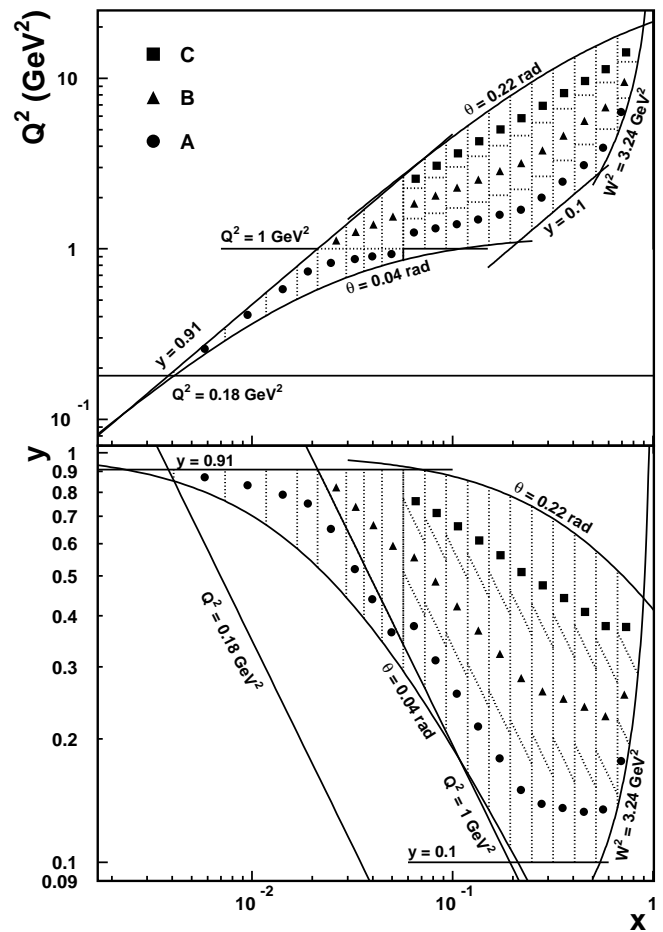}
\caption{\label{fig:xq2plane}
Kinematic   $x-Q^2$ and $x-y$  planes covered by this analysis. 
The symbols represent the average values of $(x,Q^2)$ (top panel) and
$(x,y)$  (bottom panel) in each bin.
Subdivision of the  $x$-bins along $Q^2$ is denoted by different
symbols (A, B, C).}
\end{figure}%
The aperture of the spectrometer  limits the acceptance
to scattering angles $0.04 \leq \theta \leq 0.22$~mrad.
The constraint $y>0.1$ is used to exclude regions where the momentum
resolution starts to degrade~\cite{Ackerstaff:1998av}. The constraint $y\leq 0.91$
discards the low momentum region ($E<2.5$ GeV) where the trigger
efficiencies have not yet reached a momentum plateau. 
The requirement $W^2>(1.8)^2$ GeV$^2$ suppresses the region of baryon 
resonances. 
The resulting ($x,Q^2$) region, defined by $0.0041\leq x \leq 0.9$ and $0.18$~GeV$^2$~$\leq Q^2\leq 20$~GeV$^2$, 
 was divided into 19 bins in $x$,
guided by the available statistics. Furthermore most $x$-bins are
subdivided into up to
3 bins in $Q^2$, designated A, B and C bins.  
The purpose of this is to
allow appropriate statistical weighting of these $Q^2$ bins, thereby 
exploiting 
the higher figure of merit at larger $Q^2$, which is due to both the
larger polarization transfer from the incident lepton to the virtual
photon, and the smaller kinematic smearing between $x$ bins.
Logarithmically equidistant bin boundaries  are used along  the $x$ axis in the
region $0.044<x<0.66$, and along the $Q^2$ axis in the region $x>0.06$.


\subsection{Particle Identification}\label{subs:PID}

The lepton (positron and electron) identification is 
achieved with a probability analysis based on the 
responses of the TRD, the pre-shower detector, 
the Calorimeter, and the \v{C}erenkov detectors, described in Sect.~\ref{subsec:hermes-spectr}. 

The requirement used to identify a lepton is:
\begin{equation}\label{her:eq:pid1}
PID=\log_{10}\frac{P(l|{\cal R},p,\theta)}{P(h|{\cal R},p,\theta)} \,>\, {PID}_{cut},
\end{equation}
where $P(l(h)|{\cal R},p,\theta)$ is the conditional probability that the particle is
a lepton $l$ or a hadron $h$, given the combined  response set $\cal R$ from
all the Particle Identification detectors, the momentum
$p$, and the scattering angle $\theta$. 
Bayes' theorem relates $P(l(h)|{\cal R},p,\theta)$ to the probability 
$P(l(h)|p,\theta)$ that a particle with momentum $p$, scattered at
polar angle $\theta$, is a lepton (hadron), and the probability
$P({\cal R}|l(h),p)$ that 
a lepton (hadron) with momentum $p$ causes  the combined signal ${\cal R}$.
The $PID$ can be re-written as:
\begin{equation}\label{her:eq:pid2}
PID\!=\log_{10}\frac{P({\cal R}|l,p)}{ P({\cal R}|h,p)}\cdot 
\frac{P(l|p,\theta)}{ P(h|p,\theta)} =PID'-\log_{10} \Phi(p,\theta).
\end{equation}
Here $\Phi(p,\theta)$ is the ratio between the hadron and lepton 
fluxes impinging onto the detector:
\begin{equation}
\Phi(p,\theta)=\frac{P(h|p,\theta)}{ P(l|p,\theta)}\,.
\end{equation}
The quantity $PID'$ is defined combining the responses ${\cal R}_D$  of each detector $D$
used for particle identification:
\begin{equation}
PID'=\log_{10} \prod_D \frac{P_D({\cal R}_D|l,p)}{ P_D({\cal R}_D|h,p)}.
\end{equation}
Under the approximation that the responses of the particle
identification detectors are uncorrelated, 
the distribution $P_D({\cal R}_D|l(h),p)$ of detector $D$, 
i.e. the typical detector response for leptons (hadrons), can be measured
by placing very restrictive cuts on the response of the other PID detectors
to isolate a very clean sample of a particular particle 
type~\cite{juergen_thesis}.

The \textit{hadron contamination} is defined as the number
of hadrons with $PID>PID_{cut}$ divided by the number of
identified leptons,
and the \textit{lepton efficiency} is defined as the number of identified leptons 
over the total number of leptons.
The choice $PID_{cut}=1$ used in this analysis optimizes the tradeoff between efficiency
and purity in the sample.
For this choice, hadron contaminations are less than 0.2$\%$ 
over the entire $x$ range, and  
lepton efficiencies larger than 96$\%$, assuming that the detector
responses are uncorrelated.


\subsection{\label{subsec:extraction-asy}Inclusive Asymmetries}

The yield averaged over target polarization state is:
\begin{equation}\label{g1d:n0}
N(x,Q^2)\!=\!\sigma_{UU} (x,Q^2)\!\int\dt a(x,Q^2)~\varepsilon
(t,x,Q^2)\tau(t)L(t), 
\end{equation}
where $t$ is time and $\tau(t)$ is the live-time factor,
which is typically 0.97 for this analysis,  
$\sigma_{UU}$ is the  polarization-averaged cross section,
$a(x,Q^2)$ is the detector acceptance, 
$\varepsilon (t,x,Q^2)$ is  the total detection efficiency
(tracking and trigger) and $L(t)$ is the luminosity.
In the case of double-polarized scattering this relation becomes:
\begin{multline}\label{eq:crosssect}
\hspace{-0.3cm}N^{\spar(\sant)}(x,Q^2)=a(x,Q^2)\, \sigma_{UU}(x,Q^2) 
\int\dt~\varepsilon (t,x,Q^2)~  \\
\times \tau(t) L^{\spar(\sant)}(t)[1+(-) |P_B(t)P_z(t)|~ A_{||}(x,Q^2)]\,,
\end{multline}
where $P_B$ and $P_z$ are the beam and target polarizations.
For a spin-1 target Eq.~(\ref{eq:crosssect}) contains an additional
term depending on the tensor polarization, and this case will be
treated later in this section.

The  measured asymmetry $A_{||}^m$ is therefore:
\begin{equation}\label{eq:asyratio}
A_{||}^m=\frac{N^{\spar} \int\dt \varepsilon\tau
L^{\sant}
-N^{\sant}\int\dt \varepsilon\tau
L^{\spar}}{ N^{\spar} \int\dt \varepsilon\tau
L^{\sant}P_BP_z
+N^{\sant}\int\dt \varepsilon\tau
L^{\spar}P_BP_z
},
\end{equation}
where the dependences on $x$, $Q^2$ and $t$ have been suppressed for
simplicity.
Since the target  changes every 60~s (90~s) for hydrogen
(deuterium)  between the two polarization
states, any variation  in the efficiencies  can be safely assumed
to be the same in the 
anti-aligned and aligned configurations of  beam and target polarizations,
implying that they cancel in the ratio if the measurement is fully
differential in the kinematics.  
In a simulation made with  Monte Carlo data
covering a $4\pi$ acceptance, it has been confirmed that the asymmetry is
not significantly affected by the limited acceptance 
or non-uniform efficiency  of the detector. 

The measured asymmetry $A_{||}^m$ is obtained from the number of 
events in the 
 two polarization states as:
\begin{equation}
A_{||}^m(x,Q^2)=
\frac{N^{\sant}_\tx{}(x,Q^2)~\mathcal{{L}}^{\spar}-
N^{\spar}_\tx{}(x,Q^2)~\mathcal{{L}}^{\sant}
}{ 
N^{\sant}_\tx{}(x,Q^2)~\mathcal{{L}}_P^{\spar}+
N^{\spar}_\tx{}(x,Q^2)~\mathcal{{L}}_P^{\sant}
}\,,
\end{equation}
where the luminosities
$\mathcal{{L}}^{\spar(\sant)}$ and $\mathcal{{L}}_P^{\spar(\sant)}$ 
are defined as:
\begin{eqnarray}
\mathcal{{L}}^{\spar(\sant)}&=&\int\dt{L}^{\spar(\sant)} (t)~\tau(t) ,\nn\\
\mathcal{{L}}_P^{\spar(\sant)}&=&\int\dt{L}^{\spar(\sant)}(t)~|P_B(t) P_z(t)|~ \tau(t).
\end{eqnarray}

The asymmetries are ratios of yields integrated over bin
widths.  As the yields depend nonlinearly on $x$ and $Q^2$,
a question arises about the effect of the nonzero bin widths.
Using parameterizations for unpolarized and spin structure
functions, it was confirmed that there
is no significant difference
between values of $\Apar(x_i)$ calculated from
the yields integrated over the experimental
geometric acceptance for $Q^2$, 
and $\Apar(x_i,\langle Q^2\rangle_i)$ evaluated 
at $\langle Q^2\rangle_i$.

The various stages of the analysis are now introduced
in the order in which they are applied. 

\setcounter{paragraph}{0}

\vspace{0.4cm}

\noindent{\it Charge symmetric background}

\noindent The observed event sample is  contaminated by 
background from charge symmetric (CS) processes, such as 
meson Dalitz decays or photon conversions into $e^+e^-$ pairs.
Since  these $e^+$ and $e^-$ originate from secondary processes, they typically have 
lower momenta and  are thus concentrated at high $y$.
A correction for this background is applied in each kinematic bin
by subtracting the number of leptons with the charge opposite to that
of the beam particle.  
The charge symmetric background reaches up to 25\% at low $x$ and
becomes negligible at large values of $x$, as shown in Fig.~\ref{fig:cs}.  
\begin{figure}[!b]
\includegraphics[width=\columnwidth]{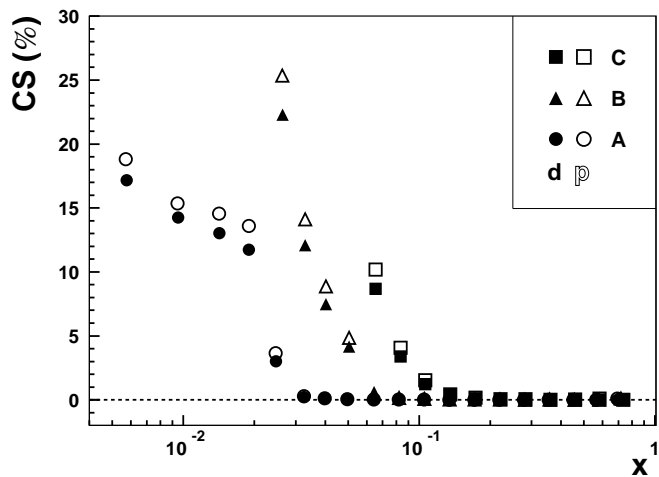}
\caption{\label{fig:cs}
Percentage of charge symmetric background in each $x-Q^2$ bin, for
the proton and the deuteron targets.   
Subdivisions A, B, and C are those defined in Fig.~\ref{fig:xq2plane}.
}
\end{figure}

\vspace{0.4cm}

\noindent{\it Hadron contamination}

\noindent The hadron contamination is less than $0.2\%$ 
over the entire $x$ range, so no
correction is required.

\vspace{0.4cm}

\noindent{\it Final data sample}

\noindent After data selection as discussed above, the numbers of events available for 
asymmetry analyses  on proton and deuteron are shown in Tabs.~\ref{table:events} 
and \ref{tab:totallumi}. 
\renewcommand{\baselinestretch}{1.3}
\begin{table}[!h]
{\renewcommand{\baselinestretch}{1.}\caption{\label{table:events}Numbers
of events in millions (corrected for charge-symmetric background) for the three data 
sets used, separated for top and bottom spectrometer halves and
various beam and target spin configurations.}}
\begin{ruledtabular}
\begin{tabular}{c|c|cccc}
Year   & Target & Top $\spar$ & Top $\sant$& Bottom $\spar$ &
Bottom  $\sant$ \\ 
\hline
1996   &   p    &     0.158   & 0.168     &  0.169   &  0.179   \\
1997   &   p    &     0.660   & 0.700     &  0.698   &  0.741   \\
2000   &   d    &     2.439   & 2.498     &  2.600   &  2.654   \\
\end{tabular}
\end{ruledtabular} 
\end{table}
\renewcommand{\baselinestretch}{1.}

\vspace{0.4cm}

\noindent{\it Normalization}

\noindent The luminosity $L$ is related to the beam current $I$, 
the electron charge $e$ and the areal target density $\rho$ by the relation
\begin{equation}
\frac{L(t)}{I(t)}=\frac{\rho}{e}.
\end{equation}
The areal target density $\rho$ was monitored to be a stable quantity 
independent of the target spin state, implying that 
the luminosity does not depend on the target polarization.
The ratio of luminosity-monitor rates to beam current was
averaged over spin states to eliminate the effect of the residual
electron polarization  in the target gas  
on the Bhabha rates measured by the luminosity-monitor. 
The average  was calculated separately for each 
data-taking period with uniform target and beam conditions, at least
for each HERA positron fill. 
The luminosity  calculated as the product of these averages with the beam current 
has been used for the extraction of the DIS asymmetry.

Tab.~\ref{tab:totallumi} shows the integrated luminosities for the
data sets used in this analysis.
\renewcommand{\baselinestretch}{1.3}
\begin{table}[th]
{
\renewcommand{\baselinestretch}{1}
\caption{\label{tab:totallumi} Integrated luminosities and total
numbers of events in millions for the three  data sets used in the analysis.
The yields per unit luminosity differ among the years because of varying 
calorimeter trigger thresholds.
}}
\begin{ruledtabular}
\begin{tabular}{c|c|c}
~~~~Year~~~~ & ~~~~Luminosity (pb$^{-1}$) ~~~~~~& Total events~~~~~\\
\hline
~~~~1996~~~~   &  ~12.6  & ~0.67~~~~~\\
~~~~1997~~~~   &  ~37.3  & ~2.80~~~~~\\
~~~~2000~~~~   & 138.7  & 10.19~~~~~\\
\end{tabular}
\end{ruledtabular}
\end{table}

\renewcommand{\baselinestretch}{1}

\vspace{0.4cm}

\noindent{\it Top-bottom asymmetries}

\noindent Since the HERMES detector consists of two symmetric halves, 
they are considered as two separate
and independent spectrometers, with individual application of data
quality criteria.
The asymmetry  $A_{||}$ is evaluated separately for the top and bottom halves.
This procedure allows polarization-independent systematic effects
present in each detector half to cancel independently.
The two asymmetries are tested to be consistent within their 
statistical uncertainties, and the final asymmetry is obtained as the weighted
average of the two.

\vspace{0.4cm}

\noindent{\it Stability checks}

\noindent The asymmetries measured on proton and deuteron have been
calculated as  
functions of time, beam current, target vertex and azimuthal angle, searching
 for possible systematic  deviations from the average value.
No significant effect was observed.

Geometrical constraints were also investigated, varying the
target vertex selection that ensures that the event
originated inside the target, and the polar angle constraint which limits
the angular acceptance. Again, no effects on the asymmetry were observed.

The helicity of the positron beam was reversed twice (six times)  during the
running periods of hydrogen (deuterium) measurement.
The asymmetry \Apar was found to be consistent within statistical
uncertainties when calculated separately for the two beam helicities.

\vspace{0.4cm}

\noindent{\it Unfolding of Radiative and Instrumental
Smearing}\label{sec:unfolding}

Radiative effects include vertex corrections to the
QED hard scattering amplitude, and kinematic migration
of DIS events due to radiation of real photons by the lepton.
Because only a fraction of the photons that may be radiated
by the initial or final state lepton can be detected in the HERMES
spectrometer, 
no attempt is made to identify and reject such radiative events.
Therefore, radiative corrections must be applied to the
experimentally observed asymmetries $A_{||}^m$.
These asymmetries are also affected by instrumental smearing
due to multiple scattering in target and detector material and
external  bremsstrahlung.
As illustrated in Fig.~\ref{binmigration},
a significant part of the events are not 
reconstructed inside the bin to which they 
belong according to their kinematics at the Born level. 
Events  migrating into other mostly adjacent bins 
 introduce a systematic
correlation between data points and may affect  the measured asymmetry.
\begin{figure}[b!]
\begin{center}
\includegraphics[width=0.6\columnwidth,angle=-90]{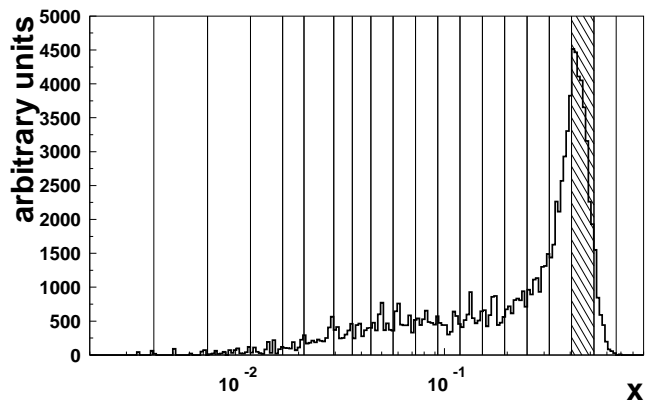}
\end{center}
\caption{\label{binmigration} 
The distribution of events originating from the arbitrarily 
selected $x$-bin shown as a shaded area, 
from a simulation of QED radiative effects and detector smearing,
using a proton target. The vertical lines indicate $x$-bin
 boundaries. }
\end{figure}

An incident positron can also radiate an energetic real photon while
scattering elastically on a proton or deuteron,
or quasi-elastically on a nucleon in the deuteron.
The final kinematics of these Bethe-Heitler (B-H) events
can be such that they pass the DIS analysis cuts. 
Such events represent a background to the usual DIS events, which 
has to be effectively subtracted.

In order to correct for these effects and retrieve the Born asymmetries, 
an unfolding algorithm has been applied (see App.~\ref{app:unfolding}). 
The radiative and detector-smearing effects are simulated  in a Monte
Carlo model yielding detailed information about how events 
migrate from one kinematic bin to another.
The background arising from B-H events is included in the simulation.
The Monte Carlo data samples used in this analysis have a statistical accuracy 
three times better than that of the measured data.

Schematically, in the unfolding algorithm the vector 
of Born asymmetries $A_{||}$ is obtained from the vector of
measured asymmetries $A_{||}^m$ by applying a
matrix that corrects for the smearing, while effectively subtracting
the radiative background. 
The expressions relating the measured and Born asymmetries are
given in  Eqs.~(\ref{eq:Bsol}--\ref{eq:Asol}).
The unfolding corrections depend on the Monte Carlo models 
for background, detector behavior and  asymmetries outside the measured region,  
and on the model for the polarization-averaged cross section. They do not depend 
on any model for the asymmetry in the measured region. 
Before unfolding, the experimental asymmetry data points 
contain events which  originate from other bins.
After unfolding, the data points are statistically correlated but
systematic correlations due to kinematic smearing have been removed,
resulting in a resolution in $x$ or $Q^2$ of a single-bin width.
This is a large improvement over the resolution function
shown in Fig.~\ref{binmigration}, which would still apply if a traditional
`iterative' method of applying radiative corrections were employed as in
Refs.~\cite{E142n,E154n,Abe:1998wq,SMC:1999pdlox,Anthony:1999rm}.
The unfolding algorithm provides the correlation matrix that should be used to 
calculate the statistical uncertainties on quantities obtained 
from the Born asymmetry, such as the weighted average 
of $g_1$ over  $Q^2$ bins  or the  integrals of $g_1$ over the measured range:
treating the uncertainties as uncorrelated  would result in overestimating
the uncertainty. 

In  the case of B-H background events,
the radiated photon  has a significant probability of
hitting the detector frames close to the beam line in the front
region of the HERMES  detector. As a consequence, an extensive
electromagnetic shower is produced causing a very high hit
multiplicity in the tracking detectors, making the track
reconstruction impossible~\cite{Emshower}.  This detector inefficiency
for B-H events was taken into account 
in order  to  not over-correct for radiative processes that
 are not observed in the spectrometer .
The efficiency $\varepsilon_{e.m.}$ for the detection of
B-H events was extracted with a dedicated Monte Carlo simulation
that includes a complete treatment of showers in material outside of
the geometric acceptance. 
\begin{figure}[!t]
\includegraphics[width=\columnwidth]{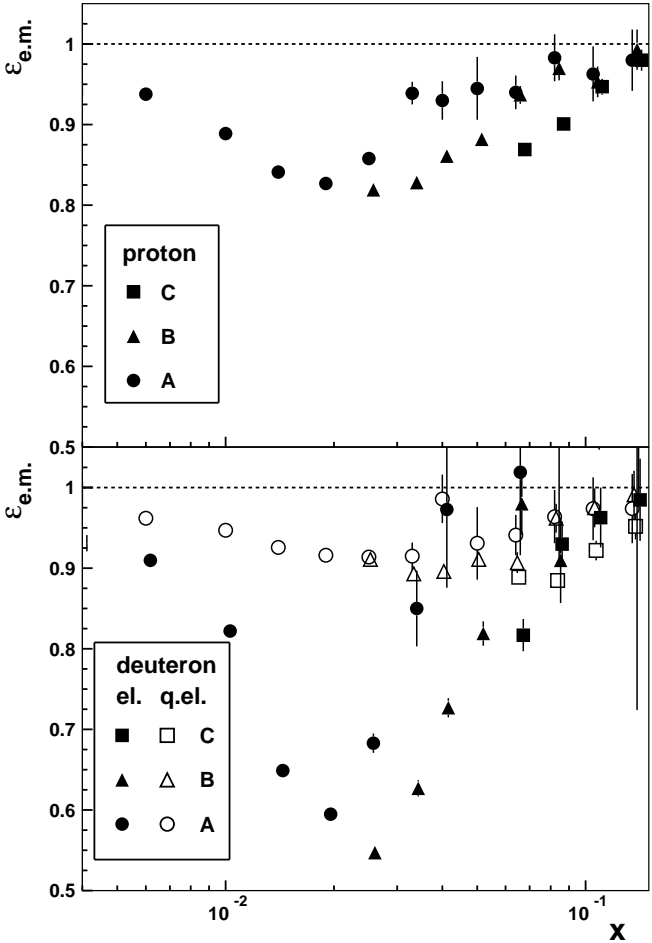} 
\caption{\label{fig:fem_epsilon}
Efficiency for the detection of B-H elastic (el.) and quasi-elastic 
(q.el.) events in the HERMES  
spectrometer, for scattering on both deuteron (d)  and proton
(p). The efficiency is set to unity for $x>0.1$.
Subdivisions A, B, and C are those defined in Fig.~\ref{fig:xq2plane}.
}
\end{figure}
Results for both proton and deuteron are shown
 in Fig.~\ref{fig:fem_epsilon}, where $\varepsilon_{e.m.}$ 
is plotted as a function of $x$.
 The lowest detector efficiency corresponds to $x$ in the 
range between $0.01$ and $0.03$ where $\varepsilon_{e.m.}$ can be as
low as 58\% for deuteron and 82\% for  proton. 
The efficiencies have been calculated for $x\leq 0.1$, 
where the contamination of B-H events cannot be  neglected, 
and they are set to unity for  $x>0.1$.
They are applied as event weights to the Bethe-Heitler events
produced by the first Monte Carlo simulation described above.

\vspace{0.4cm}

\noindent{\it Correction for non-vanishing tensor  asymmetry}\label{sec:b1d}

\noindent Generally a vector-polarized ($P_z\neq 0$) spin-1 target
such as the deuteron is also tensor-polarized.
The tensor polarization is defined as
$P_{zz} \equiv (n^+ + n^- - 2n^0)/(n^+ + n^- + n^0)$.
In this work, the tensor polarization is large because the  
vector polarization could be maximized by minimizing the
$n^0$ population (see Sec.~\ref{subsec:hermes-target}),
resulting in a tensor polarization approaching unity.
The  average target tensor polarization  in the data set used for this
analysis is  $P_{zz}=0.83\pm0.03$. 
In this case the measured yield (see Eq.~\eqref{eq:crosssect})
depends not only on the target
vector polarization and longitudinal  asymmetry, but also on the
target tensor polarization $P_{zz}$ and the corresponding
tensor asymmetry $A_{zz}^d$:
\begin{equation}
N^{\spar(\sant)} \sim \left[1+(-)~|P_BP_z|~A_{||}^d+\frac{1}{2}P_{zz}A_{zz}^d\right]\,.
\end{equation}
The tensor asymmetry 
\begin{equation}
A_{zz}^d=\frac{\sigma^{\spar}+\sigma^{\sant}-2\sigma^0} 
{\sigma^{\spar}+\sigma^{\sant}+\sigma^0}
\end{equation}
is defined in terms of  cross-sections where unpolarized leptons 
scatter off longitudinally polarized spin-1 targets with 
either non-zero ($\sigma^\spar,\sigma^\sant$) or zero 
($\sigma^0$) helicity state~\cite{Jaffe}. (It is related to the 
structure function $b_1^d$ by the relation 
$b_1^d/F_1^d=-3A_{zz}^d/2$.)
The $x$ dependence of the asymmetry $A_{zz}^d$ was measured at
HERMES~\cite{hermes:b1paper},
the magnitude of $A_{zz}^d$ being of order $10^{-2}$.
This result was applied as a correction to Eq.~\eqref{eq:asyratio}. 


\section{Evaluation of Results}\label{sec:results}


\subsection{\label{subsec:extraction-g1}Extraction of $\mathbf{g_1}$}

The structure function $g_1$ is determined from the Born
asymmetry $\Apar$ according to  Eq.~\eqref{eq:g1fromApar},
using existing spin-averaged DIS cross sections
$\sigma_{UU}$, and small corrections involving $g_2$.
Unfortunately, a self-consistent parameterization of $\sigma_{UU}$ is
unavailable.  
Nevertheless, values of $\sigma_{UU}$ are calculated using the expression:
\begin{eqnarray}
\label{eq:sigUU}
\frac{{\rm d^2}\sigma_{UU}} {{\rm d}x {\rm d}Q^2} &=&
\frac{4 {\pi}{\alpha}^2}{Q^{4}} \, 
\left[F_1(x,Q^2)\cdot y^2 \right.
\nn\\ 
 &+& \left.
\frac{F_2(x,Q^2)}{x}\left(1 - y - \frac{y^2}{4}\gamma^2 \right)
\right]
\nn\\
&=& \frac{4 {\pi}{\alpha}^2}{Q^{4}} \, \frac{F_2(x,Q^2)}{x}  
\nn\\ &\times&
\left[1 - y - \frac{y^2}{4}\gamma^2 +
      \frac{y^2\left(1+\gamma^2\right)}{2\left(1+R(x,Q^2)\right)}\right].
\end{eqnarray}
There do exist parameterizations of $F_2$ based on
all available cross section data, but they were fitted to values of
$F_2$ that were extracted from various data sets using different
values or assumptions for $R$, the ratio of longitudinal to transverse
virtual-photon cross sections.  The values
of $g_1$ extracted  at the measured values of $Q^2$ do not
depend on $F_2$ and $R$ individually, but only on how faithfully
the available combinations of $F_2$ and $R$ parameterizations
represent experimental knowledge of $\sigma_{UU}$.\footnote{As such a
combination of parameterizations of $F_2$ and $R$ should be considered 
in this context as a single
parameterization of $\sigma_{UU}$ that must be self-consistent,
it might be misleading for $F_2$ or $R$ to appear individually
in the formalism leading to the extraction of $g_1$.
}

In the case of the proton, the  parameterizations ALLM97~\cite{allm:1997}
and $R$1990~\cite{whitlow:1990}  are
used for $F_2$ and $R$, respectively.
The parameterization $R$1990
is valid only for $Q^2>0.3$~GeV$^2$. Below this value $R$ was
linearly extrapolated  to zero  to account for  the fact that  for real photons
$R=0$.
In the case of the deuteron, the NMC parameterization of
the measured ratio $F_2^n/F^p_2$~\cite{f2param} 
is used in conjunction with the ALLM97 parameterization for $F_2^p$ to
calculate $F_2^d$. 
The required values of $g_2$ are computed from a parameterization
of all available proton and deuteron 
data~\cite{Anthony:a2,Abe:e143a2,e155:2003,SMC:a2p,smc:1997}.

The results for $g_1$ in Tab.~\ref{tab:g1_46} in the 45 individual 
kinematic bins of
Fig.~\ref{fig:xq2plane} are considered to be the primary result
of this work, in that the only previously published information 
used in their {\em extraction} is the spin-averaged DIS cross section 
$\sigma_{UU}$, and small corrections involving $g_2$.
For both the convenient {\em presentation} and {\em interpretation}
of these results, some ansatz must be adopted to `evolve' these
results in $Q^2$, as it is impossible to use QCD to evolve individual
values of structure functions at diverse values of $x$.  
Two degrees of `evolution' are involved.  For the convenient presentation 
of the $x$ dependence of $g_1$, its values in the two or three $Q^2$ bins 
that may be associated with each $x_i$ bin must be evolved to their mean 
$\langle Q^2\rangle_i$  and then averaged.  
Also, in order to compute moments of $g_1$, the 
measurements in the 45 $(x_i,Q^2_i)$ bins must be brought to a common $Q^2_0$.  
One previously used ansatz is to assume that the structure
function ratio $g_1/F_1$ is independent of $Q^2$.  
This is usually justified by the observation that the non-singlet
evolution kernel is the same for $g_1$ and $F_1$, and the ansatz cannot 
be excluded by the limited available data set.
However, the singlet evolution kernels are different, and anyway both kernels
operate on different initial $x$ distributions.  Hence this ansatz seems
arbitrary, and also precludes the assignment of an appropriate
uncertainty to the evolution that is based on how well it is constrained by
QCD together with existing data.  In another approach, used
in this work, the primary $g_1$ values are `evolved'
by using an NLO QCD fit to all available $g_1$ data, assuming that
the difference between $g_1(x_i,Q^2_i)$ and  $g_1(x_i,Q^2_0)$ is the
same as obtained in the QCD fit:
\be\label{eq:evolution}
g_1(x_i,Q^2_i)-g_1(x_i,Q^2_0)=g_1^{fit}(x_i,Q^2_i)-g_1^{fit}(x_i,Q^2_0)~.
\ee

The QCD fit used here is described in detail in Ref.~\cite{bb}.  It was
extended to include the present $g_1^{p,d}$ data,
as well as the new data of Refs.~\cite{compass} and \cite{jlab:g1n}.
The uncertainty due to the evolution of $g_1$ was
propagated from the statistical and recently implemented systematic 
uncertainties in the
fit parameters to the quantities appearing in Eq.~(\ref{eq:evolution}).


\subsection{Statistical Uncertainties}\label{sec:stat}

The asymmetry values  obtained after unfolding of radiative and smearing
effects are statistically correlated  between kinematic bins.
Contributions to cov$(A_{||})$ from the  finite statistics
of the Monte Carlo data sample are typically  
one order of magnitude smaller than those of the
experimental covariance matrix.
The (statistics-based) covariance of $A_{||}$ originating from the
unfolding and the one 
from the finite statistics of the Monte Carlo are summed;
the result is called 
{\it statistical covariance} hereafter.

Eq.~(\ref{eq:g1fromApar}) implies that the covariance matrix for
$g_1$ is given by:
\begin{eqnarray}
\!\!\!\!\!\!\textrm{cov}\left(g_1\right)_{lm}\!\!&=&\!\!
C_l~C_m~\textrm{cov}(A_{||})_{lm}\,, \nn\\
C_i \!\!&\equiv &\!\!\frac{1}{1-\frac{y_i}{2}-\frac{y_i^2}{4}\gamma_i^2} 
\frac{Q^4_i}{8\pi\alpha^2y_i} \frac{\detwo\sigma_{UU}(x_i,Q^2_i)}{\de x \de Q^2} 
\end{eqnarray}
where the subscripted kinematic quantities 
correspond to the average $x$ and $Q^2$ values of the kinematic bins.


\subsection{Systematic Uncertainties}\label{sec:sys}

The systematic uncertainties originate
 from {\it i}) the experiment (beam and target polarizations, PID,
misalignment of the detector) and {\it ii})  the parameterizations 
($g_2$, $F_2$, $R$, $A_{zz}^d$, $\omega_D$).
The various contributions to the systematic uncertainty were added quadratically. 
For $Q^2\geq 1$~GeV$^2$ they are given in Tab.~\ref{tab:mom-ratios}, 
where each value represents an average over the  measured $x$ range. 

\vspace{0.4cm}

\setcounter{paragraph}{0}
\underline{{\it i)}  Experimental Sources}

\vspace{0.4cm}

\noindent{\it Polarizations}

\noindent The uncertainties on  beam and target polarization values 
(see Tabs.~\ref{tab:beampol} and \ref{tab:targetpol}) are the
dominant sources of systematic uncertainties in this measurement.
The polarization  values  are varied within their
uncertainties  and the corresponding change in the central value
of $A_{||}$ is assigned as its systematic uncertainty, which is then
propagated to $g_1$.

\vspace{0.5cm}

\noindent{\it Particle identification}\\
\noindent The hadron contamination in the DIS lepton  sample,
is less than 0.2$\%$ leading to a negligible  contribution to 
the systematic uncertainty.

\vspace{0.4cm}

\noindent{\it Detector Misalignment}\\
\noindent Studies have shown that the two halves of the HERMES detector are not
perfectly aligned symmetrically to the beam axis.
As a conservative approach, a
Monte Carlo simulation using a  misaligned detector geometry
  is compared to one with  perfect alignment,
separately for each target and detector half.  
A contribution to the systematic uncertainty of the final
 unfolded asymmetry is assigned as the difference in the central values of the
respective reconstructed Monte Carlo asymmetries.

\vspace{0.4cm}

\underline{{\it ii)} Input parameterizations}

\vspace{0.4cm}

The parameterizations enter at two different stages of the analysis: unfolding
and extraction.
The unfolding algorithm does not involve any model of the
asymmetries within the acceptance.
Nevertheless a  model dependence can arise from the 
description  outside the acceptance. 
The comparison of  different models for both the polarized and 
unpolarized Born cross sections
shows no significant  deviations in the unfolded Born asymmetry
$A_{||}$ within the statistical accuracy of the Monte
Carlo test samples used, which is about three times better than that of
the data sample. 

While calculating $g_1$ from the Born asymmetry $A_{||}$ according to
Eq.~(\ref{eq:g1fromApar}), and `evolving' the values of $g_1$ to a common
value of $Q^2$, non-negligible
systematic uncertainties occur due to various input parameterizations 
of $g_2$, $F_2$, $R$, $A_{zz}^d$ and $\omega_D$. 

\vspace{0.4cm}

\noindent{\it Structure function $g_2$}\\
\noindent The systematic uncertainty due to $g_2^{p,d}$ is obtained 
as the effect on the results of its variation over the range
corresponding to the covariance matrix from the fit to the
world data for $g_2$.

\vspace{0.4cm}

\noindent{\it Cross section $\sigma_{UU}$}\\
\noindent The systematic uncertainty due to the employed combination of
$F_2^{p,d}$ and $R$ receives contributions as follows: 
{\em(a)} those intrinsic to the original cross section measurements, 
{\em(b)} from model dependence of the parameterizations,
and 
{\em(c)} from inconsistencies between the $R$ values from its parameterization
and the $R$ values originally used to extract $F_2$ from the $\sigma_{UU}$
data. 
In an attempt to account for all of these contributions in a conservative
manner, the systematic uncertainty in $\sigma_{UU}$ was taken to be the sum 
in quadrature of the difference between the results using the ALLM and 
SMC parameterizations for $F_2$, accounting for contribution {\em(b)} above,
and half of the difference between the results using the upper and lower error bands 
in the SMC parameterization, approximately accounting for {\em(a)} and {\em(c)}\footnote{
The SMC parameterization was not adopted for the central values of this analysis
because of an apparent anomaly in its shape at $x>0.6$.  Also, in Ref.~\cite{Adeva:1998vv}, 
there is  a typographic error in a parameter for the lower error band, but this
error does not appear in the related thesis~\cite{Cuhadar:thesis} and the SMC web page.
}.

\vspace{0.4cm}

\noindent{\it Tensor asymmetry $A_{zz}^d$}\\
\noindent The contribution from the uncertainty on the published value
of $A_{zz}^d$~\cite{hermes:b1paper} is negligible, but is nevertheless
included in the systematic uncertainty. 

\vspace{0.4cm}

\noindent{\it D-wave admixture $\omega_D$}\\
\noindent The limiting values of $\omega_D=0.05\pm 0.01$ are used to
determine the $\omega_D$ contribution to the systematic uncertainty of
$g_1^n$ and of the non-singlet structure function $g_1^{NS}$. 

\vspace{0.4cm}

\noindent {\it `Evolution' to a common value of} $Q^2$\\
\noindent The uncertainties of the NLO QCD fit of all available data
that is used to `evolve' the $g_1$ values of the present work are propagated
through both the fit and the `evolution', accounting for  correlations.
These uncertainties include experimental statistical and systematic as well
as `theoretical' uncertainties influencing the fit.  The resulting total `evolution'
uncertainty contributions are presented separately in the tables of $g_1$ values
in either 19 or 15 $x$-bins, and also with the value of each reported moment. 


\section{\label{sec:finalresults}Discussion of the Results }

\subsection{Born asymmetries and uncertainty correlations} 
For the 45 ($x$, $Q^2$) bins defined  in Fig.~\ref{fig:xq2plane}, the
measured and Born asymmetries  $A_{||}^{m}$ and $A_{||}$  
are listed in Tabs.~\ref{tab:aparp} and \ref{tab:apard}
for the proton and  the deuteron, 
together with statistical  and systematic uncertainties.
The Born asymmetries $\Apar$ 
vary, for fixed values of $x$, substantially  with $Q^2$. (For more 
details see Fig.~\ref{fig:inflation} in the appendix.) This reflects
the fact that the polarization of the virtual photon, probing the 
helicity-dependent quark number densities in the nucleon as 
discussed in the Introduction, is smaller than the
polarization of the incident lepton and dependent on the lepton
kinematics~\cite{lea82}. 
At a given value of $x$, the polarisation transfer
is smaller at low values of $y$, i.e., at low $Q^2$.
This introduces an inflation of the
statistical uncertainty in the corresponding kinematic bins, when 
extracting $g_1$ (see Eq.~\eqref{eq:g1fromApar}) and $A_1$ (see Sec.~\ref{sec:Vasymmetry}).

Comparing in Fig.~\ref{fig:inflation} the statistical uncertainties of $A_{||}^m$ and $A_{||}$
at  each $x$, it is clearly seen that the 
correction for smearing and radiative background introduces 
a considerable uncertainty inflation, especially in the lowest $Q^2$ bins at
a given value of $x$. 
At low $x$, this is  due to the  subtraction of radiative background,
while at larger $x$ the uncertainty inflation arises from a substantial
instrumental and radiative smearing, which results in considerable
bin-to-bin correlations.  
Note that the statistical uncertainties quoted in Tabs.~\ref{tab:aparp} and \ref{tab:apard}
and shown in Fig.~\ref{fig:inflation} correspond only
to the diagonal elements of the covariance matrix. 
Especially at low $Q^2$ the non-diagonal  elements are large,
 as can be seen  in the correlation matrices  listed for
the  45  ($x$,$Q^2$) bins in 
Tabs.~\ref{tab:corr-46bins-hyd1} and \ref{tab:corr-46bins-deu1}, and
shown in Fig.~\ref{fig:cov} for the 
proton\footnote{The correlation matrix is related to the covariance matrix through the
statistical  uncertanties $\sigma_i $ and $\sigma_j$:
$\textrm{cov}(i,j)=\sigma_i \sigma_j \textrm{corr}(i,j)$.}.
\begin{figure}[!t]
\includegraphics[width=\columnwidth]{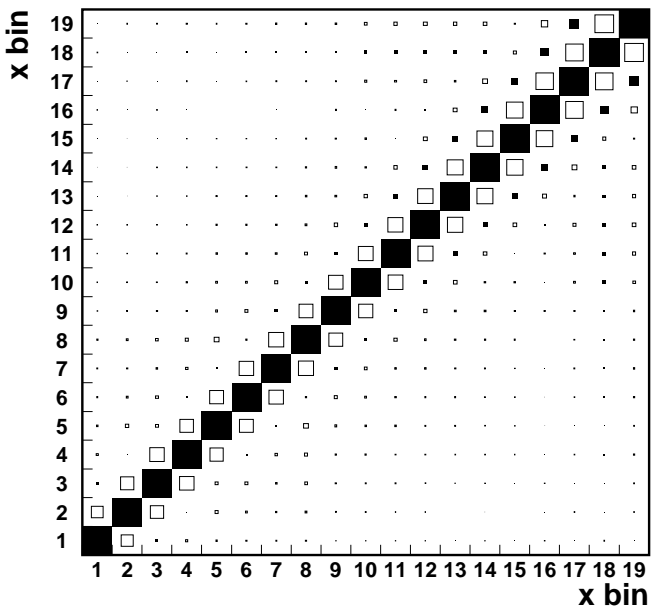}
\caption{\label{fig:cov}
Correlation matrix in 19 $x$-bins of $g_1$ for the 
proton. The closed symbols represent positive values, 
while the  open ones are for negative values. The area of the symbols
represents the size of the correlation.}
\end{figure}

\subsection{\label{sec:Vasymmetry} Virtual-photon asymmetry}

The virtual-photon asymmetry  $A_1$ is directly related
to the photon-nucleon absorption cross sections and hence to the structure functions
$g_1$, $g_2$ and $F_1$:
\begin{eqnarray}
\label{eq:formalism-A12def}
A_1  & =&  \frac{\sigma_{1/2} - \sigma_{3/2}}{\sigma_{1/2} + \sigma_{3/2}}
       =  \frac{g_1 - \gamma^2 g_2}{F_1} ~~~.~~~  
\end{eqnarray}
Here $\sigma_{1/2}$ and  $\sigma_{3/2}$ are the virtual-photon absorption 
cross
sections when the projection of the total angular momentum of the
photon-nucleon system along the incident photon direction is $1/2$ or $3/2$,
respectively.
                                                                                                                                               
The virtual-photon asymmetry  $A_1$ is obtained from the Born asymmetry
$\Apar$ according to Eq.~(\ref{eq:g1fromApar}) and Eq.~(\ref{eq:formalism-A12def}),
by using values of $\sigma_{UU}$ and $F_1$ calculated in terms of the available 
$F_2$ and $R$ parameterizations, i.e., from Eq.~(\ref{eq:sigUU}) and 
\begin{eqnarray}
\label{eq:formalism-F1def}
F_1  & =&  \frac{(1+ \gamma^2) F_2}{2x(1+R)} ~~~,
\end{eqnarray}
and by using a parameterization for $g_2$ obtained from fits to all available
proton and deuteron data~\cite{Anthony:a2,Abe:e143a2,e155:2003,SMC:a2p,SMC:a2d}.
The systematic uncertainties on the
virtual-photon asymmetry $A_1$ are determined analogously to the  case 
of $g_1$ described in Sec.~\ref{sec:sys}, taking into account any correlation between 
different systematic sources.  
The results are listed for the 45 ($x, Q^2$) bins together with their
statistical and systematic uncertainties in Tab.~\ref{tab:A1_45} for the
proton and the deuteron.

The asymmetries $A_1^{p,d}$ provide a convenient way of comparing the precision
of various experiments, as most of the dependence of $A_{||}$ 
on $Q^2$ due to the polarization 
transfer from the lepton to the virtual photon 
 is cancelled. However, a substantial additional contribution 
to the systematic uncertainty of $A_1$ due to the poor knowledge of $R$ in
the extraction of $F_1$ from measured values of  $\sigma_{UU}$ and $R$ 
is unavoidable.  The required values of $R$ as well as their
uncertainties were computed using the $R$1990 
parameterization~\cite{whitlow:1990}.

The values from the present work, averaged over the corresponding $Q^2$
bins,  are shown in 
Fig.~\ref{fig:gf-hermesonly} as a function of $x$
and are compared to the world data in the top (middle) panel of Fig.~\ref{fig:g1f1-all}
for the proton (deuteron).
Note that the low-$x$  data points of HERMES ($x < 0.03$) and
SMC ($x< 0.003$) are measured at $\langle Q^2 \rangle < 1$ GeV$^2$.

In the case of the proton, the accuracy of the HERMES measurement is
comparable to the  most precise
measurements at SLAC (E143~\cite{Abe:1998wq},
E155~\cite{Anthony:2000fn}) and at CERN (SMC~\cite{Adeva:1998vv}).
The HERMES measurement extends to lower values in $x$ than
the SLAC data points, into a region covered up to now
by SMC data only, although at higher values of $Q^2$.
In the case of the deuteron,  HERMES data provide the
most precise determination of the asymmetry $A_1^d$.
The available world data from previous
measurements~\cite{Adeva:1998vv,SMC:1999pdlox,Adeva:2000er,Abe:1998wq,Anthony:1999rm,Anthony:2000fn,compass}
are considerably less accurate than in case of the proton.

The asymmetries rise smoothly from zero with increasing $x$. For
$ x \rightarrow 1$, $A_1^p$ becomes of order unity 
(i.e. the quark carrying most of the
nucleon's momentum is also carrying most of its spin). For the deuteron, the
asymmetry for $x$ above 0.01 appears to be smaller 
than that of the proton.

Apart from the similar general trend, the asymmetries $A_1$ for proton and
deuteron are rather different in their $x$ dependence and magnitude. The
ratio $A_1^d/A_1^p$ is smaller than 0.5 over the measured $x$ range,
indicating a negative contribution to $A_1^d$ from the neutron. HERMES data 
at low $Q^2$ indicate that 
$A_1^p$ is compatible with zero within the statistical uncertainties below
$x = 0.01$ while $A_1^d$ vanishes already below $x = 0.05$.
 
\begin{figure}[t!]
\includegraphics[width=\columnwidth]{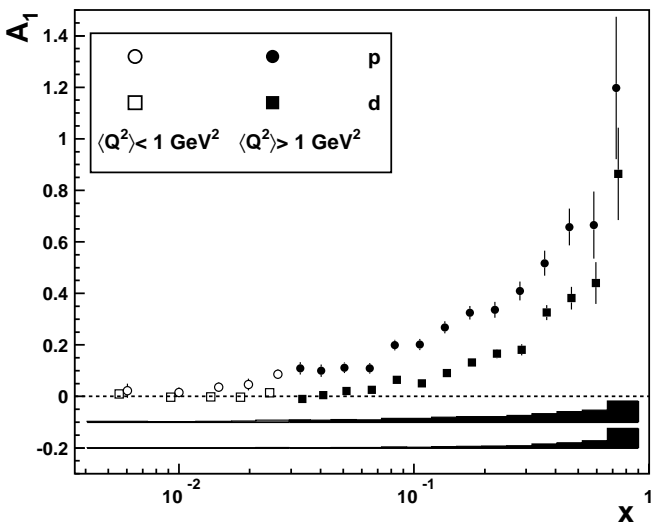}
\caption{\label{fig:gf-hermesonly}
 HERMES results on $A_1$ vs $x$ for 
 proton and deuteron.  Error bars represent statistical uncertainties
 of the data
({\it diagonal} elements of the covariance matrix) combined
 quadratically with those from 
 Monte Carlo statistics. Bands represent systematic uncertainties. 
Deuteron data points have been slightly shifted in $x$ for visual purposes. }
\end{figure}
\begin{figure*}[h!]
\includegraphics[width=0.98\textwidth]{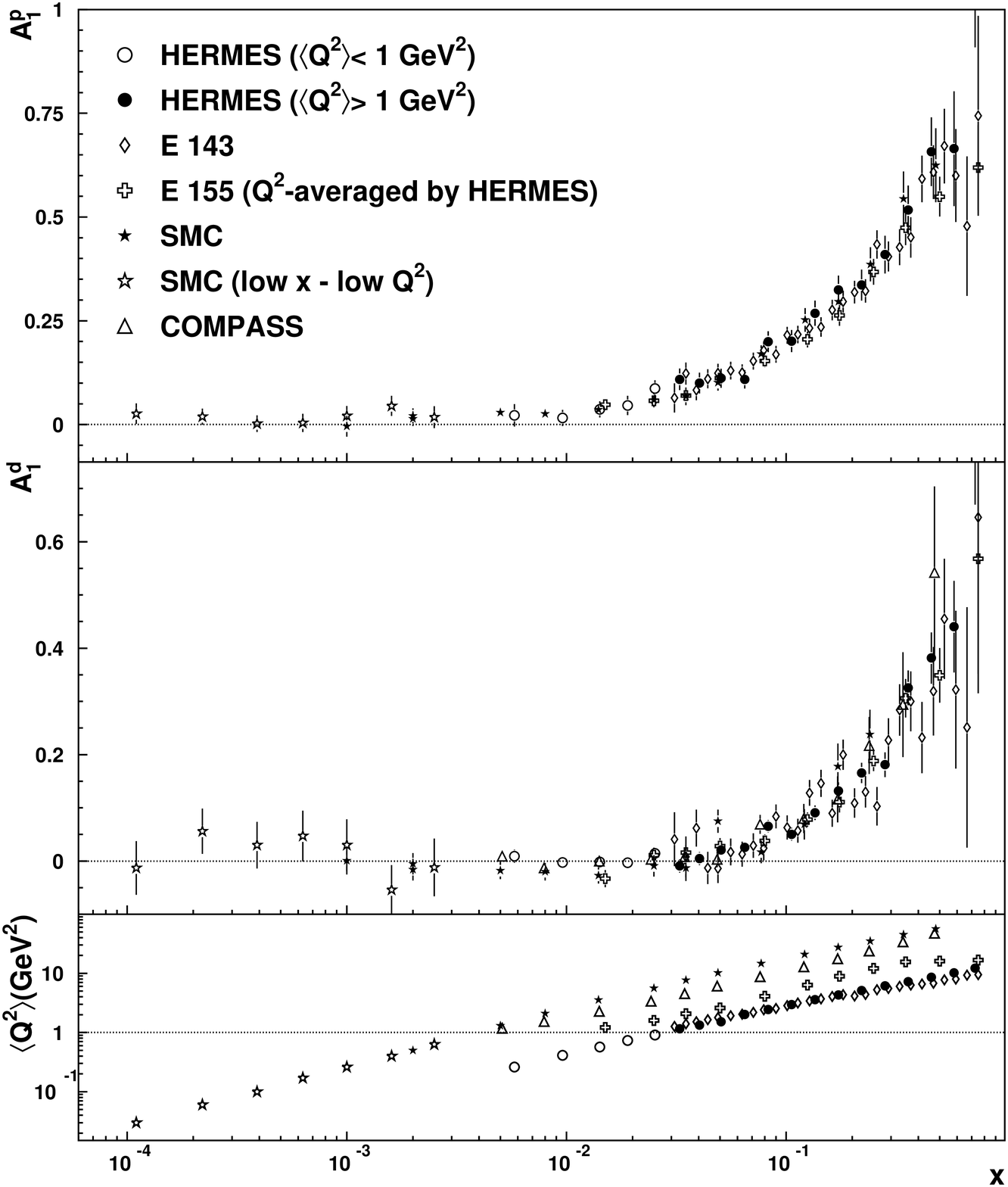}
\caption{\label{fig:g1f1-all}HERMES results on \Ap and
{\Ad}\/ vs $x$, shown on separate panels,
compared to data from  SMC~\cite{Adeva:1998vv,SMC:1999pdlox,Adeva:2000er},
E143~\cite{Abe:1998wq},  E155~\cite{Anthony:2000fn}, and
COMPASS~\cite{compass}.
 Error bars represent the sum in quadrature of statistical  and
systematic uncertainties.
For the HERMES data the closed (open) symbols represent values derived by
selecting events with $Q^2 >1$~GeV$^2$ ($ Q^2 <1$~GeV$^2$).
The HERMES data points shown are
statistically correlated (cf. Fig.~\ref{fig:cov}) by  unfolding 
QED radiative and detector smearing effects; the statistical
uncertainties shown are  obtained from   the {\it diagonal} elements of the
covariance matrix, only. The E143 and E155 data
points are correlated through QED radiative  corrections. 
The lower panel shows the $x$ dependence of  $\langle Q^2\rangle$  for
the different experiments.}
\end{figure*}

\subsection{Structure functions $g_1$}

\setcounter{paragraph}{0}
\paragraph{Proton and Deuteron.}

The  primary values for the  structure functions $g_1$
for both the proton and deuteron  
are given in Tab.~\ref{tab:g1_46} for all of the 45 bins shown in Fig.~\ref{fig:xq2plane}.
The correlation matrices are given  in Tab.~\ref{tab:corr-46bins-hyd1} for the proton and in
Tab.~\ref{tab:corr-46bins-deu1} for the deuteron. For those $x$ bins having more than one
$Q^2$ bin, the $g_1$ values were evolved to a common value of $Q^2$,
as described in sect.~\ref{subsec:extraction-g1}, and averaged. 
The results 
are shown in 19 $x$-bins in Fig.~\ref{fig:g1-all}, together with all previously
published data.
 Alternatively, the functions $xg_1^p$ and $xg_1^d$  
are shown in Fig.~\ref{fig:g-hermesonly}, and compared
to the previously published data in Fig.~\ref{fig:xg1-all}.

The numerical values for $g_1^{p,d}$  are given in
Tab.~\ref{tab:gpd_20} and the correlation matrices are 
in Tab.~\ref{tab:corr-20bins-hyd} for the proton
and in Tab.~\ref{tab:corr-20bins-deu}  for the deuteron. 
When events are selected subject to $Q^2>1$\,GeV$^2$, only 15 $x$-bins
remain (see e.g. Fig.~\ref{fig:xq2plane}).
The corresponding structure functions and  correlation matrices are
given in Tabs.~\ref{tab:gpd_15}, \ref{tab:corr-15bins-hyd}, and \ref{tab:corr-15bins-deu}.
\begin{figure}[t!]
\includegraphics[width=\columnwidth]{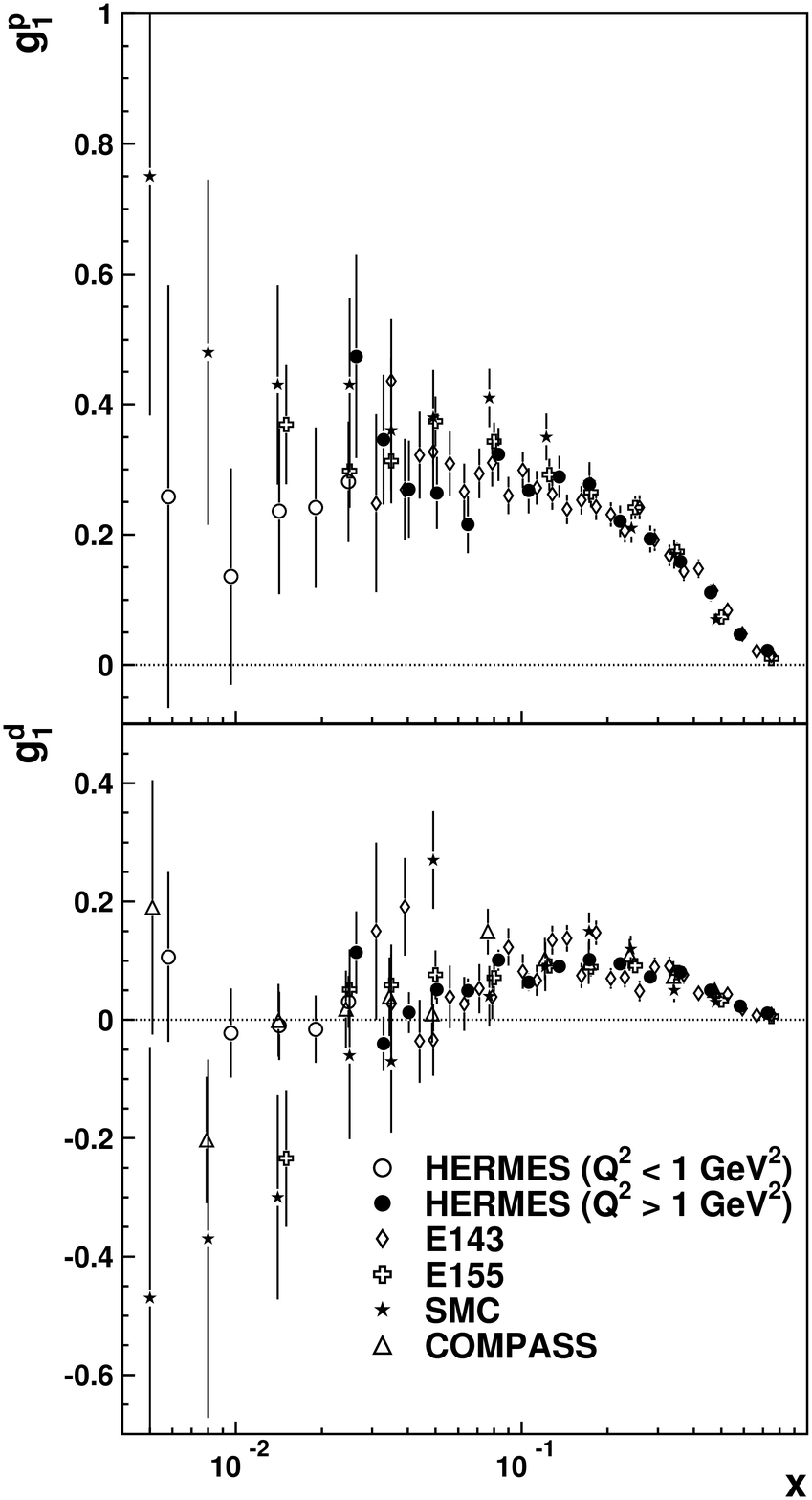}
\caption{\label{fig:g1-all}HERMES results on  $g_1^p $ and
$g_1^d$ vs $x$, shown on separate panels,
compared to data from  SMC~\cite{Adeva:1998vv,SMC:1999pdlox,Adeva:2000er},
E143~\cite{Abe:1998wq},  E155~\cite{Anthony:2000fn,Anthony:1999rm}, and
COMPASS~\cite{compass}, in the HERMES $x$-range.
 Error bars represent the sum in quadrature of statistical and
systematic uncertainties. The HERMES data points shown are
statistically correlated (cf. Fig.~\ref{fig:cov}) by  unfolding 
QED radiative and detector smearing effects; the statistical
uncertainties shown are  obtained from only the {\it diagonal} elements of the
covariance matrix. The E143 and E155 data
points are  correlated through QED radiative
corrections.  The E155 points have been averaged over their $Q^2$ bins for visibility.
For the HERMES data the closed (open) symbols represent values derived by
selecting events with $Q^2 >1$~GeV$^2$ ($ Q^2 <1$~GeV$^2$). 
}
\end{figure}
In the case of the proton,
the central values of the SMC data points are larger than those of HERMES, in the low-$x$ region.
This reflects the difference in $\langle Q^2 \rangle$ values between the two experiments, 
and is expected from the $Q^2$ evolution of $g_1$.

In the case of the deuteron, the HERMES data are compatible with zero
for  $x<0.04$. In this region  the  SMC data favor
negative values for $g_1^d$ while the COMPASS results~\cite{compass}
are also consistent with zero. 

\begin{figure}[t!]
\includegraphics[width=\columnwidth]{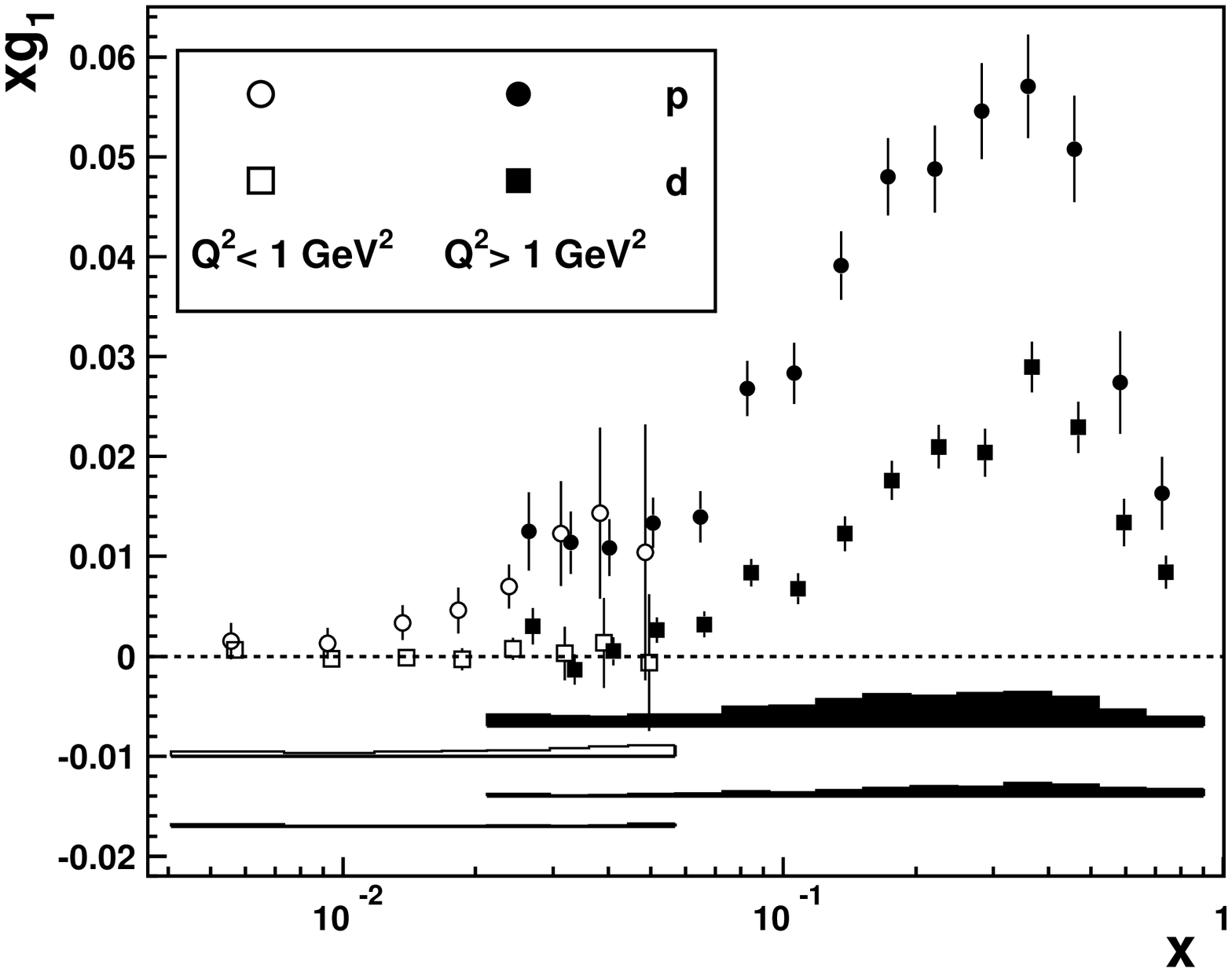}
\caption{\label{fig:g-hermesonly}HERMES results for $xg_1$ vs $x$ for the
 proton and the deuteron.  Error bars represent statistical uncertainties
 of the data (from the {\it diagonal} elements of the covariance matrix) combined
 quadratically with those from 
 Monte Carlo statistics. The upper and lower error bands represent 
the total systematic uncertainties for the proton and deuteron, respectively. 
The deuteron data points have been slightly shifted in $x$ for
visibility. The closed (open) symbols represent values derived by
selecting events with $Q^2 >1$~GeV$^2$ ($ Q^2 <1$~GeV$^2$). 
}
\end{figure}
\begin{figure*}[t!]
\includegraphics[width=\textwidth]{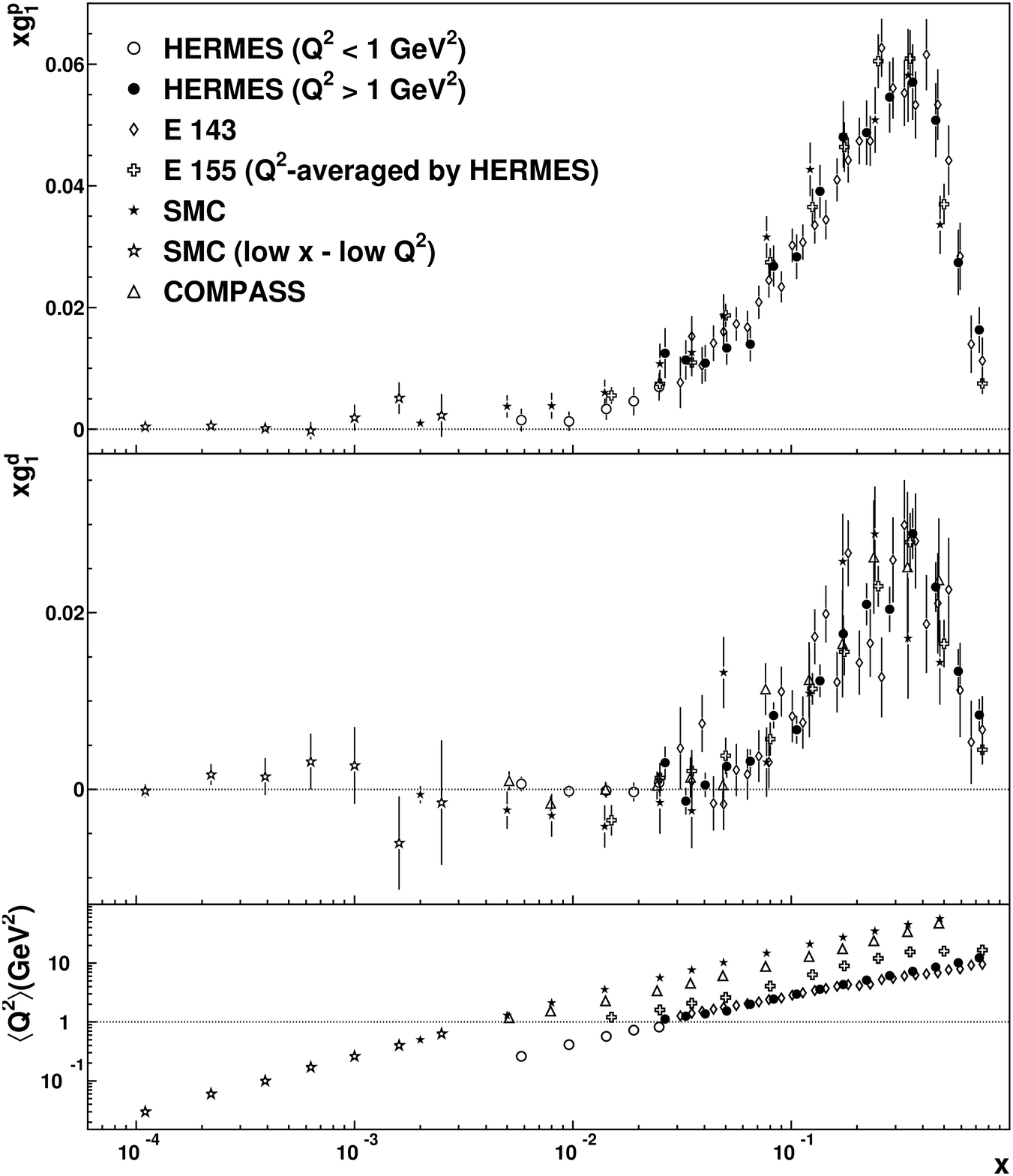}
\caption{\label{fig:xg1-all}HERMES results on  $xg_1^p $ and
$xg_1^d$ vs $x$, shown on separate panels,
compared to data from  SMC~\cite{Adeva:1998vv,SMC:1999pdlox,Adeva:2000er},
E143~\cite{Abe:1998wq},  E155~\cite{Anthony:2000fn,Anthony:1999rm}, and
COMPASS~\cite{compass}.
 The error bars represent the sum in quadrature of statistical and
systematic uncertainties. The HERMES data points shown are
statistically correlated (cf. Fig.~\ref{fig:cov}) by  unfolding 
QED radiative and detector smearing effects; the statistical
uncertainties shown are  obtained from only the {\it diagonal} elements of the
covariance matrix. The E143 and E155 data
points are correlated due to the method for correcting for QED
radiation.
For the HERMES data the closed (open) symbols represent values derived by
selecting events with $Q^2 >1$~GeV$^2$ ($ Q^2 <1$~GeV$^2$ ). 
}
\end{figure*}

\bigskip

\paragraph{Neutron.}

The neutron structure function $g_1^n$ is extracted from  $g_1^p$ and
$g_1^d$ using Eq.~(\ref{g1dfromg1pn}). 
Other nuclear effects  like shadowing and Fermi motion of the nucleons
in the deuteron are neglected.
\begin{figure}[!t]
\includegraphics[width=\columnwidth]{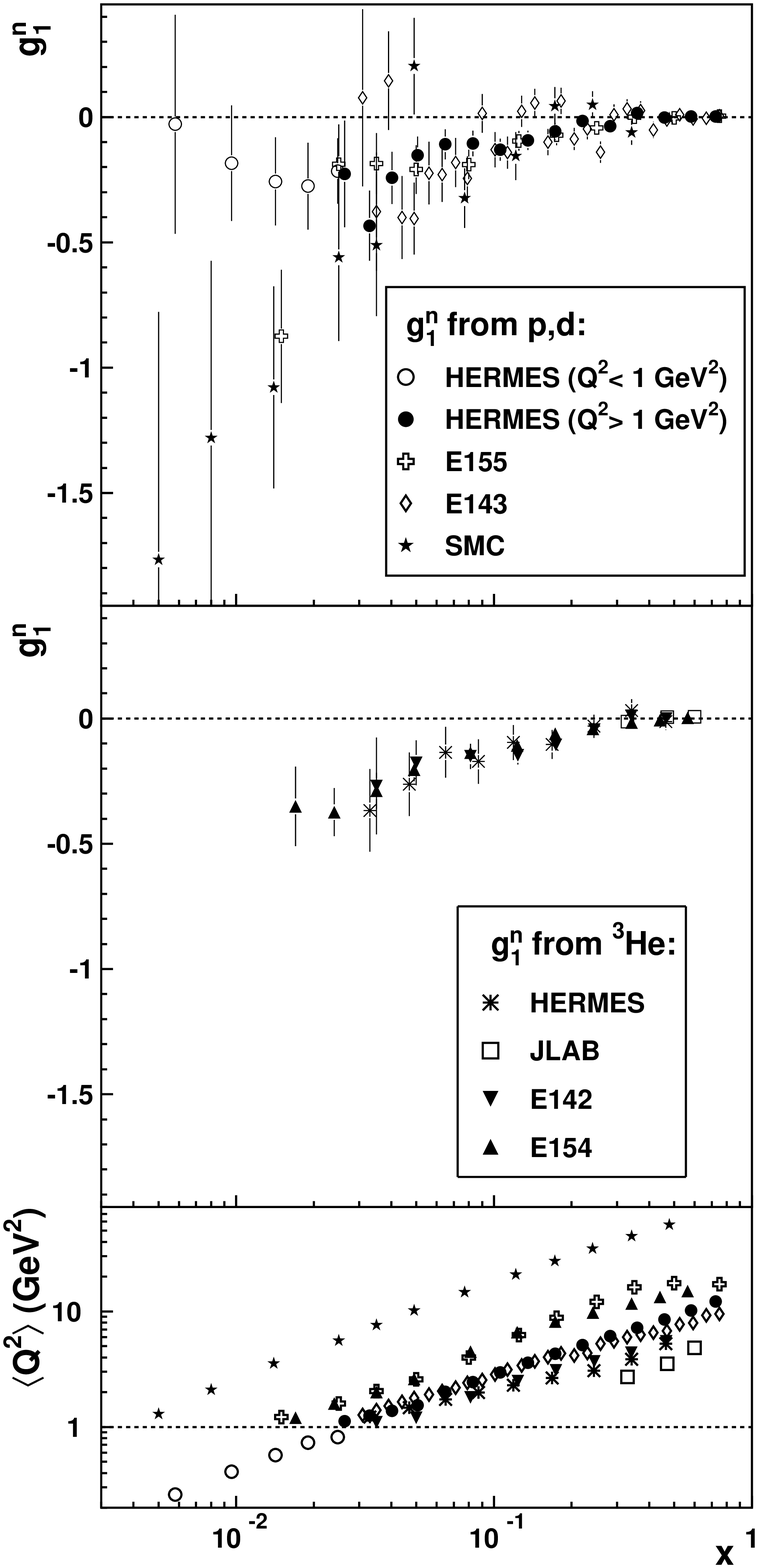}
\caption{\label{fig:gn}
Top panel: the structure function  $g_1^n$   obtained from
 $g_1^p$   and  $g_1^d$, compared with similar data from
SMC~\cite{Adeva:1998vv,SMC:1999pdlox,Adeva:2000er},
E143~\cite{Abe:1998wq}, and E155~\cite{Anthony:2000fn,Anthony:1999rm} in the HERMES
x-range. 
Second panel from the top:  $g_1^n$ as obtained from a $^3$He target by 
JLab~\cite{jlab:g1n}, HERMES~\cite{g1n_hermes}, E142~\cite{E142n}, and E154~\cite{E154n}.
Total error bars are shown, obtained by
combining statistical and systematic uncertainties in quadrature.
The bottom panel shows the $\langle Q^2\rangle$ of each data point in the top two
panels. E155 data have been averaged over their $Q^2$ bins for
visibility.
For the HERMES data the closed (open) symbols represent values derived by
selecting events with $Q^2 >1$~GeV$^2$ ($ Q^2 <1$~GeV$^2$). 
}
\end{figure}

In Fig.~\ref{fig:gn} (top panel), results on  $g_1^n(x)$, extracted from 
$g_1^p$ and $g_1^d$, are shown for HERMES in comparison to the
world data.
As can be seen from the lower panel of the figure, the average $Q^2$
values of HERMES and SLAC measurements are similar, while those of SMC are
higher by one order of magnitude at a given $x$.
Compared to previous data, the HERMES measurement restricts $g_1^n(x)$
now very well.
The structure function $g_1^n$ is negative everywhere, except for the
very high $x$ region, where it becomes slightly positive. 
For decreasing $x$ values below about $0.03$,
$g_1^n$ appears to gradually approach zero from below, complementary
to the behavior of $g_1^p$. 
While this behaviour is based on data with $Q^2\leq 1$~GeV$^2$,
it 
differs from the strong decrease of $g_1^n(x)$ for $x \rightarrow 0$
that was previously conjectured on the basis of
the E154 measurement on $^3$He~\cite{E154n}, and
also on SMC data~\cite{Adeva:1998vv}, both with $Q^2\geq 1$~GeV$^2$
as shown in Fig.~\ref{fig:gn}. 

For completeness, the HERMES results on the $x$ dependence of
$xg_1^n$ are shown  in Fig.~\ref{fig:xg1ns} (top panel), compared
to the world data.
Tabs.~\ref{tab:g1n46}, \ref{tab:g1n20} and \ref{tab:g1n15} show the
results for $g_1^n$ in 
45 bins, 19 $x$-bins (obtained after averaging over $Q^2$), and in 15
$x$-bins (obtained after applying a $Q^2>1$~GeV$^2$ cut to data 
and then averaging in $Q^2$).

\bigskip

\paragraph{Non-singlet.}
\begin{figure}[!t]
\includegraphics[width=\columnwidth]{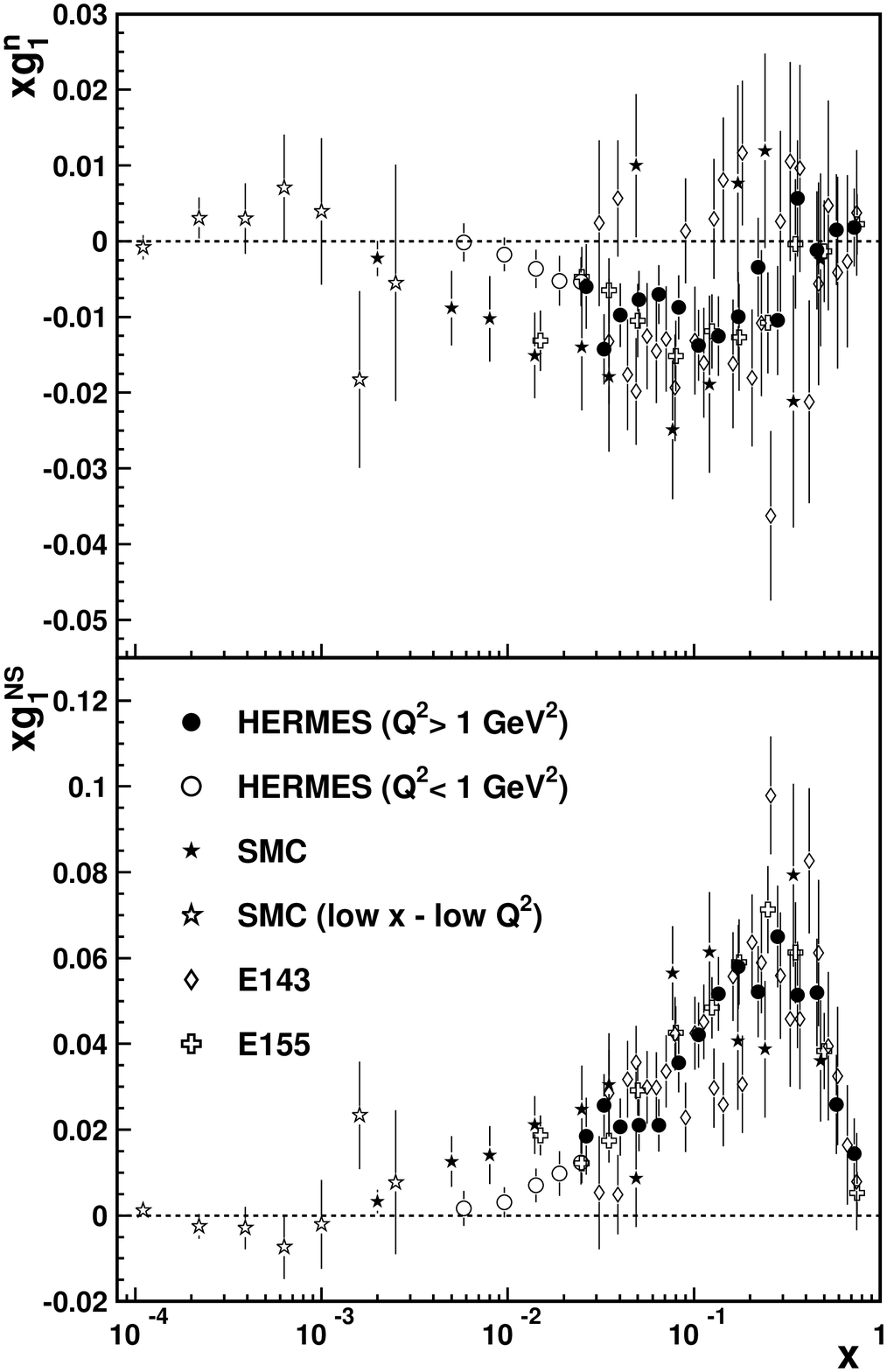}
\caption{\label{fig:xg1ns}
Top panel: $xg_1^n$ from data for $g_1^p$ and $g_1^d$.
Bottom panel: the $x$-weighted non-singlet spin structure
 function $xg_1^{NS}$,  
 compared to data from SMC~\cite{Adeva:1998vv,SMC:1999pdlox,Adeva:2000er}, 
E143~\cite{Abe:1998wq}, and E155 \cite{Anthony:2000fn,Anthony:1999rm}.
All data are presented at their measured values of $\langle Q^2\rangle$, except 
that the E155 data in each $x$ bin have been averaged over $Q^2$ for visual purposes.
Total uncertainties are shown as bars.
For the HERMES data the closed (open) symbols represent values derived by
selecting events with $Q^2 >1$~GeV$^2$ ($ Q^2 <1$~GeV$^2$). 
}
\end{figure}

The non-singlet spin structure function $g_1^{NS}(x,Q^2)$ is
defined as:  
\begin{equation}\label{g1ns}
g_1^{NS}\equiv g_1^p-g_1^n=2\left[g_1^p-\frac{g_1^d}{1-\frac{3}{2}\omega_D}\right]\,.
\end{equation}

In Fig.~\ref{fig:xg1ns} (bottom panel), the $x$ dependence of
$xg_1^{NS}$ as measured by HERMES is shown in comparison with data
from E143, E155 and SMC.
The non-singlet structure function shows a behavior similar to that
of the deuteron and neutron, which was discussed above: the HERMES
data constrain the $x$ dependence of $xg_1^{NS}$
 much better than earlier measurements;
the above-mentioned difference between the
HERMES and SMC deuteron data for $x<0.03$ is  reflected also in
$g_1^{NS}$.

\subsection{Integrals of $g_1$}

Important information about the spin structure of the nucleon can be
obtained from the first moment of $g_1$, in particular when combining
results on proton, deuteron and neutron.
Experimentally, only a limited range in $x$ is accessible.
The integrals for proton and deuteron over a certain
$x$ range and at a given $Q^2_0$ are obtained as:
\be\label{mom1}
{\widetilde{\Gamma}_1}(Q^2_0)=\sum_{i} 
\frac{g_1(\langle x\rangle_i,Q^2_0)}{g_1^{fit}(\langle x\rangle_i, Q^2_0) }
\int_{x_{i}}^{x_{i+1}}\!\dx g_1^{fit}(x,Q^2_0),
\ee
where $\langle x\rangle_i$ are the average values of $x$ in 
$x$-bin $i$ with boundaries $x_i$ and $x_{i+1}$;
the integral of $g_1^{fit}$ accounts for the non-linear $x$-dependence of 
$g_1$. 
The integrals for $g_1^n$ and $g_1^{NS}$ are obtained by linearly
combining the ones for proton and deuteron.

The statistical uncertainty is calculated as:
\begin{eqnarray}
\sigma^2_{{\widetilde{\Gamma}_1}}&=&\sum_{ij}
\int_{x_{i}}^{x_{i+1}}\!\dx
g_1^{fit}(x,Q^2_0) \int_{x_{j}}^{x_{j+1}}\!\dx g_1^{fit}(x,Q^2_0)\nn\\
&&\times \frac{\textrm{cov}\left({g_1}\right)_{ij}}{g_1^{fit}(\langle x \rangle_i,Q^2_0)~g_1^{fit}(\langle x \rangle_j,Q^2_0)}
.
\end{eqnarray}
The systematic uncertainties for the integrals are
determined analogously to the above described case of structure
functions.  
The correlations between systematic
uncertainties of $g_1^p$ and $g_1^d$ were taken into account
in the calculation of the systematic uncertainty for $g_1^n$ and
$g_1^{NS}$.  

The integrals  for $g_1^p$, $g_1^d$, $g_1^n$, and $g_1^{NS}$, calculated
at $Q_0^2=2.5$ and $5$~GeV$^2$,  are given in Tab.~\ref{tab:mom} together with the
statistical, systematic and evolution uncertainties.
They are shown for the $x$ range
$0.021 \leq x \leq 0.9$,  corresponding to
the event selection $Q^2\geq 1$\,GeV$^2$.	
(For $x>0.0568$ the integrals over the
regions $A$, $B$ and $C$ in Fig.~\ref{fig:xq2plane} were calculated 
separately, found to be consistent, and then averaged.)
The precision of the integrals given in Tab.~\ref{tab:mom} is less affected 
by the unfolding procedure since all inter-bin correlations from
the unfolding procedure are taken into account.
The statistical uncertainty is smaller by about 25$\%$   
compared to the case when only diagonal elements of the covariance
matrices are considered. 
Note that the error bars displayed in Figs.~\ref{fig:g1-all} to \ref{fig:xg1ns} 
are derived only from the diagonal elements of the covariance matrix, 
and the data points are statistically correlated.
The individual contributions to the systematic uncertainties are
displayed in Tab.~\ref{tab:mom-ratios}.
The systematic uncertainty is dominated by the
uncertainty on the polarization measurements. 

\renewcommand{\baselinestretch}{1.3}
\begin{table}[!t]
{\renewcommand{\baselinestretch}{1.}\caption{
\label{tab:mom}Integrals of $g_1^p$, $g_1^d$, $g_1^n$, and $g_1^{NS}$.
The uncertainties are separated into statistical, systematic components
coming from the experiment (misalignment, particle
identification and polarizations)
 and from the  parameterizations ($R$, $F_2$, $A_2$, $A_{zz}^d$,
$\omega_D$), and evolution uncertainty.  
}}
\begin{ruledtabular}
\begin{tabular}{c|c|c|rrrr}
& $\displaystyle\int_{0.021}^{0.9}\!\dx g_1$ &\multicolumn{4}{c}{uncertainties}\\ 
\cline{3-6}
    &             &    stat.    & syst.       & par.        & evol.\\
\hline
\multicolumn{6}{c}{$Q^2$=2.5 GeV$^2$}\\
\hline
 p  &~0.1201 & 0.0025 & 0.0068 & 0.0028 & 0.0046 \\
 d  &~0.0428 & 0.0011 & 0.0018 & 0.0008 & 0.0027 \\
 n  &-0.0276 & 0.0035 & 0.0079 & 0.0031 & 0.0017 \\ 
NS  &~0.1477 & 0.0055 & 0.0142 & 0.0055 & 0.0039 \\ 
\hline
\multicolumn{6}{c}{$Q^2$=5 GeV$^2$}\\
\hline
p  &~0.1211 & 0.0025 & 0.0068 & 0.0028 & 0.0050\\
d  &~0.0436 & 0.0012 & 0.0018 & 0.0008 & 0.0026\\
n  &-0.0268 & 0.0035 & 0.0079 & 0.0031 & 0.0018\\
NS &~0.1479 & 0.0055 & 0.0142 & 0.0055 & 0.0049\\
\end{tabular}
\end{ruledtabular}
\end{table}
\renewcommand{\baselinestretch}{1.3}
\begin{table}[!t]
{\renewcommand{\baselinestretch}{1.}
\caption{\label{tab:mom-ratios}Contributions to the
total systematic uncertainty of the integrals listed in Tab.~\ref{tab:mom},
over the measured $x$ range $0.021  <x<0.9$, calculated at
$Q^2_0=5$~GeV$^2$. The horizontal line separates the  
sources (experiment and parameterizations).}}
\begin{ruledtabular}
\begin{tabular}{l|c|c|c|c}
source & p   & d  & n & NS \\
\hline
\hline
Polarizations & 0.0066  &  0.0017 &  0.0076 &   0.0137\\
$PID_{cut}$   & 0.0002  &  0.0000 &  0.0002 &   0.0003\\
Misalignment  & 0.0016  &  0.0006 &  0.0020 &   0.0034\\
\hline
$A_2$         & 0.0002  &  0.0001 &  0.0003 &   0.0005 \\
$\sigma_{UU}$ & 0.0028  &  0.0008 &  0.0027 &   0.0053 \\
$A_{zz}^d$    &   -     &  0.0001 &  0.0002 &   0.0002 \\
$\omega_D$    &   -     &     -   &  0.0015 &   0.0015 \\
\end{tabular}
\end{ruledtabular}
\end{table}

A comparison of the integrals over the common measured $x$ range shows 
agreement with E143, as seen  in Tab.~\ref{tab:integrals_comparisons}. 
Comparisons with E155, SMC and E142 also show good agreement within   
uncertainties. 
SMC and E143 used the hypothesis that $g_1/F_1$ is
independent of $Q^2$ to perform the
evolution to a common $Q^2$, while E142 used the hypothesis of 
 $Q^2$-independence  of $A_1$ and E155 used QCD fits.
HERMES and E143 have almost identical $\langle Q^2\rangle$ values at
the same $x$.  
\renewcommand{\baselinestretch}{1.3}
\begin{table*}[ht]
{\renewcommand{\baselinestretch}{1}
\caption{\label{tab:integrals_comparisons}Comparisons of $g_1$ integrals
over different measured $x$ ranges from this experiment (including the
SIDIS measurement~\cite{hermesdeltaq} and the measurement of $g_1^n$ from a  $^3$He target~\cite{g1n_hermes}) and
SMC~\cite{Adeva:1998vv}, EMC~\cite{emc-new}, E143~\cite{Abe:1998wq},  E155~\cite{Anthony:1999rm}, E142~\cite{E142n}, and E154~\cite{E154n}.
In the case of E143, the normalization uncertainties, not included in the original result,
have been added in quadrature to the systematic uncertainties. 
The results from SMC, originally in the $x$ range $0.003 \leq x \leq 0.7$, 
as well as the results from E154, originally in the  $x$ range $0.014 \leq x \leq 0.7$,
the results from EMC, originally in the  $x$ range $0.01 \leq x \leq 0.7$, 
 and those of E155, originally in the range $0.01 \leq x \leq 0.9$,
have all been recalculated in the HERMES range $0.021 \leq x \leq 0.7$ 
from the $g_1$ values,
following the procedure used in this paper for the calculation of the 
HERMES moments (see Eq.~(\ref{mom1}))
and are indicated with an asterisk.
In the case of EMC it was not possible to calculate uncertainties from the evolution, as the $g_1$ values were already 
at $Q^2=10.7$~GeV$^2$, and no evolution uncertainty was given.
}}
\begin{ruledtabular}
\begin{tabular}{l|c|c|c|lllll}
Exp.        & $Q^2_0$  & $x$ range & type     &\multicolumn{5}{c}{Integral}    \\
\cline{5-9}
           & (GeV$^2$)&           &        &    value  &   stat. & syst. & param.  & evol.    \\
\hline
E143   & \multirow{2}{*}{5} & \multirow{2}{*}{0.03 - 0.8}& \multirow{2}{*}{p} & ~0.117 &0.003 &\multicolumn{2}{c}{0.007}  & -   \\
HERMES &                    &                            &                    & ~0.115 &0.002 &0.006 &0.003               &0.004\\
\hline
SMC (*)& \multirow{2}{*}{10}& \multirow{2}{*}{0.021-0.7} & \multirow{2}{*}{p} & ~0.120 &0.005 & \multicolumn{2}{c}{0.007} &0.002\\
HERMES &                    &                            &                    & ~0.119 &0.003 &0.007  &0.003              &0.005\\
\hline
EMC (*)&\multirow{2}{*}{10.7}& \multirow{2}{*}{0.021-0.7}&\multirow{2}{*}{p}  & ~0.110 &0.011 & \multicolumn{2}{c}{0.019} & -   \\
HERMES &                    &                            &                    & ~0.119 &0.003 &0.007 & 0.003              &0.005\\
\hline
E155 (*)& \multirow{2}{*}{5}& \multirow{2}{*}{0.021-0.9} & \multirow{2}{*}{p} & ~0.124 &0.002 & \multicolumn{2}{c}{0.009} &0.005\\
HERMES &                    &                            &                    & ~0.121 &0.002 &0.007  &0.003              &0.005\\
\hline
E143   & \multirow{2}{*}{5} & \multirow{2}{*}{0.03 - 0.8}& \multirow{2}{*}{d} & ~0.043 &0.003 & \multicolumn{2}{c}{0.003} & -   \\
HERMES &                    &                            &                    & ~0.042 &0.001 &0.002 &0.001               &0.002\\ 
\hline
SMC (*)&\multirow{2}{*}{10} & \multirow{2}{*}{0.021-0.7} & \multirow{2}{*}{d} & ~0.042 &0.005 & \multicolumn{2}{c}{0.004} &0.001\\
HERMES &                    &                            &                    & ~0.043 &0.001 &0.002  &0.001              &0.002\\
\hline
E155 (*)& \multirow{2}{*}{5}& \multirow{2}{*}{0.021-0.9} & \multirow{2}{*}{d} & ~0.043 &0.002 & \multicolumn{2}{c}{0.003} &0.003\\
HERMES &                    &                            &                    & ~0.044 &0.001 &0.002  &0.001              &0.003\\
\hline 
E142   & \multirow{2}{*}{2} & \multirow{2}{*}{0.03-0.6}  & n ($^3$He)         &-0.028  &0.006 & \multicolumn{2}{c}{0.006} & -   \\
HERMES &                    &                            & n (p,d)            &-0.025  &0.003 &0.007 &0.002               &0.001\\
\hline
E154 (*)& \multirow{2}{*}{2}& \multirow{2}{*}{0.021-0.7} & n ($^3$He)         &-0.032  &0.003 &\multicolumn{2}{c}{0.005}  &0.003\\
HERMES &                    &                            & n (p,d)            &-0.027  &0.004 &0.008 &0.003               &0.002\\
\hline
HERMES&\multirow{2}{*}{2.5}& \multirow{2}{*}{0.023-0.6} &  n ($^3$He)         & -0.034 & 0.013 &\multicolumn{2}{c}{0.005} & - \\
HERMES &                    &                           &  n (p,d)            & -0.027 & 0.003 &0.007 &0.003              & 0.001\\
\hline
HERMES/&\multirow{3}{*}{2.5}& \multirow{3}{*}{0.023-0.6} & \multirow{3}{*}{NS}& \multirow{2}{*}{~0.147}& \multirow{2}{*}{0.008} &\multicolumn{2}{c}{\multirow{2}{*}{0.019}} &\multirow{2}{*}{-}\\
SIDIS  & & & & & &  \\
HERMES &                       &                         &                    &~0.138 &0.005 &0.013&0.005&0.003 \\
\end{tabular}
\end{ruledtabular}
\end{table*}
\renewcommand{\baselinestretch}{1}

 For $x\rightarrow 1$, $g_1$ almost vanishes and a possible remaining small
contribution from the region $0.9\leq x \leq 1$ to the full  
integral was estimated for $Q^2=5$~GeV$^2$ to be $0.0003\pm
0.0003$ for the proton and $0.00006\pm 0.00005$ for the deuteron,
assuming a functional dependence of $g_1$ of the form
$(1-x)^{\alpha}(1+\beta x)$ in the high-$x$ region,  
and fitting HERMES data alone. From these values the high-$x$
contributions to the neutron and non-singlet integrals were estimated 
to be $-0.0002\pm0.0003$ and $0.0005\pm0.0006$, respectively.
 
Figure~\ref{fig:integrals} shows the cumulative integral of $g_1^{
p,d,n,NS}$ as a function of the lower integration limit in $x$.
\begin{figure}[!t]
\includegraphics[width=\columnwidth]{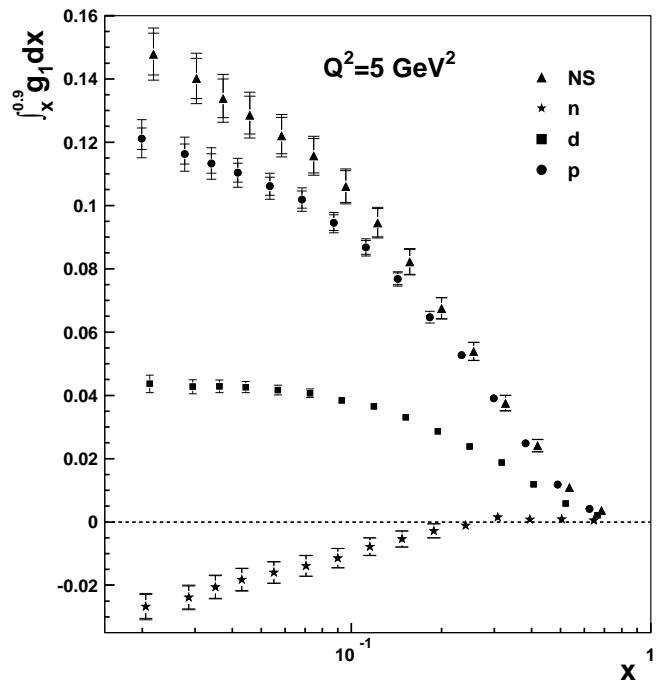}
\caption{\label{fig:integrals}
Integrals of $g_1^{p,d,n,NS}$ over the
 range $0.021\leq x \leq 0.9$ as a function of the low-$x$ limit of
integration, evaluated at  $Q^2=5$~GeV$^2$. 
Inner error bars represent total uncertainties excluding the normalization
systematic uncertainty from beam and target; 
outer error bars include the 
contribution coming from the $Q^2$ evolution.
Three of the four sets of points are slightly shifted in $x$ for visibility.
}
\end{figure}
For $x < 0.04$, $g_1^d(x)$ becomes
 compatible with zero (see also Fig.~\ref{fig:xg1-all}) and its measured 
integral shows saturation, while the other integrals
still show the tendency of a small rise in magnitude towards lower $x$. 
Also, the partial first moment of $g_1^d(x)$ calculated over the range 
$0<x<0.021$ at $Q^2 = 5$\,GeV$^2$ from the NLO QCD fit of all available data, 
used here for `evolution', was found to be consistent with zero within one 
statistical standard deviation. 
Hence, in the remaining discussion {\it it will be assumed that the deuteron first
moment $g_1^d$  saturates for $x < 0.04$}.
Under this assumption, conclusions
 can be drawn on the values of the singlet axial charge $a_0$ and the
 first moment of the singlet quark helicity distribution $\Delta \Sigma$ 
as well as on
 the strange-quark helicity distribution $\Delta s + \Delta \bar{s}$. 
 Using the Bjorken Sum Rule, the saturation of the integral of 
 $g_1^d$  allows
 an estimate of the possible contribution of the excluded region  
$0 < x < 0.021$ to the proton moment $\Gamma_1^p$.

In the following, results are given for $Q^2_0 = 5 $~GeV$^2$ and
in $\mathcal{O}(\alpha_s^3)$, unless otherwise noted. The right hand side
of Eq.~(\ref{bjsr}) yields for the
Bjorken Sum a value of  
$0.1821 \pm 0.0004 \pm 0.0019$, where the first uncertainty arises from 
$g_A/g_V$ and the second one from varying $\alpha_s$  within its
limits given by the value at the $Z^0$ mass: 
$ \alpha_s (Z^0) = 0.1187 \pm 0.002$~\cite{pdg}. 
An unambiguous test of the Bjorken Sum Rule
requires the measurement of both the proton and the deuteron
(neutron) integrals over the whole $x$-range. However, the proton
integral does not yet saturate in the measured region, as discussed
above, and therefore some uncertainty remains due to the
required extrapolation into the unmeasured small-$x$ region.

The HERMES integral for the non-singlet distribution (Eq.~\ref{g1ns}) in
the range $0.021 < x < 1$ has a value of 
$0.1484 \pm 0.0055\mathrm{(stat.)}\pm 0.0142\mathrm{(exp.)}\pm0.0055\mathrm{(param.)}
\pm 0.0049$(evol.). This partial moment is significantly 
smaller than the value for the Bjorken Sum, given for various orders in  
Table~\ref{tab:bjsr}, presumably because of the contribution to the proton
integral from the unmeasured region at small $x$. Assuming the validity of the 
Bjorken Sum Rule and ignoring possible higher-twist terms~\cite{bjsr-ht}, the
contribution from the unmeasured region $0 < x < 0.021$ to
$\Gamma_1^p$ has been evaluated as the difference between the inferred value
$\Gamma_1^p$({\it inferred}) and the measured value $\widetilde{\Gamma}^p_1$ 
in LO to NNNLO. Its central value ranges from 0.0316 in LO to 0.0169 in NNNLO
(see Table~\ref{tab:bjsr}). The NLO value is  in good agreement
with earlier estimates based on  %
QCD fits~\cite{ball-ridolfi}.

By combining Eqs.~(\ref{a3-deltaq}) and (\ref{bjsr}) (with $\Delta C_{NS}=1$),
this non-singlet integral can be directly compared in LO to the recently
published value for  $(\Delta u +\Delta \bar{u}) - (\Delta d+\Delta
\bar{d})$  obtained from semi-inclusive HERMES
data~\cite{hermesdeltaq} (see Table~\ref{tab:integrals_comparisons}). 
The partial non-singlet moment is calculated at $Q^2_0 = 2.5 $~GeV$^2$, 
for the sub-sample of the present data in the same kinematic
range as the semi-inclusive analysis, i.e. $0.023 < x < 0.6$ and $Q^2 > 1$~GeV$^2$. 
The resulting value
$6~\widetilde{\Gamma}_1^{NS}= 0.828 \pm 0.030\mathrm{(stat.)}\pm 0.078\mathrm{(syst.)}
\pm 0.030\mathrm{(param.)}\pm 0.018$(evol.)  is in
agreement with the published value $ (\Delta u +\Delta \bar{u}) - (\Delta d+\Delta
\bar{d})= 0.880\pm  0.045\mathrm{(stat.)}\pm 0.107 $(syst.) within the statistical
uncertainties. (The experimental systematic uncertainties are highly correlated). 

For convenience  in the following, the argument $\alpha_s(Q^2)$ of the
moments of the coefficient functions will be suppressed. 
The deuteron first moment is given by:
\begin{equation}\label{Gammad-a0}
\!\Gamma_1^d (Q^2_0)\!=\! \left(1-\frac{3}{2}\omega_D\right)\frac{1}{36} \left[
a_8\Delta C_{NS}^{\overline{MS}}+4 a_0 \Delta C_S^{\overline{MS}}\right],
\end{equation}
and one obtains: 
\begin{equation}\label{eq:a0}
a_0 (Q_0^2) = \frac{1}{\Delta C_S^{\overline{MS}}} \left[\frac{9 \Gamma_1^d }{\left(1
-\frac{3}{2}\omega_D\right)} -
\frac{1}{4} a_8\Delta C_{NS}^{\overline{MS}} \right].
\end{equation}
Note that the Bjorken Sum Rule is not employed here.
With $\alpha_s = 0.29 \pm 0.01$ for $Q^2_0 = 5 $~GeV$^2$, and
the  values for $\Delta C_{NS}^{\overline{MS}}$ and 
$\Delta C_S^{\overline{MS}}$ to $\mathcal{O}(\alpha_s^2)$,  
together with the values for $\widetilde{\Gamma}_1^d$  and $a_8$, 
the singlet axial charge yields a value 
$a_0 = 0.330 \pm0.011\mathrm{(theo.)}\pm0.025\mathrm{(exp.)}\pm 0.028$(evol.).  

\renewcommand{\baselinestretch}{1.3}
\begin{table*}[!t]
{\renewcommand{\baselinestretch}{1.}
\caption{\label{tab:bjsr}Expected Bjorken Sum (BJS) calculated in LO through
NNNLO, at $Q^2=5$~GeV$^2$, with uncertainties coming from $g_A/g_V$ and $\alpha_s$.
For the proton, the inferred  value of $\Gamma_1^p$ in the table  
is obtained using the BJS and the measured $\Gamma_1^d$ (assuming $\Gamma_1^d=\widetilde{\Gamma}^d_1$);
its first uncertainty  comes from $g_A/g_V$, $\alpha_s$, and $\omega_D$, 
the second from  $\Gamma_1^d$ (statistical and  systematic), while the last one comes from  the $Q^2$ evolution.
 Using the measured value for $\widetilde{\Gamma}_1^p$ and assuming the validity of the Bjorken Sum Rule, 
 the missing part of the proton integral due to the low-$x$ region is estimated
in the last column; 
its first uncertainty  comes from $g_A/g_V$, $\alpha_s$ and $\omega_D$, 
the second from  $\Gamma_1^p$ (statistical and  systematic), the third from  $\Gamma_1^d$ (statistical and  systematic), 
while the last one comes from  the $Q^2$ evolution on both  $\Gamma_1^p$  and  $\Gamma_1^d$.
}
\begin{ruledtabular}
\begin{tabular}{l|l|l|l|l}
     &                      &      BJS          &  $\Gamma_1^p$({\it inferred})                   & Estimated $\Gamma_1^p-\widetilde{\Gamma}^p_1$ \\
\hline
LO   &${\cal O}(\alpha_s^0)$&$ 0.2116\pm 0.0005$&$0.1530\pm 0.0008 \pm 0.0025\pm 0.0028$&$0.0316 \pm 0.0008 \pm 0.0025 \pm 0.0079\pm 0.0025$\\
NLO  &${\cal O}(\alpha_s^1)$&$ 0.1923\pm 0.0009$&$0.1434\pm 0.0008 \pm 0.0025\pm 0.0028$&$0.0219 \pm 0.0008 \pm 0.0025 \pm 0.0079\pm 0.0025$\\
NNLO &${\cal O}(\alpha_s^2)$&$ 0.1856\pm 0.0015$&$0.1400\pm 0.0009 \pm 0.0025\pm 0.0028$&$0.0186 \pm 0.0009 \pm 0.0025 \pm 0.0079\pm 0.0025$\\
NNNLO &${\cal O}(\alpha_s^3)$&$ 0.1821\pm 0.0019$&$0.1383\pm 0.0013 \pm 0.0025\pm 0.0028$&$0.0169 \pm 0.0013 \pm 0.0025 \pm 0.0079\pm 0.0025$\\
\end{tabular}
\end{ruledtabular}
}
\end{table*}

It is interesting to note that the deuteron target has an intrinsic
advantage over the proton target with respect to the precision that can
be achieved in the determination of $a_0$ from $\Gamma_1$,
in the typical case that the uncertainty in $\Gamma_1$ is dominated
by scale uncertainties associated with beam and target polarization.
This is because the magnitude of $\Gamma_1^d$ is only about 30\% of
that of $\Gamma_1^p$, so that a similar scale uncertainty produces
a correspondingly smaller absolute uncertainty in $\Gamma_1^d$,
and as a consequence also in $a_0$.  
This advantage of deuterium can be
expected to be also reflected in the impact of such data on 
the precision of global QCD fits.

The interpretation of quark distributions and their moments 
extracted in NLO is subject to some ambiguity due to their scheme
dependence.
Nevertheless some conclusions can be drawn, especially for the sake of
comparison to earlier results, if one stays within the $\overline{MS}$
scheme. 
In this scheme $a_0$ can be identified with $\Delta \Sigma$: $a_0 \MSeq
\Delta \Sigma = 
 (\Delta u  +\Delta \bar{u})+ (\Delta d +\Delta \bar{d}) +( \Delta s
+\Delta \bar{s}) $. The
value obtained for  $\Delta \Sigma(Q_0^2 = 5$~GeV$^2)$ is 
compared with other results 
from both experimental analyses  and QCD fits in Tab.~\ref{tab:deltasigma}. 
The HERMES value obtained from $\widetilde{\Gamma}_1^d$ 
is in good agreement with the earlier E143
result~\cite{Abe:1998wq} on the same target at $Q^2 = 3 $~GeV$^2$,
but less so with the SMC result~\cite{Adeva:1998vv} at $Q^2 = 10 $~GeV$^2$.
The HERMES value
represents about 55$\%$ of the relativistic QPM expectation of 0.6.
The HERMES data therefore suggest that the quark helicities
contribute a substantial fraction to the nucleon helicity, but
there is still need for a considerable contribution from gluons and/or
orbital angular momenta. 

From Tab.~\ref{tab:deltasigma} it can be seen that, among results 
for $\Delta \Sigma$ 
extracted from the data of individual experiments, those using estimates of 
the contribution from small $x$ based on NLO QCD fits tend to give
systematically smaller values that are more consistent with the QCD-fit values 
of $\Delta \Sigma$ in the lower section of the table.
This emphasizes the importance of the unmeasured region at small $x$.
We note that the partial first moment of $\Delta \Sigma(x)$
calculated over the range $0<x<0.021$
at $Q^2 = 5$\,GeV$^2$ from the NLO QCD fit of all available data that
is used here for `evolution' is found to be 
$-0.1334\pm 0.1104\mathrm{(stat.)}\pm 0.0273\mathrm{(syst.)}\pm 0.0604$(theo.).
It may also be relevant that all of the NLO QCD fits
listed here differ from the direct extractions in that they
impose symmetry among the sea flavours, and invoke the
Bjorken Sum Rule.

Under the assumption of SU(3) symmetry, the first moment
of the strange-quark helicity distribution can be obtained from the
relation 
\begin{multline}\label{eq:deltasextraction} 
  \Delta s +\Delta \bar{s} = \frac{1}{3} \left(a_0 - a_8\right) \\
= \frac{1}{\Delta C_S^{\overline{MS}}} \!\left[ \frac{3\Gamma_1^d}{(1-\frac{3}{2}\omega_d)}
  - \frac{a_8\left(4 \Delta C_S^{\overline{MS}} + \Delta
C_{NS}^{\overline{MS}}\right)}{12}\right]\,.
\end{multline}
The value 
$\Delta s  +\Delta \bar{s} = -0.085 \pm 0.013\mathrm{(theo.)}\pm0.008\mathrm{(exp.)}\pm 0.009$(evol.)
is obtained from HERMES deuteron data, in order ${\cal{O}}(\alpha_s^2)$.
This value is negative and different from zero by about 4.7~$\sigma$. 

The above results are
based on the assumption of the validity of SU(3) 
flavor symmetry in hyperon $\beta$-decays. 
The validity of this assumption is open to question. 
A recent analysis ~\cite{Ratcliffe04} of such data
leads, however, to the conclusion that such symmetry breaking effects
are small. Even if  
we assume that SU(3) symmetry is broken by up to 20$\%$ and that $a_8$
lies in the range $0.47 \leq a_8 \leq 0.70$, the above conclusions
are little changed and one obtains 
$0.358 \geq a_0 \geq 0.302$ and $-0.037 \leq \Delta s +\Delta \bar{s}\leq -0.133$.
It is interesting to note that a recent global NLO fit of the previous
world data for polarized inclusive and semi-inclusive deep-inelastic
scattering~\cite{dFS:2005}, which was made without the assumption of SU(3)
flavor symmetry, results in a small SU(3) breaking of -1\% to -8\%, a
value of $\Delta s$ between -0.045 and -0.051 and a value of $\Delta \Sigma$
between 0.284 and 0.311, depending on the set of NLO fragmentation
functions used in the analysis.

Values for $\Delta u  +\Delta \bar{u}$ and $\Delta d  +\Delta \bar{d}$ 
can be determined in the $\overline{MS}$ scheme 
using as experimental input the value of $a_0$
extracted from the  deuteron data and as
additional  input the values for the matrix elements $a_3$  and $a_8$:  
\begin{eqnarray}\label{eq:deltaudextraction}
\Delta u+\Delta \bar{u}& = &\frac{1}{6}[2a_0 + a_8 + 3 a_3]\,,\nn\\
\Delta d  +\Delta \bar{d}&=&\frac{1}{6}[2a_0 + a_8 - 3 a_3].
\end{eqnarray}
The values
$\Delta u +\Delta \bar{u}=  0.842 \pm 0.004\mathrm{(theo.)}\pm0.008\mathrm{(exp.)}\pm 0.009$(evol.) 
and 
$\Delta d +\Delta \bar{d}= -0.427 \pm 0.004\mathrm{(theo.)}\pm0.008\mathrm{(exp.)}\pm 0.009$(evol.) are
obtained. These results are essentially free from uncertainties due to
extrapolations  of $g_1$ towards $x = 0$,  provided the integral for $g_1^d$
really saturates at  $x < 0.04$,
but they invoke the validity of the Bjorken Sum Rule.

The values for $a_0$, $\Delta u +\Delta \bar{u}$, 
$\Delta d +\Delta \bar{d}$, and $\Delta s +\Delta \bar{s}$ 
have been calculated in LO, NLO and  NNLO from the deuteron integral
and are shown in Table~\ref{tab:a0}. 
\begin{table}[!t]
{\renewcommand{\baselinestretch}{1.}
\caption{\label{tab:a0} Values for $a_0$, $\Delta u +\Delta \bar{u}$, 
$\Delta d +\Delta \bar{d}$, and $\Delta s +\Delta \bar{s}$ 
calculated according to Eqs.~(\ref{eq:a0}), 
(\ref{eq:deltasextraction}), and (\ref{eq:deltaudextraction}) 
at $Q^2=5$~GeV$^2$.
Uncertainties are separated into theoretical ($a_3$, $a_8$ and $\alpha_s$)
and those coming from the  uncertainty of $\widetilde{\Gamma}_1^d$ 
(where 'exp.' includes statistical, systematic, and parameterizations, while 'evol.' 
comes from the evolution).
}}
\begin{ruledtabular}
\begin{tabular}{l|c|c|ccc}
     &                          & central    &\multicolumn{3}{c}{uncertainties}\\
\cline{4-6}
     &                          & value      & theor.      &   exp. & evol.\\
\hline
                                \multicolumn{6}{c}{$a_0$}  \\
\hline

LO   &   ${\cal O}(\alpha_s^0)$ & 0.278  &  0.010  &  0.022 & 0.025 \\
NLO  &   ${\cal O}(\alpha_s^1)$ & 0.321  &  0.011  &  0.024 & 0.028\\ 
NNLO &   ${\cal O}(\alpha_s^2)$ & 0.330  &  0.011  &  0.025 & 0.028\\
\hline
                                \multicolumn{6}{c}{$\Delta u +\Delta \bar{u}$}  \\
\hline
LO   &   ${\cal O}(\alpha_s^0)$ &0.825 & 0.004 & 0.007& 0.008\\
NLO  &   ${\cal O}(\alpha_s^1)$ &0.839 & 0.004 & 0.008& 0.009\\
NNLO &   ${\cal O}(\alpha_s^2)$ &0.842 & 0.004 & 0.008& 0.009\\
\hline
                                \multicolumn{6}{c}{$\Delta d +\Delta \bar{d}$ }\\  
\hline
LO   &   ${\cal O}(\alpha_s^0)$ &-0.444 & 0.004 & 0.007& 0.008\\  
NLO  &   ${\cal O}(\alpha_s^1)$ &-0.430 & 0.004 & 0.008& 0.009\\ 
NNLO &   ${\cal O}(\alpha_s^2)$ &-0.427 & 0.004 & 0.008& 0.009\\ 
\hline
                                \multicolumn{6}{c}{$\Delta s +\Delta \bar{s}$}\\
\hline
LO   &   ${\cal O}(\alpha_s^0)$ &-0.103 & 0.013 & 0.007& 0.008\\
NLO  &   ${\cal O}(\alpha_s^1)$ &-0.088 & 0.013 & 0.008& 0.009\\
NNLO &   ${\cal O}(\alpha_s^2)$ &-0.085 & 0.013 & 0.008& 0.009\\
\end{tabular}
\end{ruledtabular}
\end{table}
They are somewhat different from the 
partial moments that were previously extracted in LO by HERMES from semi-inclusive 
DIS data: $\int_{0.023}^{0.6} (\Delta u+\Delta \bar{u})\,\mathrm{d}x = 0.599\pm 0.022\pm0.065$,  
$\int_{0.023}^{0.6} (\Delta d+\Delta \bar{d})\,\mathrm{d}x = -0.280\pm 0.026\pm0.057$, and
$\int_{0.023}^{0.3} \Delta s\,\mathrm{d}x = 0.028\pm 0.033\pm0.009$~\cite{hermesdeltaq}. 
These partial moments for the up- and down-quark helicity 
distributions 
are smaller in magnitude than the full moments given in Table~\ref{tab:bjsr}, 
due to the 
restricted $x$-range. They also do not invoke the validity of the
Bjorken Sum Rule.
One possible explanation of the difference between the above quoted
partial moment
of the strange-quark helicity distribution and the value for 
$\frac{1}{2}(\Delta s +\Delta \bar{s})$ derived in this analysis could be
a substantial
negative contribution to it at small $x$.


\section{\label{sec:summary}Conclusions}

HERMES has measured the spin structure function
$g_1(x, Q^2)$ of the proton and the deuteron in the kinematic
range $0.0041 <  x < 0.90$ and 
$0.18 $~GeV$^2$$< Q^2 < 20 $~GeV$^2$. 
By combining the proton and the deuteron data,  
the neutron spin
structure function $g_1^n(x, Q^2)$ is extracted. 

In the HERMES analysis, 
the measured asymmetries are corrected for detector smearing
and QED radiative effects by applying an unfolding algorithm. 
As a consequence, the resulting data points are 
no longer correlated systematically but instead statistically.
The full information on the statistical
correlations is contained in the  covariance matrix.  
In order to avoid overestimating the statistical uncertainties,
 it is mandatory to take into account its off-diagonal 
elements when using the HERMES data for  any further analysis.
A fair comparison of the statistical power of different experiments
is given by the accuracy of integrals of structure functions.

The statistical precision of the HERMES proton data is
comparable to that of the hitherto most precise data from SLAC and
CERN in the same $x$ range.  
The HERMES deuteron data provide the most precise published determination of the 
spin structure function $g_1^d(x, Q^2)$, compared to
previous measurements. 
In the region $x<0.03$ the SMC data favor negative values, while
the HERMES deuteron data are compatible with zero, 
as are the recent COMPASS data.

The combination of the HERMES measurements of $g_1^p$ and $g_1^d$ constrains
the neutron spin structure function $g_1^n(x)$ well. Its accuracy is comparable to 
the E154 result that is obtained from a $^3He$ target.
Its behavior in the region $x<0.03$ with $\langle Q^2 \rangle <
1$~GeV$^2$ differs from the dramatic drop-off of $g_1^n(x)$ for $x \rightarrow 0$, 
earlier conjectured on the basis of previous data.

Integrals of the spin structure function $g_1$ of proton, 
deuteron and neutron have been calculated in various $x$-ranges and 
values of $Q^2_0$. The HERMES proton integral agrees
with E143 and within statistical uncertainties also with 
SMC, at their respective $Q^2_0$-values. The HERMES deuteron 
integral improves on the accuracy (statistical combined with systematic) 
of previous results.
It appears to saturate for $x <0.04$. 
Assuming this saturation, the 
first moment of the deuteron structure function $g_1(x,Q^2)$ can be 
determined with small uncertainties due to the extrapolation of $g_1^d$ 
towards $x = 0$.
It yields $\Gamma_1^d(Q^2 = 5$~GeV$^2) = 0.0437 \pm 0.0035$(total).
The difference of twice the proton integral in the measured range,
$\widetilde{\Gamma}_1^p = 0.1214 \pm 0.0093$(total) and the 
first deuteron moment, corrected for the D-wave admixture to the deuteron 
wave function, amounts to $\widetilde{\Gamma}_1^{NS} = 0.1484 \pm 0.0170$(total). 
It agrees within 2\,$\sigma$ with the Bjorken Sum Rule,
$\Gamma_1^{NS} = 0.1821 \pm 0.0019$, calculated in QCD 
in  $\mathcal{O}(\alpha_s^3)$.
From the difference of the theoretical value and the measured value, 
the contribution of the excluded region $ 0 \leq x < 0.021$ to 
the first proton moment $\Gamma_1^p$
is  estimated to yield only $0.017\pm 0.009$.

Based on the assumed saturation of the integral of $g_1^d$ and with the 
assumption of SU(3) flavor symmetry in the decays of hyperons in the
spin-$\frac {1} {2}$ baryon octet, the 
flavor-singlet axial charge $a_0$ has been determined 
in the $\overline{MS}$ scheme using  $\Gamma_1^d$ and the axial 
charge $a_8$ as inputs, in order $\alpha_s^2$. 
It amounts to 
\begin{multline} 
\bm{a_0}(Q^2 = 5\,\mathrm{GeV}^2) \bm{=} \\ 
\bm{0.330 \pm 0.011}\mathrm{(theo.)}\bm{\pm0.025}\mathrm{(exp.)}\bm{\pm 0.028}\mathrm{(evol.).}
\end{multline}
In this factorization scheme, this result can be
interpreted as the contribution $\Delta{\Sigma}$ 
of quark helicities to the nucleon helicity.
The HERMES data therefore suggest that the  quark helicities
contribute a substantial fraction to the nucleon helicity, but
there is still need for a major contribution from gluons and/or
orbital angular momenta of more than half of the sum of Eq.~(\ref{spinsum}).
Under the same assumptions, a negative value for the first moment of the
helicity distribution for strange quarks 
$\Delta s + \Delta \bar{s} = - 0.085 \pm 0.013\mathrm{(theo.)}\pm0.008\mathrm{(exp.)}\pm 0.009$(evol.)
is obtained from the deuteron data, in order $\alpha_s^2$.
 
Values for $\Delta u + \Delta \bar{u}$ and $\Delta d + \Delta \bar{d}$
have been determined in the  $\overline{MS}$ scheme using  $\Gamma_1^d$
and the axial charges $a_3$ and $a_8$ as inputs, and the values 
$\Delta u + \Delta \bar{u} = 0.842 \pm 0.004\mathrm{(theo.)}\pm0.008\mathrm{(exp.)}\pm 0.009$(evol.) 
and 
$\Delta d + \Delta \bar{d} =-0.427 \pm 0.004\mathrm{(theo.)}\pm0.008\mathrm{(exp.)}\pm 0.009$(evol.)
are obtained.


\begin{acknowledgments}
We gratefully acknowledge the DESY management for its support and the staff
at DESY and the collaborating institutions for their significant effort.
This work was supported by the FWO-Flanders, Belgium;
the Natural Sciences and Engineering Research Council of Canada;
the National Natural Science Foundation of China;
the Alexander von Humboldt Stiftung;
the German Bundesministerium f\"ur Bildung und Forschung (BMBF);
the Deutsche Forschungsgemeinschaft (DFG);
the Italian Istituto Nazionale di Fisica Nucleare (INFN);
the MEXT, JSPS, and COE21 of Japan;
the Dutch Foundation for Fundamenteel Onderzoek der Materie (FOM);
the U. K. Engineering and Physical Sciences Research Council, the
Particle Physics and Astronomy Research Councili and the
Scottish Universities Physics Alliance;
the U. S. Department of Energy (DOE) and the National Science Foundation (NSF)
and the Ministry of Trade and Economical Development and the Ministry
of Education and Science of Armenia.
\end{acknowledgments}

\begin{appendix}


\section{Unfolding algorithm }\label{app:unfolding}

\def\be{\begin{equation}}
\def\ee{\end{equation}}
\def\bea{\begin{eqnarray}}
\def\eea{\end{eqnarray}}
\def\bml{\begin{multline}}
\def\eml{\end{multline}}
\def\b{N^B}            
\def\bd{{\cal N}^B}    
\def\B{B}              
\def\x{N^X}            
\def\X{X}              
\def\bh{N^{bg}}        
\def\BH{X^{bg}}        
\def\K{{\cal K}}       
\def\Kb{{\cal K}^B}    
\def\sb{\sigma^B}      
\def\sx{\sigma^X}      
\def\sp{\sigma_{\frac{3}{2}}}
\def\sa{\sigma_{\frac{1}{2}}}
\def\bx{\bm{x}}        
\def\bt{\bm{t}}        
\def\UP{\left\{\begin{array}{c}U\\P\end{array}\right\}}
\def\XU{\left\{\x_u:A^m_{||}\x_u\right\}}

An unfolding algorithm was used to correct the asymmetries for 
the kinematic smearing effects of higher order QED 
processes and the spectrometer resolution.
The complexity of such an algorithm stems from the fact
that the calculable absolute cross section of the spin-dependent radiative 
background processes, {\em e.g.}, `elastic tails', must be effectively 
normalized to the data by comparing simulated and
measured unpolarized yields based on world data on $\sigma_{UU}$.

Radiative effects include vertex corrections to the
QED hard scattering amplitude, and kinematic migration
due to radiation of real photons by the lepton in
either the initial (ISR) or final (FSR) state.
Also ISR may in principle flip the helicity of the incident lepton
before it emits the virtual photon~\cite{helicityflipbrems}, 
reducing the effective
`photon depolarization parameter' to a degree that is
correlated with kinematic migration.  Thus this additional
`spin-state mixing' doubles the size of the smearing
matrix, as the kinematic bins for the two spin states
are coupled.  
This generality is retained here
for pedagogical reasons in spite of the suppression of
`helicity-flip bremsstrahlung' by an additional power of 
$\alpha$.

Our basic assumption here is that radiative effects
on the observed experimental cross sections $\sx(\bx)$
($X$ signifying experimental),
depending on various kinematic parameters collectively
designated $\bx$, can be represented by a pair of coupled 
integral equations containing elements of a 
matrix of radiative kernels $\K$:
\bea  \label{eq:Kernels1}
\!\!\!\!\!\!\!\!\sx_\sant(\bx) \!\!\!&=&\!\!\! \int\!\!\!\de\bt 
\left[\K_{\sant\sant}(\bx,\bt)\, \sa(\bt) + \K_{\sant\spar}(\bx,\bt)\, \sp(\bt)\right] \\ \label{eq:Kernels2}
\!\!\!\!\!\!\!\!\sx_\spar(\bx) \!\!\!&=&\!\!\! \int\!\!\!\de\bt 
\left[\K_{\spar\sant}(\bx,\bt)\, \sa(\bt) + \K_{\spar\spar}(\bx,\bt)\, \sp(\bt)\right].
\eea
Here the radiative (quasi-)elastic (Bethe-Heitler) background is 
included by continuing the integrals up to $x_{Bj}=1$.  
(We neglect the small contribution from the transverse asymmetry $A_2$.)
Because parity is conserved in QED,
we have 
\bea  \label{eq:symm}
\K_{\sant\sant} &=& \K_{\spar\spar} \ \equiv\  \K_{diag} \\ 
\K_{\sant\spar} &=& \K_{\spar\sant} \ \equiv\  \K_{off}\,.
\eea
In the absence of radiative effects, these equations simplify 
to represent the Born-approximation cross section $\sb(\bx)$ as:
\bea \label{eq:BornKernels1}
\sb_\sant(\bx) &=& 
\left[\Kb_{diag}(\bx)\, \sa(\bx) + \Kb_{off}(\bx)\, \sp(\bx)\right] \\
\label{eq:BornKernels2}
\sb_\spar(\bx) &=& 
\left[\Kb_{off}(\bx)\, \sa(\bx) + \Kb_{diag}(\bx)\, \sp(\bx)\right].
\eea
In this case, we can recognize the coefficients as representing the
effective `depolarization parameter':
\be
D(\bx) = \frac{\Kb_{diag}(\bx)-\Kb_{off}(\bx)} 
              {\Kb_{diag}(\bx)+\Kb_{off}(\bx)}\,.  
\ee
\begin{figure*}[t!]
\begin{center}
\centerline{
\includegraphics[width=\columnwidth]{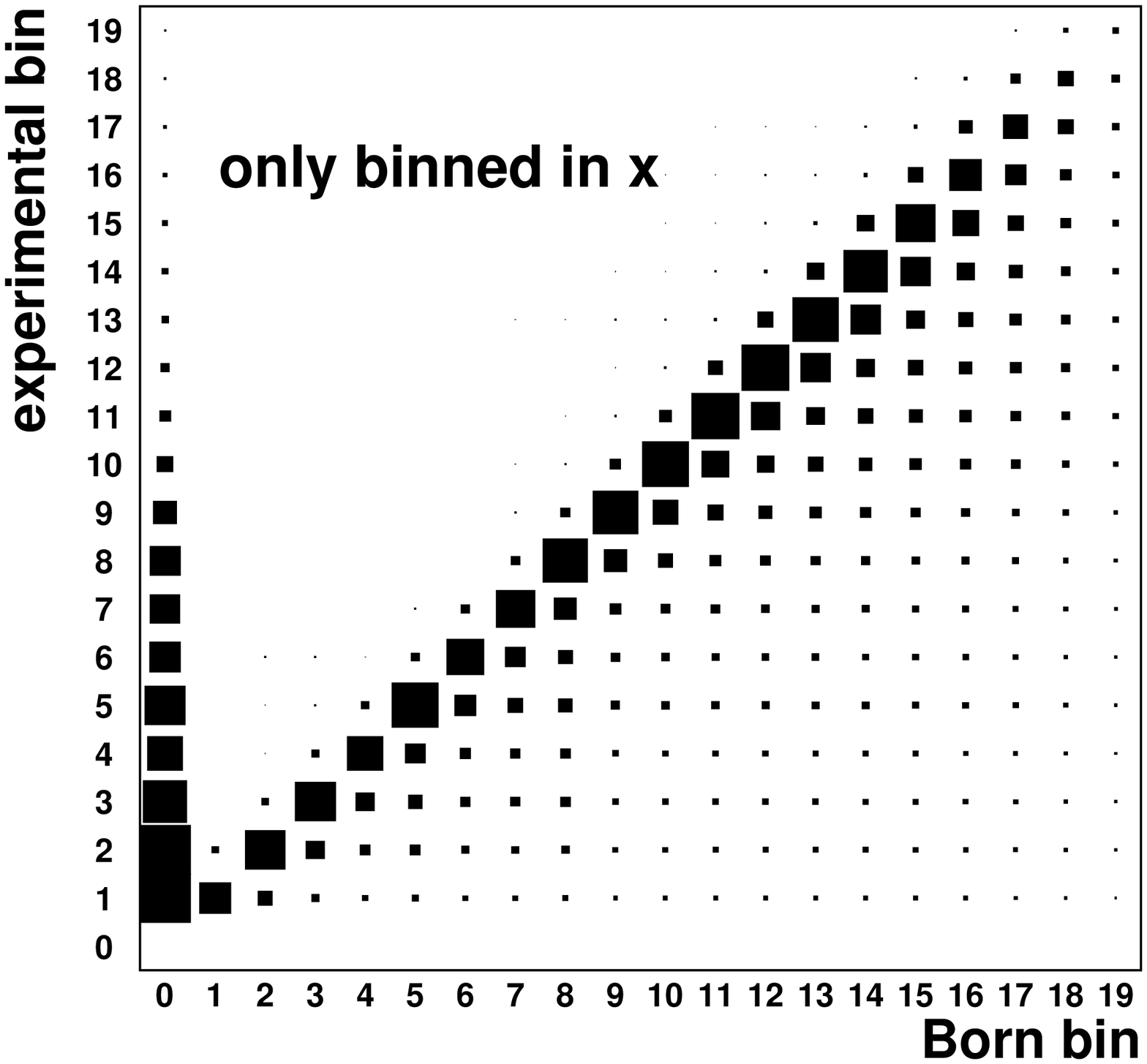}
\includegraphics[width=\columnwidth]{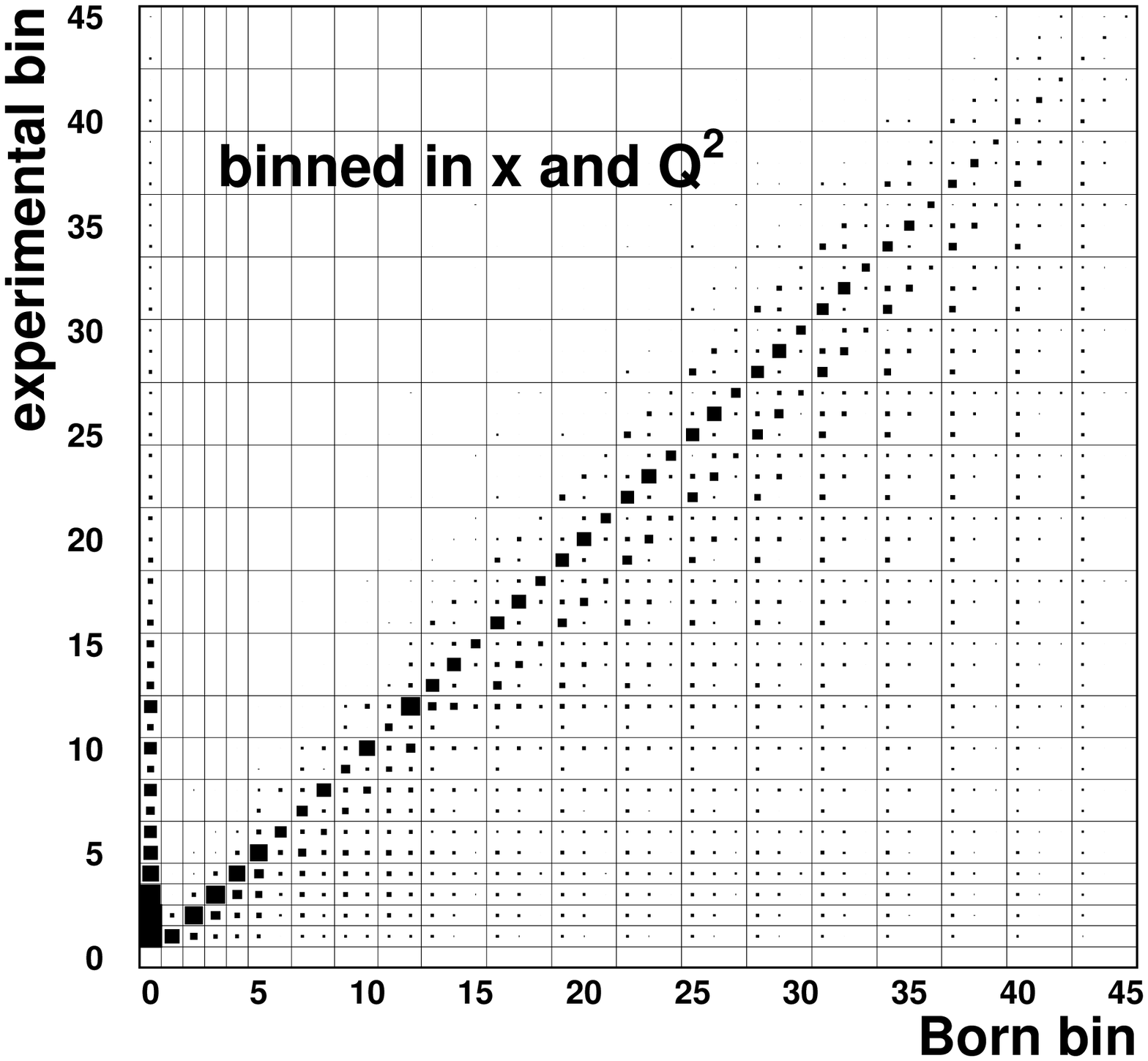}}
{\renewcommand{\baselinestretch}{1}
\caption[Migration matrices]{\label{migrationmatrices} 
{Migration matrices $M_{\spar}$ 
(the $M_{\sant}$ matrices look similar)
for two different binnings. Left panel: migration matrix for a 
pure $x$ binning. Right panel: migration matrix for the 
combined $x$ and $Q^2$ binning used in this analysis. 
The horizontal
and vertical lines divide the $x$-bins; within each $x$-bin the $Q^2$
bins are arranged in increasing $Q^2$.
The matrices are extracted from a fully reconstructed Monte Carlo data
set simulating both QED radiative and detector effects for inclusive
DIS on a proton target.}} }
\end{center}
\end{figure*}

Inverting Eqs.~(\ref{eq:BornKernels1})--(\ref{eq:BornKernels2}) and 
substituting in Eqs.~(\ref{eq:Kernels1})--(\ref{eq:Kernels2}),
we obtain equations relating Born and measured yields, of the form:
\bea  \label{eq:Ksmear1}
\!\!\!\!\!\!\!\!\sx_\sant(\bx) \!\!\!&=&\!\!\! \int\!\!\!\de\bt
\left[\K'_{diag}(\bx,\bt)\, \sb_\sant(\bt) + \K'_{off}(\bx,\bt)\, \sb_\spar(\bt)\right] \\ \label{eq:Ksmear2}
\!\!\!\!\!\!\!\!\sx_\spar(\bx) \!\!\!&=&\!\!\! \int\!\!\!\de\bt
\left[\K'_{off}(\bx,\bt)\, \sb_\sant(\bt) + \K'_{diag}(\bx,\bt)\, \sb_\spar(\bt)\right].
\eea
For the analysis of experimental data, 
the kernels in these equations are modified to also account for
the smearing effects of instrumental resolution.  The resulting equations 
are discretized by dividing the experimental (Born) kinematic space
into $n_i$ ($n_j$) bins, with $n_i \le n_j$.  It is convenient to work in
the $2 n_i$ ($2 n_j$) dimensional space containing vectors consisting 
of $n_i$ ($n_j$) event yields for the $\sant$ spin state followed 
by the same number for the $\spar$ state.  
The yield vectors binned in experimental and Born variables are related 
via the discretization of Eqs.~(\ref{eq:Ksmear1})--(\ref{eq:Ksmear2}):
\be   \label{eq:mcsmear}
\x = S \b + \bh\,, 
\ee
where $\x$ ($\b$) is a vector of length $2 n_i$ ($2 n_j$)
representing experimental (Born) yields, and $S$ is a $2 n_i \times 2 n_j$
smearing matrix.  The two $n_i \times n_j$ equal off-diagonal quarters 
of $S$ are designated $S_{off}$, and correspond to
$\K'_{off}$ in Eqs.~(\ref{eq:Ksmear1})--(\ref{eq:Ksmear2}), 
while the diagonal quarters $S_{diag}$ correspond to $\K'_{diag}$.  
As none of the kinematic bins are assumed to include the value $x_{Bj}=1$, 
the additional $2 n_i$-vector $\bh$ appears to account 
for the background contribution from the Bethe-Heitler process
as well as kinematic migration into the acceptance from outside.

The smearing matrix $S$ is extracted from two Monte Carlo simulations.
One simulation accounts for QED radiative effects and includes
complete detector simulation, followed by standard event reconstruction.
Hence kinematic smearing from both radiative and instrumental effects are included.
Radiative processes and vertex corrections
have been computed using the method outlined in
Ref.~\cite{radgen}. The elastic and quasi-elastic radiative backgrounds have
been evaluated using parameterizations of the nucleon form factors 
~\cite{gari} and the deuteron form factors ~\cite{locher}. For the deuteron,
the reduction of the bound nucleon cross section with respect to that for the
free nucleon (quasi-elastic suppression) was evaluated using the 
results from Ref.~\cite{bernabeu}. 
The DIS event generator uses parameterizations representing
the world data for $\sigma_{UU}$~\cite{allm:1997,whitlow:1990}.
The events were generated in the angular range
$|\theta_x| < 180$ mrad, $35$ mrad $< |\theta_y| < 150$ mrad
of {\itshape observed} kinematics.
The electrons are tracked through a detailed GEANT~\cite{geant} model
of the HERMES spectrometer and
reconstructed with the same HRC~\cite{Ackerstaff:1998av}
reconstruction algorithm as for real data. 

 For each simulated event, the 
calculation provides both the kinematic values reconstructed from
the spectrometer as well as the `Born' values describing the primary 
QED vertex where the incident lepton emits the virtual photon. 
Hence those Monte Carlo events for spin state, {\it e.g.}, $\sant$
that are selected under the
same criteria that are applied to the analysis of experimental
data can be sorted into a $n_i \times n_j$ `migration matrix' $M_\sant$,
where $M_\sant(i,j)$ contains the yield of events originating in Born bin $j$
and smeared into experimental bin $i$.  Background events originating
from the Bethe-Heitler process or from outside the acceptance
are {\em not} included in $M$.

As an example, Fig.~\ref{migrationmatrices} illustrates the migration
matrix $M_{\spar}(i,j)$. An $x$ distribution like the one in
Fig.~\ref{binmigration} represents a slice along a given Born bin $j$
in the matrix that is shown in the left panel of Fig.~\ref{migrationmatrices}.
In the right panel of Fig.~\ref{migrationmatrices}, a migration matrix
for the two-dimensional $x$-$Q^2$ binning is displayed. As the
unfolding procedure is oblivious to the kinematic proximity  of the
bins, it can be used for any kind of kinematic binning. 
Therefore a two-dimensional smearing problem can simply be reduced to
a one-dimensional problem.

The migration matrices are related
to the smearing matrix via the vector $\b$ of Born yields.
The vector $\b$ is not available from the same Monte Carlo simulation, 
as some events from inside the nominal kinematic acceptance migrate 
outside and are lost.
Hence $\b$ is obtained from a second Monte Carlo simulation with the
same luminosity but with both radiative and instrumental effects omitted. 
We designate $\bd_\sant$ and $\bd_\spar$ to be the $n_j \times n_j$ diagonal 
matrices with diagonal elements that are respectively the first and second 
halves of $\b$.  Then the migration matrices can be written as
\bea   \label{eq:migrate}
M_\sant &=& S_{diag}\, \bd_\sant + S_{off}\, \bd_\spar \\
M_\spar &=& S_{diag}\, \bd_\spar + S_{off}\, \bd_\sant \,. 
\eea
If we define a discriminant 
${\cal D} \equiv \bd_\sant \bd_\sant - \bd_\spar \bd_\spar$,
the above equations can be solved to yield:
\bea   \label{eq:msoln}
S_{diag} &=& \left[M_\sant\, \bd_\sant - M_\spar\, \bd_\spar\right] {\cal D}^{-1} \\
S_{off}  &=& \left[M_\spar\, \bd_\sant - M_\sant\, \bd_\spar\right] {\cal D}^{-1} \,.
\eea
Then $S$ can be assembled from $S_{diag}$ and $S_{off}$.
The off-diagonal terms give rise to some dependence on
the Monte Carlo model for $\b$.
The smearing matrix for polarization-averaged data is 
given by $S_u = S_{diag} + S_{off}$.

The experimental yields need have no absolute normalization
in order to subtract the calculable Bethe-Heitler background
because the polarization-averaged cross section has been
measured with high precision.  Those two known cross sections 
are encoded in the Monte Carlo event generator.  Hence 
effectively normalizing the polarization-averaged Monte Carlo
yields to the observed yields results in properly normalized 
Bethe-Heitler background yields from the Monte Carlo.
Representing this scheme here requires transformation of the 
$\sant$ and $\spar$ variables into polarization-averaged and 
polarization-difference variables. 
The $n_i$-vector of polarization-averaged yields $\x_u$ can be obtained 
through application of an $n_i \times 2 n_i$ projector matrix $U$
(signifying Unpolarized).  $U$ is constructed by juxtaposing
two $n_i \times n_i$ diagonal matrices, with all diagonal
elements of both left and right half equal to $\frac{1}{2}$.
Similarly, the vector of polarization-difference yields $\x_p$ 
can be obtained by application of the projector $P$, which differs
from $U$ only in that the diagonal elements of the right-hand half
are $-\frac{1}{2}$.  Thus we have
\bea  \label{eq:projmc}
\x_u &=& U (S \b + \bh) = S_u\b_u + U\bh  \\
\x_p &=& P (S \b + \bh)\,.
\eea

Eq.~(\ref{eq:mcsmear}) applies to the results of the Monte Carlo simulation,
which is assumed to describe reality with the possible exception of 
unsimulated polarization-independent detector inefficiencies.  A similar equation
relates the vectors representing the measured experimental yields $\X$
and the corresponding unknown Born yields $\B$, the polarization
asymmetry $A_{||}$ of which is the goal of the analysis.
We represent the detector efficiencies in this equation
as the elements of an $n_i \times n_i$ diagonal matrix $K_n$.
It is convenient to construct a $2 n_i \times 2 n_i$ diagonal matrix 
$K_{2n}$ by repeating $K_n$ as the diagonal quarters.  These constructions
are related by $U K_{2n} = K_n U$ and similarly for $P$.
The smearing of the experimental data is then given by:
\bea  \label{eq:smearx}
\X &=& K_{2n} S \B + \BH \\ \label{eq:smearxu}
\X_u &=& U (K_{2n} S \B + \BH) = K_n S_u \B_u + U\BH\ \\ \label{eq:smearxp}
\X_p &=& P (K_{2n} S \B + \BH)\,.
\eea

The Monte Carlo normalization relative to the 
experimental data is arbitrary.  This is represented here in terms of
an arbitrary constant $c$ relating the polarization-averaged Born yields
$\b_u$ from the Monte Carlo generator to $\B$ corresponding to the 
experimental observations: 
\be   \label{eq:normc}
\B_u = c \b_u\,.
\ee
The Monte Carlo also correctly describes the entire polarization-dependent
background $\bh$ and its relationship to the polarization-averaged
yields.  Hence the same normalization constant applies to the background terms:
\be   \label{eq:cbh}
\BH = c K_{2n} \bh\,.
\ee
Combining this relation with Eqs.~(\ref{eq:projmc}),
(\ref{eq:smearxu}) and (\ref{eq:normc}), we have: 
\begin{equation} \label{eq:ckNX}
\X_u = c K_n \x_u\,.
\end{equation}
Note that such a relationship does not generally hold for 
$\X_p$,  as the Monte Carlo model for $\x_p$ need not describe reality.

Rearranging Eq.~(\ref{eq:projmc}) and applying Eq.~(\ref{eq:normc}),
we get:
\be \label{eq:USB}
U S \B  = c \left[\x_u - U \bh\right]\,.
\ee
The diagonal $n_i\times n_i$ matrix $A^m_{||}$ containing
the measured asymmetries is defined as 
\be
A^m_{||}(i,i) \equiv \frac{\X_p(i)}{\X_u(i)} \,.
\ee
Then applying Eq.~(\ref{eq:smearxp}), we get:
\be
A^m_{||} \X_u = \X_p = P (K_{2n} S \B + \BH)\,.
\ee
Rearranging, applying Eqs.~(\ref{eq:cbh}), (\ref{eq:ckNX}) and 
$P K_{2n} = K_n P$, and then multiplying both sides by $K_n^{-1}$,
we get
\bea \nonumber
P K_{2n} S \B &=& A^m_{||} \X_u - P \BH \\ \label{eq:PSB}
P S \B  &=&  c  \left[A^m_{||} \x_u - P \bh\right]\,,
\eea
where we have used $A^m_{||} K_n = K_n A^m_{||}$, as both 
$K_n$ and $A^m_{||}$ are diagonal.
We now stack the two $n_i \times 2 n_i$ projector matrices $U$ and $P$
to form the $2 n_i \times 2 n_i$ projector $\UP$. 
We also concatinate the two vectors $\x_u$ and $A^m_{||} \x_u$ to form
the $2 n_i$-vector $\XU$.  With these combinations, we can unify
Eqs.~(\ref{eq:USB}) and (\ref{eq:PSB}):
\be  \label{eq:bigeq}
\UP S \B = c \left[\XU - \UP \bh\right]\,.
\ee
This system may be solved for the unknown vector $\B$.
If $n_i=n_j\equiv n$, this may be done by
multiplying both sides by 
$\left[\UP S\right]^{-1}$ to get:
\be \label{eq:Bsol}
\B =  c \left[\UP S\right]^{-1}\left[\XU - \UP \bh\right]\,.
\ee
Finally, we form the Born asymmetry of interest:
\be \label{eq:Asol}
A_{||}(j) = \frac{[P \B](j)}{[U \B](j)}\,,\quad j = 1\ldots\, n\,,
\ee
where the constant $c$ cancels.
The covariance matrix that follows from Eqs.~(\ref{eq:Bsol})--(\ref{eq:Asol})
describes the statistical correlation between the unfolded asymmetry values
for any two kinematic bins.  It is given as
\be
\textrm{cov}( A_{||})(i,j) =
\sum_{k=1}^{n}{ D(i,k) D(j,k)\, \sigma^2( A_{||}^m(k)) }\,,
\ee
where $\sigma( A_{||}^m(k))$ is the statistical uncertainty of the
measured asymmetry $A_{||}^m(k)$, and the matrix $D$ is defined as:
\bea \nonumber
D(j,i) &\equiv& \frac{\partial A_{||}(j)}{\partial A_{||}^m(i)} \\
&=& \frac{\left[P \left[\UP S\right]^{-1}\right](j,i+n)\, \x_u(i)}
         {\b_u(j)} \,,
\eea
where we have used Eq.~(\ref{eq:normc}) and $U \B=\B_u$.

Unfolding causes an inflation of the statistical uncertainty of the asymmetry,
driven by
the elastic contamination at low $x$ and by detector smearing at large $x$.
This last effect depends on the momentum resolution and is thus
larger at lower values of $y$ (and hence lower $Q^2$). Thus the $Q^2$ binning
separates regions with different degrees of smearing.
The amount  of smearing in the variable $Q^2$ is much smaller than that in $x$.
In Fig.~\ref{fig:inflation} (top two panels), the measured and
Born asymmetries are shown, while in the bottom panel the uncertainty inflation for
$A_{\|}$  is depicted. 
\begin{figure}[!t]
\includegraphics[width=0.93\columnwidth]{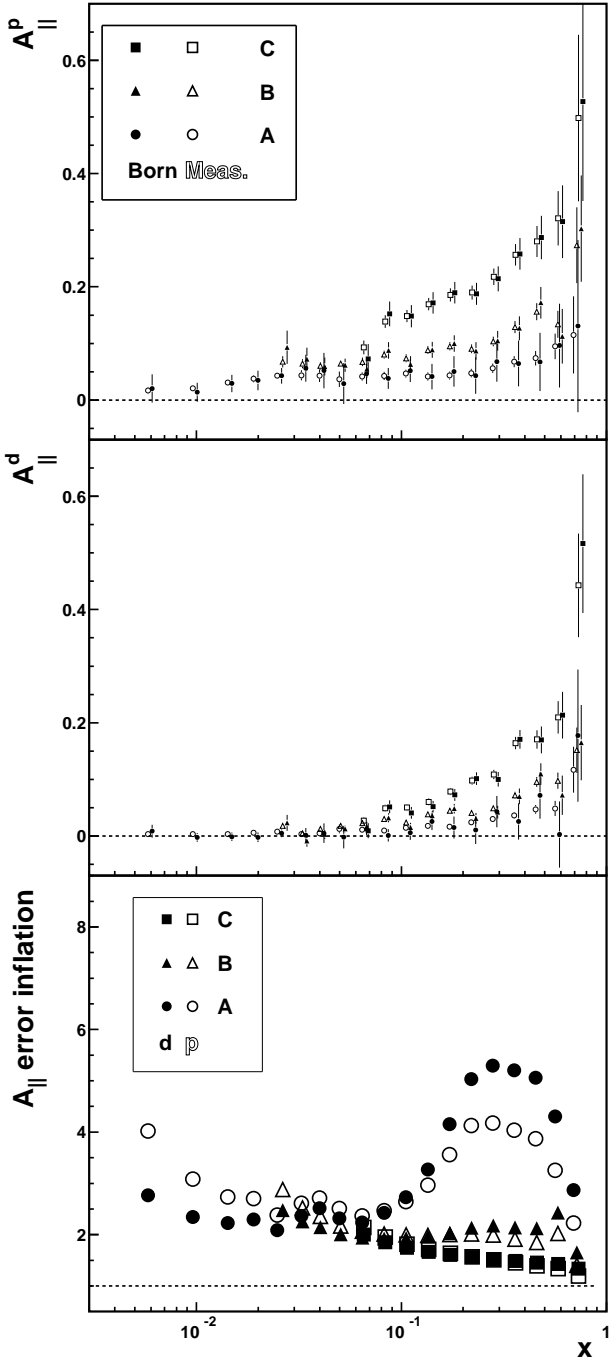}
{\renewcommand{\baselinestretch}{1}
\caption{\label{fig:inflation}Top two panels: measured and unfolded
Born asymmetries for proton 
and deuteron. The error bars of the Born asymmetries are the square root of 
the diagonal elements of the covariance matrix obtained after
unfolding, with the  inclusion of  the uncertainty due to the 
 statistics of the Monte Carlo. 
Bottom panel: uncertainty inflation when going from measured to Born
asymmetry due to the unfolding procedure. The symbols $A$, $B$ and $C$
refer to the $Q^2$-bin sets defined in Fig.~\ref{fig:xq2plane}.}}
\rule{0mm}{7mm}
\end{figure}

The statistical uncertainties of the Monte Carlo data enter mainly via the
simulated experimental count rates in the migration matrices.
In order to propagate these statistical uncertainties through the unfolding
algorithm with its matrix inversion and to evaluate the influence on the
unfolded {Born} level asymmetry $A_{||}$, a numerical approach is used. 
The measured asymmetry is unfolded $N=10000$ times using modified
migration matrices.  
Each time all elements are selected from a Gaussian
distribution corresponding to their respective statistical
uncertainty.
The standard deviation in each bin and the correlation
between different bins of the unfolded results are calculated.
The (statistics-based) covariance of $A_{||}$ coming from the
unfolding and the one coming from the finite statistics of the Monte
Carlo are summed; 
the result is called {\it statistical uncertainty}.


\section{Averages}\label{app:averages}

\subsubsection{Matrix calculation}

To obtain an expression for the weighted average of correlated
quantities, it is necessary to start from the $\chi^2$ definition:
\begin{equation}
\chi^2
=(\textbf{y}-F~\textbf{ I})^T ~\mathcal{ C}^{-1}~(\textbf{y}-F~\textbf{ I})\,,
\end{equation}
where $\textbf{y}$ is the vector of $n$ measurements of a given quantity, $F$
is the functional expression (in our case a constant) 
to be fitted,  $\mathcal{ C}$ is the covariance matrix of the measurements
 and $\textbf{I}$ is the vector with all components equal to 1.

The best value for $F$ is obtained by minimizing the $\chi^2$ with respect to $F$:
\be
\frac{\partial \chi^2}{\partial F}=-\textbf{ I}^T \mathcal{ C}^{-1}
\textbf{ y} -\textbf{ y}^T~  \mathcal{ C}^{-1}\textbf{ I}+2F ~\textbf{
I}^{T}\mathcal{ C}^{-1} ~ \textbf{ I}=0\,, 
\ee
so that:
\be\label{eq:b3}
F=\left(\textbf{ I}^{T}\mathcal{ C}^{-1} ~ \textbf{ I} \right)^{-1}\textbf{ I}^T \mathcal{
C}^{-1} \textbf{ y}\equiv \textbf{ A} \textbf{ y}\,.
\ee
The quantity $\textbf{ I}^{T}\mathcal{ C}^{-1} \textbf{
I} =\textrm{Tr}~{\cal C}^{-1}$ is a scalar. Therefore,
the statistical uncertainty of $F$ is given by:
\beqa\label{eq:b4}
\sigma^2_F&=&
\left( \frac{\partial F}{\partial \textbf{ y}}\right) \mathcal{
C}\left( \frac{\partial F}{\partial \textbf{ y}}\right)^T
=
\textbf{ A} ~\mathcal{ C}~ \textbf{ A}^T=\nn\\
&=&
\left(\textbf{ I}^{T}\mathcal{ C}^{-1} ~ \textbf{ I} \right)^{-2} 
\left( \textbf{ I}^{T}\mathcal{ C}^{-1} \right)~\mathcal{ C} ~ \left( \textbf{
I}^{T}\mathcal{ C}^{-1} \right)^T=\nn\\
&=&\left(\textbf{ I}^{T}\mathcal{ C}^{-1} ~ \textbf{ I} \right)^{-1}\,.
\eeqa
Only the following trivial cases of Eqs.~(\ref{eq:b3}) and
(\ref{eq:b4}) are required for this analysis.

\paragraph{Case $n$=2.}In the case of a  weighted average between two measurements, $F$ and
$\sigma^2_F$ are given  by the expressions:
\begin{eqnarray}\label{cov2}
F&=&\displaystyle{\frac{
y_1\left(\displaystyle{\mathcal{ C}_{22}-\mathcal{ C}_{12}}\right)+
y_2\left(\displaystyle{\mathcal{ C}_{11}-\mathcal{ C}_{12}}\right)
}{\displaystyle
{\mathcal{ C}_{11}+\mathcal{ C}_{22}-2\mathcal{ C}_{12}}
}}\nonumber\\
\sigma^2_F&=& \displaystyle\frac{\mathcal{ C}_{11}\mathcal{ C}_{22}-\mathcal{ C}_{12}^2}{\mathcal{ C}_{11}+\mathcal{ C}_{22}-2\mathcal{ C}_{12}}\,.
\end{eqnarray}

\paragraph{Case $n$=3.}In the case of three measurements the expression is more complicated.
One can define $a_1$, $a_2$ and $a_3$ as:
\beqa
a_1\!&=&\!\mathcal{ C}_{22}\mathcal{ C}_{33}-\mathcal{ C}_{23}^2-\mathcal{ C}_{12}\mathcal{ C}_{33}+\mathcal{ C}_{13}\mathcal{ C}_{23}+\mathcal{ C}_{12}\mathcal{ C}_{23}-\mathcal{ C}_{22}\mathcal{ C}_{13}\nn\\
a_2\!&=&\!\mathcal{ C}_{11}\mathcal{ C}_{33}-\mathcal{ C}_{13}^2-\mathcal{ C}_{12}\mathcal{ C}_{33}+\mathcal{ C}_{13}\mathcal{ C}_{23}+\mathcal{ C}_{12}\mathcal{ C}_{13}-\mathcal{ C}_{11}\mathcal{ C}_{23}\nn\\
a_3\!&=&\!\mathcal{ C}_{11}\mathcal{ C}_{22}-\mathcal{ C}_{12}^2-\mathcal{ C}_{13}\mathcal{ C}_{22}+\mathcal{ C}_{12}\mathcal{ C}_{23}+\mathcal{ C}_{12}\mathcal{ C}_{13}-\mathcal{ C}_{11}\mathcal{ C}_{23}\,,\nn\\
\eeqa
so that:
\be\label{cov3}
F=\frac{a_1y_1+a_2y_2+a_3y_3}{a_1+a_2+a_3}~~~~,~~~~\sigma^2_F=\frac{\textrm{det}~\mathcal{ C} }{a_1+a_2+a_3}\,.
\ee
It is easy to check that these expressions become the usual weighted average 
among independent measurements when $\mathcal{ C}_{ij}=0$ with $i\neq j$, and $\mathcal{ C}_{ii}=\sigma^2_i$ .

\subsubsection{Covariance matrix of the average.}

The expression for $g_1$ averaged over the $Q^2$ bins can be generically written
as:
\be
\left(g_1\right)_{n_x}=\textbf{ A} \left(g_1\right)_{45}\,,
\ee 
where   $n_x$ is the number of  $x$-bins
 taken into account (19 when no $Q^2$ cut is applied, and 15 for $Q^2\geq 1$~ GeV$^2$), 
and   $\textbf{ A}$ is a $n_x\times$45 matrix.
The covariance matrix of  $g_1$ in $n_x$ $x$-bins is then given by:
\be
\mathcal{ C}_{n_x}=\textbf{ A} ~\mathcal{ C}_{45}^{-1}~\textbf{ A}^T\,.
\ee


\bibliography{g1paper}

\section{Tables of Results}
\renewcommand{\baselinestretch}{1.2}
\begin{table*}
{\renewcommand{\baselinestretch}{1.}\caption{\label{tab:aparp}The measured  and Born
 asymmetries, $A_{||}^{m,p}$ and $A_\|^{p}$, at the average values of
$\langle x \rangle $, $\langle y \rangle $ and $\langle Q^2 \rangle $ in 45 bins, shown with
 statistical and systematic uncertainties. 
A normalization uncertainty of 5.2\%  has been included in the column `syst'.
}}
\begin{ruledtabular}
\begin{tabular}{ccccc|ccc|ccc}
bin & \!$x$ range\!&$\langle x\rangle$ & $\langle y\rangle$ & $\!\langle Q^2\rangle/$GeV$^2\!$ &
$~A_{||}^{m,p}$ & $\pm$ stat. & $\pm$ syst. & $A^{p}_\|$ & $\pm$stat.  & $\pm$ syst. \\
\hline
1 &0.0041 - 0.0073& 0.0058 & 0.866 & 0.26 &     0.0167 & 0.0060 & 0.0028 & 0.0205 & 0.0249 & 0.0026 \\
2 &0.0073 - 0.0118& 0.0096 & 0.824 & 0.41 &     0.0209 & 0.0053 & 0.0024 & 0.0138 & 0.0167 & 0.0022 \\
3 &0.0118 - 0.0168& 0.0142 & 0.778 & 0.57 &     0.0308 & 0.0055 & 0.0019 & 0.0294 & 0.0153 & 0.0027 \\
4 &0.0168 - 0.0212& 0.0190 & 0.738 & 0.73 &     0.0376 & 0.0063 & 0.0041 & 0.0347 & 0.0173 & 0.0031 \\
5 &0.0212 - 0.0295& 0.0248 & 0.642 & 0.82 &     0.0432 & 0.0056 & 0.0028 & 0.0428 & 0.0136 & 0.0032 \\
6 &0.0212 - 0.0295& 0.0264 & 0.818 & 1.12 &     0.0675 & 0.0101 & 0.0136 & 0.0931 & 0.0294 & 0.0074 \\
7 &0.0295 - 0.0362& 0.0325 & 0.515 & 0.87 &     0.0436 & 0.0089 & 0.0087 & 0.0561 & 0.0239 & 0.0031 \\
8 &0.0295 - 0.0362& 0.0329 & 0.732 & 1.25 &     0.0641 & 0.0078 & 0.0059 & 0.0724 & 0.0199 & 0.0057 \\
9 &0.0362 - 0.0444& 0.0399 & 0.435 & 0.90 &     0.0430 & 0.0112 & 0.0054 & 0.0522 & 0.0313 & 0.0031 \\
10 &0.0362 - 0.0444& 0.0403 & 0.659 & 1.38 &    0.0603 & 0.0066 & 0.0043 & 0.0592 & 0.0156 & 0.0048 \\
11 &0.0444 - 0.0568& 0.0498 & 0.361 & 0.93 &    0.0367 & 0.0138 & 0.0063 & 0.0291 & 0.0360 & 0.0029 \\
12 &0.0444 - 0.0568& 0.0506 & 0.586 & 1.54 &    0.0645 & 0.0053 & 0.0047 & 0.0613 & 0.0116 & 0.0051 \\
13 &0.0568 - 0.0727& 0.0643 & 0.375 & 1.25 &    0.0415 & 0.0074 & 0.0064 & 0.0466 & 0.0179 & 0.0028 \\
14 &0.0568 - 0.0727& 0.0645 & 0.555 & 1.85 &    0.0673 & 0.0076 & 0.0076 & 0.0549 & 0.0158 & 0.0044 \\
15 &0.0568 - 0.0727& 0.0655 & 0.761 & 2.58 &    0.0930 & 0.0122 & 0.0077 & 0.0724 & 0.0263 & 0.0069 \\
16 &0.0727 - 0.0929& 0.0823 & 0.308 & 1.31 &    0.0425 & 0.0072 & 0.0034 & 0.0382 & 0.0185 & 0.0026 \\
17 &0.0727 - 0.0929& 0.0824 & 0.484 & 2.06 &    0.0802 & 0.0072 & 0.0063 & 0.0876 & 0.0147 & 0.0053 \\
18 &0.0727 - 0.0929& 0.0835 & 0.712 & 3.08 &    0.1386 & 0.0111 & 0.0101 & 0.1521 & 0.0218 & 0.0104 \\
19 &0.0929 - 0.119& 0.1051 & 0.253 & 1.38 &     0.0467 & 0.0073 & 0.0059 & 0.0518 & 0.0200 & 0.0033 \\
20 &0.0929 - 0.119& 0.1054 & 0.420 & 2.29 &     0.0737 & 0.0070 & 0.0069 & 0.0631 & 0.0143 & 0.0043 \\
21 &0.0929 - 0.119& 0.1064 & 0.662 & 3.65 &     0.1484 & 0.0108 & 0.0083 & 0.1481 & 0.0197 & 0.0098 \\
22 &0.119 - 0.152& 0.1344 & 0.210 & 1.46 &      0.0415 & 0.0073 & 0.0071 & 0.0414 & 0.0226 & 0.0026 \\
23 &0.119 - 0.152& 0.1347 & 0.366 & 2.56 &      0.0883 & 0.0070 & 0.0049 & 0.0888 & 0.0141 & 0.0055 \\
24 &0.119 - 0.152& 0.1358 & 0.612 & 4.30 &      0.1693 & 0.0110 & 0.0095 & 0.1717 & 0.0189 & 0.0108 \\
25 &0.152 - 0.194& 0.1719 & 0.176 & 1.56 &      0.0433 & 0.0072 & 0.0077 & 0.0505 & 0.0270 & 0.0029 \\
26 &0.152 - 0.194& 0.1722 & 0.321 & 2.87 &      0.0950 & 0.0072 & 0.0070 & 0.0999 & 0.0146 & 0.0061 \\
27 &0.152 - 0.194& 0.1734 & 0.563 & 5.05 &      0.1857 & 0.0115 & 0.0104 & 0.1896 & 0.0189 & 0.0116 \\
28 &0.194 - 0.249& 0.2191 & 0.147 & 1.67 &      0.0476 & 0.0075 & 0.0046 & 0.0433 & 0.0324 & 0.0033 \\
29 &0.194 - 0.249& 0.2200 & 0.279 & 3.18 &      0.0906 & 0.0076 & 0.0060 & 0.0867 & 0.0154 & 0.0056 \\
30 &0.194 - 0.249& 0.2213 & 0.512 & 5.87 &      0.1900 & 0.0124 & 0.0105 & 0.1875 & 0.0195 & 0.0113 \\
31 &0.249 - 0.318& 0.2786 & 0.137 & 1.98 &      0.0561 & 0.0082 & 0.0086 & 0.0676 & 0.0354 & 0.0039 \\
32 &0.249 - 0.318& 0.2810 & 0.259 & 3.77 &      0.1035 & 0.0086 & 0.0065 & 0.1047 & 0.0174 & 0.0064 \\
33 &0.249 - 0.318& 0.2824 & 0.475 & 6.94 &      0.2178 & 0.0147 & 0.0123 & 0.2142 & 0.0222 & 0.0127 \\
34 &0.318 - 0.406& 0.3550 & 0.134 & 2.46 &      0.0678 & 0.0098 & 0.0041 & 0.0646 & 0.0405 & 0.0055 \\
35 &0.318 - 0.406& 0.3585 & 0.249 & 4.62 &      0.1289 & 0.0109 & 0.0074 & 0.1268 & 0.0210 & 0.0080 \\
36 &0.318 - 0.406& 0.3603 & 0.442 & 8.25 &      0.2565 & 0.0192 & 0.0142 & 0.2581 & 0.0278 & 0.0151 \\
37 &0.406 - 0.520& 0.4520 & 0.131 & 3.08 &      0.0738 & 0.0131 & 0.0080 & 0.0675 & 0.0514 & 0.0060 \\
38 &0.406 - 0.520& 0.4567 & 0.237 & 5.61 &      0.1562 & 0.0150 & 0.0130 & 0.1719 & 0.0278 & 0.0103 \\
39 &0.406 - 0.520& 0.4589 & 0.409 & 9.72 &      0.2801 & 0.0276 & 0.0161 & 0.2870 & 0.0384 & 0.0165 \\
40 &0.520 - 0.665& 0.5629 & 0.134 & 3.90 &      0.0948 & 0.0226 & 0.0079 & 0.0962 & 0.0739 & 0.0085 \\
41 &0.520 - 0.665& 0.5798 & 0.225 & 6.77 &      0.1338 & 0.0237 & 0.0126 & 0.1126 & 0.0481 & 0.0077 \\
42 &0.520 - 0.665& 0.5823 & 0.377 & 11.36 &     0.3212 & 0.0481 & 0.0179 & 0.3151 & 0.0643 & 0.0182 \\
43 &0.665 - 0.9& 0.6921 & 0.176 & 6.32 &        0.1150 & 0.0683 & 0.0137 & 0.1307 & 0.1522 & 0.0108 \\
44 &0.665 - 0.9& 0.7173 & 0.257 & 9.56 &        0.2733 & 0.0669 & 0.0196 & 0.3028 & 0.0939 & 0.0175 \\
45 &0.665 - 0.9& 0.7311 & 0.377 & 14.29 &       0.4981 & 0.1470 & 0.0298 & 0.5274 & 0.1757 & 0.0298 \\
\end{tabular}
\end{ruledtabular}
\end{table*}

\renewcommand{\baselinestretch}{1.2}
\begin{table*}
{\renewcommand{\baselinestretch}{1.}\caption{\label{tab:apard}The measured  and Born
 asymmetries, $A_{||}^{m,d}$ and $A_\|^{d}$, at the average values of
$\langle x \rangle $, $\langle y \rangle $ and $\langle Q^2 \rangle $ in 45 bins, shown with
 statistical and systematic uncertainties. 
A normalization uncertainty of 5\% has been included in the column `syst'.
}}
\begin{ruledtabular}
\begin{tabular}{ccccc|ccc|ccc}
bin & \!$x$ range\!&$\langle x\rangle$ & $\langle y\rangle$ & $\!\langle Q^2\rangle/$GeV$^2\!$ &
$~A_{||}^{m,d}$ & $\pm$stat. & $\pm$syst. & $A^{d}_\|$ & $\pm$stat.  & $\pm$syst. \\
\hline
1 &0.0041 - 0.0073& 0.0058 & 0.866 & 0.26 &     0.0032 & 0.0039 & 0.0041 & 0.0086 & 0.0114 & 0.0004 \\
2 &0.0073 - 0.0118& 0.0096 & 0.824 & 0.41 &     0.0032 & 0.0032 & 0.0031 & -0.0024 & 0.0078 & 0.0004 \\
3 &0.0118 - 0.0168& 0.0142 & 0.778 & 0.57 &     0.0034 & 0.0032 & 0.0031 & -0.0013 & 0.0074 & 0.0003 \\
4 &0.0168 - 0.0212& 0.0190 & 0.738 & 0.73 &     0.0058 & 0.0035 & 0.0052 & -0.0024 & 0.0085 & 0.0004 \\
5 &0.0212 - 0.0295& 0.0248 & 0.642 & 0.82 &     0.0077 & 0.0031 & 0.0013 & 0.0047 & 0.0069 & 0.0005 \\
6 &0.0212 - 0.0295& 0.0264 & 0.818 & 1.12 &     0.0180 & 0.0056 & 0.0028 & 0.0231 & 0.0141 & 0.0015 \\
7 &0.0295 - 0.0362& 0.0325 & 0.515 & 0.87 &     0.0034 & 0.0050 & 0.0010 & 0.0013 & 0.0127 & 0.0003 \\
8 &0.0295 - 0.0362& 0.0329 & 0.732 & 1.25 &     0.0050 & 0.0043 & 0.0079 & -0.0090 & 0.0100 & 0.0004 \\
9 &0.0362 - 0.0444& 0.0399 & 0.435 & 0.90 &     0.0046 & 0.0063 & 0.0042 & 0.0050 & 0.0172 & 0.0003 \\
10 &0.0362 - 0.0444& 0.0403 & 0.659 & 1.38 &    0.0125 & 0.0036 & 0.0049 & 0.0027 & 0.0080 & 0.0007 \\
11 &0.0444 - 0.0568& 0.0498 & 0.361 & 0.93 &    0.0124 & 0.0078 & 0.0101 & -0.0019 & 0.0200 & 0.0008 \\
12 &0.0444 - 0.0568& 0.0506 & 0.586 & 1.54 &    0.0180 & 0.0029 & 0.0032 & 0.0124 & 0.0061 & 0.0011 \\
13 &0.0568 - 0.0727& 0.0643 & 0.375 & 1.25 &    0.0109 & 0.0041 & 0.0009 & 0.0121 & 0.0100 & 0.0005 \\
14 &0.0568 - 0.0727& 0.0645 & 0.555 & 1.85 &    0.0227 & 0.0042 & 0.0035 & 0.0169 & 0.0084 & 0.0012 \\
15 &0.0568 - 0.0727& 0.0655 & 0.761 & 2.58 &    0.0269 & 0.0066 & 0.0079 & 0.0094 & 0.0135 & 0.0016 \\
16 &0.0727 - 0.0929& 0.0823 & 0.308 & 1.31 &    0.0096 & 0.0041 & 0.0049 & 0.0011 & 0.0109 & 0.0003 \\
17 &0.0727 - 0.0929& 0.0824 & 0.484 & 2.06 &    0.0299 & 0.0040 & 0.0019 & 0.0321 & 0.0080 & 0.0016 \\
18 &0.0727 - 0.0929& 0.0835 & 0.712 & 3.08 &    0.0491 & 0.0061 & 0.0037 & 0.0514 & 0.0114 & 0.0027 \\
19 &0.0929 - 0.119& 0.1051 & 0.253 & 1.38 &     0.0146 & 0.0041 & 0.0042 & 0.0058 & 0.0124 & 0.0005 \\
20 &0.0929 - 0.119& 0.1054 & 0.420 & 2.29 &     0.0238 & 0.0039 & 0.0046 & 0.0151 & 0.0079 & 0.0011 \\
21 &0.0929 - 0.119& 0.1064 & 0.662 & 3.65 &     0.0502 & 0.0060 & 0.0060 & 0.0409 & 0.0105 & 0.0023 \\
22 &0.119 - 0.152& 0.1344 & 0.210 & 1.46 &      0.0181 & 0.0041 & 0.0057 & 0.0256 & 0.0152 & 0.0009 \\
23 &0.119 - 0.152& 0.1347 & 0.366 & 2.56 &      0.0384 & 0.0039 & 0.0024 & 0.0360 & 0.0081 & 0.0019 \\
24 &0.119 - 0.152& 0.1358 & 0.612 & 4.30 &      0.0601 & 0.0061 & 0.0045 & 0.0517 & 0.0103 & 0.0025 \\
25 &0.152 - 0.194& 0.1719 & 0.176 & 1.56 &      0.0163 & 0.0041 & 0.0072 & 0.0152 & 0.0194 & 0.0005 \\
26 &0.152 - 0.194& 0.1722 & 0.321 & 2.87 &      0.0443 & 0.0041 & 0.0026 & 0.0482 & 0.0086 & 0.0023 \\
27 &0.152 - 0.194& 0.1734 & 0.563 & 5.05 &      0.0785 & 0.0064 & 0.0045 & 0.0726 & 0.0105 & 0.0033 \\
28 &0.194 - 0.249& 0.2191 & 0.147 & 1.67 &      0.0240 & 0.0043 & 0.0070 & 0.0104 & 0.0244 & 0.0012 \\
29 &0.194 - 0.249& 0.2200 & 0.279 & 3.18 &      0.0409 & 0.0043 & 0.0057 & 0.0311 & 0.0096 & 0.0021 \\
30 &0.194 - 0.249& 0.2213 & 0.512 & 5.87 &      0.0977 & 0.0070 & 0.0044 & 0.1013 & 0.0110 & 0.0043 \\
31 &0.249 - 0.318& 0.2786 & 0.137 & 1.98 &      0.0298 & 0.0048 & 0.0082 & 0.0431 & 0.0277 & 0.0017 \\
32 &0.249 - 0.318& 0.2810 & 0.259 & 3.77 &      0.0489 & 0.0050 & 0.0043 & 0.0422 & 0.0112 & 0.0024 \\
33 &0.249 - 0.318& 0.2824 & 0.475 & 6.94 &      0.1086 & 0.0084 & 0.0068 & 0.0999 & 0.0129 & 0.0043 \\
34 &0.318 - 0.406& 0.3550 & 0.134 & 2.46 &      0.0363 & 0.0058 & 0.0073 & 0.0257 & 0.0320 & 0.0023 \\
35 &0.318 - 0.406& 0.3585 & 0.249 & 4.62 &      0.0718 & 0.0063 & 0.0062 & 0.0699 & 0.0139 & 0.0032 \\
36 &0.318 - 0.406& 0.3603 & 0.442 & 8.25 &      0.1641 & 0.0111 & 0.0073 & 0.1706 & 0.0165 & 0.0072 \\
37 &0.406 - 0.520& 0.4520 & 0.131 & 3.08 &      0.0471 & 0.0078 & 0.0132 & 0.0715 & 0.0410 & 0.0043 \\
38 &0.406 - 0.520& 0.4567 & 0.237 & 5.61 &      0.0956 & 0.0089 & 0.0077 & 0.1094 & 0.0190 & 0.0047 \\
39 &0.406 - 0.520& 0.4589 & 0.409 & 9.72 &      0.1705 & 0.0163 & 0.0070 & 0.1696 & 0.0238 & 0.0072 \\
40 &0.520 - 0.665& 0.5629 & 0.134 & 3.90 &      0.0486 & 0.0133 & 0.0245 & 0.0027 & 0.0584 & 0.0053 \\
41 &0.520 - 0.665& 0.5798 & 0.225 & 6.77 &      0.0977 & 0.0142 & 0.0134 & 0.0720 & 0.0348 & 0.0032 \\
42 &0.520 - 0.665& 0.5823 & 0.377 & 11.36 &     0.2096 & 0.0287 & 0.0084 & 0.2134 & 0.0411 & 0.0088 \\
43 &0.665 - 0.9& 0.6921 & 0.176 & 6.32 &        0.1170 & 0.0405 & 0.0318 & 0.1774 & 0.1168 & 0.0106 \\
44 &0.665 - 0.9& 0.7173 & 0.257 & 9.56 &        0.1516 & 0.0401 & 0.0087 & 0.1650 & 0.0665 & 0.0066 \\
45 &0.665 - 0.9& 0.7311 & 0.377 & 14.29 &       0.4427 & 0.0916 & 0.0420 & 0.5164 & 0.1224 & 0.0210 \\
\end{tabular}
\end{ruledtabular}
\end{table*}
\begin{table*}[H]
{\renewcommand{\baselinestretch}{1.}\caption{\label{tab:A1_45}Virtual photon asymmetries $A_1^p$ 
and $A_1^d$ at the average $\langle x \rangle $ and $\langle Q^2 \rangle $ in 45 bins,
including statistical and systematic uncertainties.
A normalization uncertainty of 5.2\% for the proton and 5\% for the deuteron has been included in the column `syst'.
}}
\begin{ruledtabular}
\begin{tabular}{lcc|cccc|cccc}
$~~$bin$~~$ & $~~\langle x\rangle~~$ & $~\langle Q^2\rangle/$GeV$^2~$ &
 $~\displaystyle{A_1^p}~$ &$~\pm$stat.$~$ & $~\pm$syst.$~$ & $~\pm$par.$~$  &
 $~\displaystyle{A_1^d}~$ &$~\pm$stat.$~$ & $~\pm$syst.$~$ & $~\pm$par.$~$ \\
\hline
  1 &  0.0058 &  0.26 &  0.0221 &  0.0270 &  0.0028 &  0.0011 &  0.0092 &  0.0122 &  0.0005 &  0.0005 \\
  2 &  0.0096 &  0.41 &  0.0158 &  0.0192 &  0.0025 &  0.0007 & -0.0028 &  0.0090 &  0.0004 &  0.0001 \\
  3 &  0.0142 &  0.57 &  0.0364 &  0.0190 &  0.0034 &  0.0016 & -0.0017 &  0.0091 &  0.0004 &  0.0001 \\
  4 &  0.0190 &  0.73 &  0.0459 &  0.0229 &  0.0042 &  0.0022 & -0.0033 &  0.0111 &  0.0005 &  0.0002 \\
  5 &  0.0248 &  0.82 &  0.0667 &  0.0212 &  0.0050 &  0.0038 &  0.0072 &  0.0107 &  0.0007 &  0.0004 \\
  6 &  0.0264 &  1.12 &  0.1113 &  0.0351 &  0.0088 &  0.0036 &  0.0275 &  0.0167 &  0.0018 &  0.0009 \\
  7 &  0.0325 &  0.87 &  0.1137 &  0.0486 &  0.0063 &  0.0075 &  0.0022 &  0.0256 &  0.0006 &  0.0002 \\
  8 &  0.0329 &  1.25 &  0.0975 &  0.0268 &  0.0078 &  0.0039 & -0.0123 &  0.0133 &  0.0005 &  0.0005 \\
  9 &  0.0399 &  0.90 &  0.1289 &  0.0775 &  0.0078 &  0.0089 &  0.0119 &  0.0423 &  0.0007 &  0.0009 \\
 10 &  0.0403 &  1.38 &  0.0897 &  0.0236 &  0.0072 &  0.0039 &  0.0039 &  0.0120 &  0.0011 &  0.0002 \\
 11 &  0.0498 &  0.93 &  0.0885 &  0.1108 &  0.0090 &  0.0062 & -0.0068 &  0.0609 &  0.0025 &  0.0005 \\
 12 &  0.0506 &  1.54 &  0.1061 &  0.0201 &  0.0089 &  0.0046 &  0.0210 &  0.0104 &  0.0018 &  0.0009 \\
 13 &  0.0643 &  1.25 &  0.1358 &  0.0525 &  0.0082 &  0.0079 &  0.0344 &  0.0289 &  0.0015 &  0.0021 \\
 14 &  0.0645 &  1.85 &  0.1002 &  0.0290 &  0.0080 &  0.0038 &  0.0305 &  0.0152 &  0.0021 &  0.0012 \\
 15 &  0.0655 &  2.58 &  0.0918 &  0.0334 &  0.0089 &  0.0024 &  0.0116 &  0.0170 &  0.0020 &  0.0003 \\
 16 &  0.0823 &  1.31 &  0.1377 &  0.0673 &  0.0097 &  0.0076 &  0.0023 &  0.0389 &  0.0011 &  0.0005 \\
 17 &  0.0824 &  2.06 &  0.1871 &  0.0314 &  0.0120 &  0.0072 &  0.0678 &  0.0168 &  0.0034 &  0.0026 \\
 18 &  0.0835 &  3.08 &  0.2057 &  0.0295 &  0.0141 &  0.0079 &  0.0690 &  0.0153 &  0.0036 &  0.0027 \\
 19 &  0.1051 &  1.38 &  0.2318 &  0.0900 &  0.0149 &  0.0095 &  0.0236 &  0.0554 &  0.0023 &  0.0013 \\
 20 &  0.1054 &  2.29 &  0.1563 &  0.0355 &  0.0107 &  0.0056 &  0.0365 &  0.0194 &  0.0028 &  0.0014 \\
 21 &  0.1064 &  3.65 &  0.2131 &  0.0284 &  0.0146 &  0.0086 &  0.0585 &  0.0151 &  0.0033 &  0.0024 \\
 22 &  0.1344 &  1.46 &  0.2222 &  0.1233 &  0.0147 &  0.0078 &  0.1378 &  0.0821 &  0.0053 &  0.0046 \\
 23 &  0.1347 &  2.56 &  0.2539 &  0.0404 &  0.0160 &  0.0107 &  0.1022 &  0.0227 &  0.0056 &  0.0044 \\
 24 &  0.1358 &  4.30 &  0.2627 &  0.0289 &  0.0171 &  0.0094 &  0.0785 &  0.0156 &  0.0039 &  0.0028 \\
 25 &  0.1719 &  1.56 &  0.3247 &  0.1765 &  0.0208 &  0.0105 &  0.0941 &  0.1260 &  0.0035 &  0.0035 \\
 26 &  0.1722 &  2.87 &  0.3231 &  0.0474 &  0.0202 &  0.0135 &  0.1553 &  0.0274 &  0.0077 &  0.0065 \\
 27 &  0.1734 &  5.05 &  0.3095 &  0.0311 &  0.0198 &  0.0097 &  0.1179 &  0.0170 &  0.0055 &  0.0037 \\
 28 &  0.2191 &  1.67 &  0.3267 &  0.2539 &  0.0267 &  0.0118 &  0.0719 &  0.1899 &  0.0097 &  0.0038 \\
 29 &  0.2200 &  3.18 &  0.3160 &  0.0571 &  0.0211 &  0.0118 &  0.1113 &  0.0345 &  0.0079 &  0.0043 \\
 30 &  0.2213 &  5.87 &  0.3312 &  0.0349 &  0.0202 &  0.0101 &  0.1786 &  0.0194 &  0.0077 &  0.0054 \\
 31 &  0.2786 &  1.98 &  0.5419 &  0.2916 &  0.0327 &  0.0182 &  0.3470 &  0.2266 &  0.0141 &  0.0108 \\
 32 &  0.2810 &  3.77 &  0.4005 &  0.0681 &  0.0251 &  0.0137 &  0.1591 &  0.0429 &  0.0095 &  0.0056 \\
 33 &  0.2824 &  6.94 &  0.4020 &  0.0425 &  0.0252 &  0.0126 &  0.1866 &  0.0243 &  0.0082 &  0.0059 \\
 34 &  0.3550 &  2.46 &  0.5030 &  0.3315 &  0.0459 &  0.0191 &  0.1936 &  0.2598 &  0.0187 &  0.0079 \\
 35 &  0.3585 &  4.62 &  0.4901 &  0.0840 &  0.0342 &  0.0169 &  0.2720 &  0.0548 &  0.0130 &  0.0092 \\
 36 &  0.3603 &  8.25 &  0.5151 &  0.0570 &  0.0305 &  0.0170 &  0.3436 &  0.0337 &  0.0148 &  0.0112 \\
 37 &  0.4520 &  3.08 &  0.5114 &  0.4187 &  0.0491 &  0.0220 &  0.5691 &  0.3347 &  0.0358 &  0.0179 \\
 38 &  0.4567 &  5.61 &  0.6858 &  0.1155 &  0.0423 &  0.0240 &  0.4414 &  0.0782 &  0.0194 &  0.0149 \\
 39 &  0.4589 &  9.72 &  0.6137 &  0.0846 &  0.0359 &  0.0216 &  0.3654 &  0.0522 &  0.0158 &  0.0128 \\
 40 &  0.5629 &  3.90 &  0.6971 &  0.5745 &  0.0663 &  0.0289 & -0.0105 &  0.4548 &  0.0419 &  0.0085 \\
 41 &  0.5798 &  6.77 &  0.4508 &  0.2072 &  0.0328 &  0.0193 &  0.2927 &  0.1494 &  0.0139 &  0.0113 \\
 42 &  0.5823 & 11.36 &  0.7261 &  0.1526 &  0.0451 &  0.0277 &  0.4970 &  0.0976 &  0.0209 &  0.0187 \\
 43 &  0.6921 &  6.32 &  0.6805 &  0.8498 &  0.0603 &  0.0292 &  0.9701 &  0.6553 &  0.0600 &  0.0330 \\
 44 &  0.7173 &  9.56 &  1.0606 &  0.3406 &  0.0633 &  0.0415 &  0.5813 &  0.2416 &  0.0240 &  0.0225 \\
 45 &  0.7311 & 14.29 &  1.2060 &  0.4092 &  0.0692 &  0.0503 &  1.1976 &  0.2867 &  0.0493 &  0.0493 \\
\end{tabular}
\end{ruledtabular}
\end{table*}

\begin{table*}[H]
{\renewcommand{\baselinestretch}{1.}\caption{\label{tab:g1_46}Structure functions $g_{1}^{p}$ 
and $g_1^d$ at the average $\langle x \rangle $ and $\langle Q^2 \rangle $ in 45 bins,
including statistical and systematic uncertainties.
A normalization uncertainty of 5.2\% for the proton and 5\% for the deuteron has been included in the column `syst'.
}}
\begin{ruledtabular}
\begin{tabular}{lcc|cccc|cccc}
$~~$bin$~~$ & $~~\langle x\rangle~~$ & $~\langle Q^2\rangle/$GeV$^2~$ & $~\displaystyle{g_1^p}~$ &$~\pm$stat.$~$ & 
$~\pm$syst.$~$ & $~\pm$par.$~$  &
 $~g_1^d$  &$~\pm$stat.$~$ & $~\pm$syst.$~$
 & $~\pm$par.$~$ \\
\hline
 1 &  0.0058 &  0.26 &  0.2584 &  0.3138 &  0.0331 &  0.0697 &  0.1062 &  0.1400 &  0.0055 &  0.0286 \\
 2 &  0.0096 &  0.41 &  0.1357 &  0.1632 &  0.0213 &  0.0220 & -0.0224 &  0.0751 &  0.0037 &  0.0036 \\
 3 &  0.0142 &  0.57 &  0.2361 &  0.1229 &  0.0219 &  0.0231 & -0.0098 &  0.0575 &  0.0027 &  0.0010 \\
 4 &  0.0190 &  0.73 &  0.2416 &  0.1200 &  0.0217 &  0.0152 & -0.0158 &  0.0570 &  0.0027 &  0.0010 \\
 5 &  0.0248 &  0.82 &  0.2812 &  0.0890 &  0.0207 &  0.0126 &  0.0305 &  0.0437 &  0.0030 &  0.0014 \\
 6 &  0.0264 &  1.12 &  0.4736 &  0.1490 &  0.0375 &  0.0182 &  0.1140 &  0.0689 &  0.0074 &  0.0045 \\
 7 &  0.0325 &  0.87 &  0.3781 &  0.1609 &  0.0208 &  0.0135 &  0.0087 &  0.0824 &  0.0018 &  0.0003 \\
 8 &  0.0329 &  1.25 &  0.3459 &  0.0949 &  0.0273 &  0.0109 & -0.0405 &  0.0458 &  0.0018 &  0.0012 \\
 9 &  0.0399 &  0.90 &  0.3591 &  0.2147 &  0.0216 &  0.0121 &  0.0335 &  0.1132 &  0.0019 &  0.0010 \\
10 &  0.0403 &  1.38 &  0.2696 &  0.0706 &  0.0215 &  0.0079 &  0.0125 &  0.0346 &  0.0030 &  0.0004 \\
11 &  0.0498 &  0.93 &  0.2059 &  0.2534 &  0.0205 &  0.0069 & -0.0122 &  0.1335 &  0.0054 &  0.0004 \\
12 &  0.0506 &  1.54 &  0.2640 &  0.0498 &  0.0220 &  0.0078 &  0.0510 &  0.0248 &  0.0043 &  0.0014 \\
13 &  0.0643 &  1.25 &  0.2637 &  0.1011 &  0.0153 &  0.0091 &  0.0648 &  0.0532 &  0.0027 &  0.0019 \\
14 &  0.0645 &  1.85 &  0.2063 &  0.0591 &  0.0163 &  0.0064 &  0.0603 &  0.0295 &  0.0041 &  0.0016 \\
15 &  0.0655 &  2.58 &  0.1949 &  0.0702 &  0.0185 &  0.0048 &  0.0244 &  0.0338 &  0.0040 &  0.0006 \\
16 &  0.0823 &  1.31 &  0.2177 &  0.1049 &  0.0149 &  0.0080 &  0.0066 &  0.0573 &  0.0016 &  0.0003 \\
17 &  0.0824 &  2.06 &  0.3104 &  0.0520 &  0.0192 &  0.0106 &  0.1062 &  0.0262 &  0.0052 &  0.0029 \\
18 &  0.0835 &  3.08 &  0.3535 &  0.0504 &  0.0240 &  0.0090 &  0.1116 &  0.0246 &  0.0058 &  0.0026 \\
19 &  0.1051 &  1.38 &  0.2942 &  0.1133 &  0.0182 &  0.0107 &  0.0309 &  0.0649 &  0.0026 &  0.0008 \\
20 &  0.1054 &  2.29 &  0.2104 &  0.0472 &  0.0141 &  0.0075 &  0.0469 &  0.0241 &  0.0034 &  0.0013 \\
21 &  0.1064 &  3.65 &  0.2989 &  0.0396 &  0.0200 &  0.0086 &  0.0766 &  0.0195 &  0.0043 &  0.0018 \\
22 &  0.1344 &  1.46 &  0.2283 &  0.1240 &  0.0144 &  0.0072 &  0.1280 &  0.0754 &  0.0047 &  0.0026 \\
23 &  0.1347 &  2.56 &  0.2724 &  0.0430 &  0.0166 &  0.0090 &  0.1003 &  0.0223 &  0.0053 &  0.0024 \\
24 &  0.1358 &  4.30 &  0.2977 &  0.0325 &  0.0189 &  0.0093 &  0.0816 &  0.0160 &  0.0039 &  0.0019 \\
25 &  0.1719 &  1.56 &  0.2623 &  0.1396 &  0.0159 &  0.0063 &  0.0702 &  0.0889 &  0.0024 &  0.0011 \\
26 &  0.1722 &  2.87 &  0.2741 &  0.0398 &  0.0166 &  0.0074 &  0.1174 &  0.0209 &  0.0056 &  0.0023 \\
27 &  0.1734 &  5.05 &  0.2768 &  0.0274 &  0.0173 &  0.0080 &  0.0942 &  0.0135 &  0.0043 &  0.0021 \\
28 &  0.2191 &  1.67 &  0.2075 &  0.1545 &  0.0158 &  0.0040 &  0.0433 &  0.1003 &  0.0050 &  0.0008 \\
29 &  0.2200 &  3.18 &  0.2076 &  0.0367 &  0.0133 &  0.0043 &  0.0644 &  0.0196 &  0.0043 &  0.0010 \\
30 &  0.2213 &  5.87 &  0.2245 &  0.0232 &  0.0133 &  0.0052 &  0.1045 &  0.0113 &  0.0044 &  0.0020 \\
31 &  0.2786 &  1.98 &  0.2548 &  0.1329 &  0.0143 &  0.0047 &  0.1352 &  0.0866 &  0.0052 &  0.0034 \\
32 &  0.2810 &  3.77 &  0.1925 &  0.0318 &  0.0114 &  0.0036 &  0.0647 &  0.0171 &  0.0037 &  0.0011 \\
33 &  0.2824 &  6.94 &  0.1936 &  0.0199 &  0.0116 &  0.0037 &  0.0752 &  0.0096 &  0.0032 &  0.0011 \\
34 &  0.3550 &  2.46 &  0.1683 &  0.1049 &  0.0140 &  0.0036 &  0.0534 &  0.0660 &  0.0046 &  0.0016 \\
35 &  0.3585 &  4.62 &  0.1576 &  0.0259 &  0.0102 &  0.0030 &  0.0697 &  0.0138 &  0.0032 &  0.0016 \\
36 &  0.3603 &  8.25 &  0.1594 &  0.0171 &  0.0089 &  0.0031 &  0.0843 &  0.0081 &  0.0036 &  0.0014 \\
37 &  0.4520 &  3.08 &  0.1070 &  0.0808 &  0.0092 &  0.0030 &  0.0857 &  0.0489 &  0.0051 &  0.0035 \\
38 &  0.4567 &  5.61 &  0.1266 &  0.0203 &  0.0071 &  0.0031 &  0.0609 &  0.0105 &  0.0026 &  0.0020 \\
39 &  0.4589 &  9.72 &  0.1043 &  0.0139 &  0.0055 &  0.0023 &  0.0465 &  0.0065 &  0.0020 &  0.0013 \\
40 &  0.5629 &  3.90 &  0.0760 &  0.0578 &  0.0064 &  0.0017 &  0.0017 &  0.0325 &  0.0030 &  0.0001 \\
41 &  0.5798 &  6.77 &  0.0389 &  0.0164 &  0.0024 &  0.0013 &  0.0175 &  0.0084 &  0.0008 &  0.0008 \\
42 &  0.5823 & 11.36 &  0.0517 &  0.0105 &  0.0027 &  0.0015 &  0.0246 &  0.0047 &  0.0010 &  0.0011 \\
43 &  0.6921 &  6.32 &  0.0261 &  0.0300 &  0.0021 &  0.0009 &  0.0234 &  0.0154 &  0.0014 &  0.0013 \\
44 &  0.7173 &  9.56 &  0.0257 &  0.0079 &  0.0014 &  0.0011 &  0.0092 &  0.0037 &  0.0004 &  0.0006 \\
45 &  0.7311 & 14.29 &  0.0204 &  0.0068 &  0.0010 &  0.0006 &  0.0131 &  0.0031 &  0.0005 &  0.0013 \\
\end{tabular}
\end{ruledtabular}
\end{table*}

\clearpage
\begingroup
\squeezetable
\begin{table*}[H]
{\renewcommand{\baselinestretch}{1.}\caption{\label{tab:corr-46bins-hyd1}{Correlation matrix for 
$A_{||}^p$, $A_1^p$ and $g_1^p$ in 45 $x$-bins for proton (left).
 For $\langle x\rangle$ and $\langle Q^2\rangle$ of each bin, see e.g. 
Tabs.~\ref{tab:aparp} and \ref{tab:g1_46}.
}}}
\begin{ruledtabular}
\begin{tabular}{r|rrrrrrrrrrrrrrrrr}
  & 1 & 2 & 3 & 4 & 5 & 6 & 7 & 8 & 9 & 10 & 11 & 12 & 13 & 14 & 15 \\
\hline
 1 & \bf{1.000} &-0.159 & 0.005 &-0.005 &-0.004 & 0.002 &-0.003 & 0.001 &-0.003 & 0.000 &-0.004 & 0.000 &-0.002 & 0.001 & 0.000 \\
 2 &-0.159 & \bf{1.000} &-0.196 & 0.000 &-0.016 & 0.004 &-0.008 & 0.000 &-0.003 & 0.000 &-0.013 &-0.001 &-0.002 &-0.001 & 0.000 \\
 3 & 0.005 &-0.196 & \bf{1.000} &-0.230 &-0.014 & 0.005 &-0.018 & 0.001 &-0.012 & 0.001 &-0.038 & 0.001 & 0.000 &-0.003 & 0.001 \\
 4 &-0.005 & 0.000 &-0.230 & \bf{1.000} &-0.226 &-0.010 & 0.006 &-0.001 &-0.018 & 0.000 &-0.042 & 0.000 & 0.001 &-0.003 & 0.000 \\
 5 &-0.004 &-0.016 &-0.014 &-0.226 & \bf{1.000} &-0.127 &-0.254 &-0.004 & 0.006 &-0.005 &-0.068 &-0.004 &-0.003 &-0.003 & 0.002 \\
 6 & 0.002 & 0.004 & 0.005 &-0.010 &-0.127 & \bf{1.000} &-0.011 &-0.160 &-0.005 & 0.009 &-0.030 &-0.004 & 0.005 &-0.005 &-0.004 \\
 7 &-0.003 &-0.008 &-0.018 & 0.006 &-0.254 &-0.011 & \bf{1.000} &-0.062 &-0.331 &-0.005 & 0.004 &-0.003 &-0.014 & 0.004 & 0.001 \\
 8 & 0.001 & 0.000 & 0.001 &-0.001 &-0.004 &-0.160 &-0.062 & \bf{1.000} &-0.017 &-0.201 &-0.048 & 0.011 & 0.003 &-0.012 &-0.004 \\
 9 &-0.003 &-0.003 &-0.012 &-0.018 & 0.006 &-0.005 &-0.331 &-0.017 & \bf{1.000} &-0.041 &-0.383 & 0.001 &-0.003 & 0.004 & 0.001 \\
10 & 0.000 & 0.000 & 0.001 & 0.000 &-0.005 & 0.009 &-0.005 &-0.201 &-0.041 & \bf{1.000} &-0.053 &-0.223 & 0.023 & 0.002 &-0.001 \\
11 &-0.004 &-0.013 &-0.038 &-0.042 &-0.068 &-0.030 & 0.004 &-0.048 &-0.383 &-0.053 & \bf{1.000} &-0.061 &-0.049 & 0.010 & 0.002 \\
12 & 0.000 &-0.001 & 0.001 & 0.000 &-0.004 &-0.004 &-0.003 & 0.011 & 0.001 &-0.223 &-0.061 & \bf{1.000} &-0.225 &-0.134 &-0.052 \\
13 &-0.002 &-0.002 & 0.000 & 0.001 &-0.003 & 0.005 &-0.014 & 0.003 &-0.003 & 0.023 &-0.049 &-0.225 & \bf{1.000} &-0.032 & 0.008 \\
14 & 0.001 &-0.001 &-0.003 &-0.003 &-0.003 &-0.005 & 0.004 &-0.012 & 0.004 & 0.002 & 0.010 &-0.134 &-0.032 & \bf{1.000} &-0.070 \\
15 & 0.000 & 0.000 & 0.001 & 0.000 & 0.002 &-0.004 & 0.001 &-0.004 & 0.001 &-0.001 & 0.002 &-0.052 & 0.008 &-0.070 & \bf{1.000} \\
16 &-0.001 &-0.001 & 0.000 &-0.001 &-0.004 & 0.001 &-0.005 & 0.001 &-0.009 &-0.006 &-0.007 & 0.033 &-0.364 &-0.044 &-0.006 \\
17 & 0.000 &-0.001 &-0.001 &-0.002 &-0.002 &-0.002 & 0.001 &-0.005 & 0.002 &-0.008 & 0.002 & 0.002 & 0.017 &-0.217 &-0.051 \\
18 & 0.000 & 0.000 & 0.000 & 0.000 & 0.001 &-0.002 & 0.001 &-0.002 & 0.000 &-0.003 & 0.001 &-0.006 & 0.000 & 0.015 &-0.132 \\
19 &-0.001 &-0.001 &-0.001 & 0.000 &-0.003 & 0.001 &-0.005 & 0.001 &-0.003 & 0.000 &-0.010 &-0.012 & 0.068 & 0.002 &-0.008 \\
20 & 0.000 & 0.000 &-0.001 &-0.001 &-0.002 &-0.001 & 0.001 &-0.004 & 0.001 &-0.004 & 0.002 &-0.008 &-0.002 & 0.007 & 0.017 \\
21 & 0.000 & 0.000 & 0.000 & 0.000 & 0.001 &-0.002 & 0.000 &-0.001 & 0.000 &-0.001 & 0.000 &-0.004 & 0.001 & 0.000 &-0.012 \\
22 &-0.001 &-0.001 &-0.001 &-0.001 &-0.001 & 0.000 &-0.002 & 0.001 &-0.005 &-0.001 &-0.004 & 0.001 &-0.036 &-0.007 &-0.001 \\
23 & 0.000 & 0.000 &-0.001 &-0.001 &-0.001 &-0.001 & 0.001 &-0.003 & 0.001 &-0.003 & 0.001 &-0.004 & 0.002 &-0.015 &-0.002 \\
24 & 0.000 & 0.000 & 0.000 & 0.000 & 0.001 &-0.001 & 0.000 &-0.001 & 0.000 &-0.001 & 0.000 &-0.002 & 0.000 & 0.002 &-0.009 \\
25 &-0.001 & 0.000 & 0.000 & 0.000 &-0.001 & 0.001 &-0.003 & 0.001 & 0.000 & 0.000 &-0.004 &-0.002 & 0.009 & 0.000 & 0.000 \\
26 & 0.000 & 0.000 &-0.001 &-0.001 &-0.001 &-0.001 & 0.001 &-0.002 & 0.000 &-0.002 & 0.000 &-0.002 & 0.001 &-0.007 & 0.001 \\
27 & 0.000 & 0.000 & 0.000 & 0.000 & 0.000 &-0.001 & 0.000 &-0.001 & 0.000 &-0.001 & 0.000 &-0.001 & 0.000 & 0.001 &-0.006 \\
28 &-0.001 &-0.001 &-0.001 &-0.001 &-0.001 & 0.000 &-0.001 & 0.000 &-0.003 & 0.000 & 0.000 &-0.001 &-0.010 & 0.000 & 0.000 \\
29 & 0.000 & 0.000 & 0.000 &-0.001 & 0.000 &-0.001 & 0.001 &-0.002 & 0.000 &-0.001 & 0.002 &-0.002 & 0.001 &-0.005 & 0.000 \\
30 & 0.000 & 0.000 & 0.000 & 0.000 & 0.000 & 0.000 & 0.000 &-0.001 & 0.000 & 0.000 & 0.000 &-0.001 & 0.000 & 0.001 &-0.004 \\
31 & 0.000 &-0.001 & 0.001 & 0.000 &-0.001 & 0.001 &-0.001 & 0.001 & 0.001 & 0.000 &-0.005 &-0.001 & 0.002 & 0.000 & 0.000 \\
32 & 0.001 & 0.000 & 0.000 & 0.000 & 0.000 &-0.001 & 0.001 &-0.001 & 0.001 & 0.000 &-0.001 &-0.001 & 0.002 &-0.003 & 0.000 \\
33 & 0.000 & 0.000 & 0.000 & 0.000 & 0.000 & 0.000 & 0.000 & 0.000 & 0.000 & 0.000 & 0.000 &-0.001 & 0.000 & 0.001 &-0.002 \\
34 &-0.002 & 0.000 &-0.002 & 0.000 &-0.001 & 0.000 &-0.001 &-0.001 &-0.002 & 0.000 & 0.004 &-0.001 &-0.004 &-0.001 & 0.000 \\
35 & 0.000 & 0.001 & 0.000 & 0.000 & 0.001 &-0.001 & 0.001 & 0.000 & 0.000 & 0.000 & 0.003 & 0.000 & 0.000 &-0.001 &-0.001 \\
36 & 0.000 & 0.000 & 0.000 & 0.000 & 0.000 & 0.000 & 0.000 & 0.000 & 0.000 & 0.000 & 0.000 & 0.000 & 0.000 & 0.000 &-0.001 \\
37 & 0.001 &-0.001 & 0.002 &-0.002 & 0.000 & 0.001 &-0.001 & 0.000 & 0.000 &-0.001 &-0.010 & 0.001 & 0.002 & 0.000 & 0.000 \\
38 & 0.001 & 0.000 & 0.001 &-0.001 & 0.001 & 0.000 & 0.001 &-0.001 & 0.001 &-0.001 &-0.003 & 0.000 & 0.001 &-0.001 &-0.001 \\
39 & 0.000 & 0.000 & 0.000 & 0.000 & 0.000 & 0.000 & 0.000 & 0.000 & 0.000 &-0.001 & 0.000 &-0.001 & 0.000 & 0.000 & 0.000 \\
40 &-0.002 & 0.000 &-0.002 & 0.002 &-0.002 & 0.000 &-0.002 & 0.001 &-0.003 & 0.002 & 0.007 &-0.001 &-0.003 & 0.001 & 0.001 \\
41 &-0.001 & 0.001 & 0.001 & 0.002 & 0.001 & 0.000 & 0.000 & 0.003 &-0.002 & 0.005 & 0.000 & 0.003 & 0.003 & 0.004 &-0.001 \\
42 & 0.000 & 0.000 & 0.000 & 0.000 & 0.000 & 0.000 & 0.000 & 0.000 & 0.000 & 0.000 & 0.000 & 0.000 & 0.000 & 0.000 & 0.000 \\
43 & 0.001 &-0.002 &-0.002 &-0.003 &-0.001 &-0.001 & 0.000 &-0.005 & 0.002 &-0.008 & 0.000 &-0.006 &-0.004 &-0.006 & 0.000 \\
44 & 0.000 & 0.000 & 0.000 & 0.000 & 0.000 & 0.000 & 0.000 & 0.000 & 0.000 &-0.001 & 0.000 & 0.000 & 0.000 & 0.000 & 0.000 \\
45 & 0.000 & 0.000 & 0.000 & 0.000 & 0.000 & 0.000 & 0.000 & 0.000 & 0.000 & 0.000 & 0.000 & 0.000 & 0.000 & 0.000 & 0.000 \\
\end{tabular}
\end{ruledtabular}
\end{table*}
\endgroup
%
%
\addtocounter{table}{-1}
\begingroup
\squeezetable
\begin{table*}[H]
\caption{ -- continued.}
\begin{ruledtabular}
\begin{tabular}{r|rrrrrrrrrrrrrrrrrrr}
  &16 & 17 & 18 & 19 & 20 & 21 & 22 & 23 & 24 & 25 & 26 & 27 & 28 & 29 & 30 \\
\hline
 1 &-0.001 & 0.000 & 0.000 &-0.001 & 0.000 & 0.000 &-0.001 & 0.000 & 0.000 &-0.001 & 0.000 & 0.000 &-0.001 & 0.000 & 0.000 \\
 2 &-0.001 &-0.001 & 0.000 &-0.001 & 0.000 & 0.000 &-0.001 & 0.000 & 0.000 & 0.000 & 0.000 & 0.000 &-0.001 & 0.000 & 0.000 \\
 3 & 0.000 &-0.001 & 0.000 &-0.001 &-0.001 & 0.000 &-0.001 &-0.001 & 0.000 & 0.000 &-0.001 & 0.000 &-0.001 & 0.000 & 0.000 \\
 4 &-0.001 &-0.002 & 0.000 & 0.000 &-0.001 & 0.000 &-0.001 &-0.001 & 0.000 & 0.000 &-0.001 & 0.000 &-0.001 &-0.001 & 0.000 \\
 5 &-0.004 &-0.002 & 0.001 &-0.003 &-0.002 & 0.001 &-0.001 &-0.001 & 0.001 &-0.001 &-0.001 & 0.000 &-0.001 & 0.000 & 0.000 \\
 6 & 0.001 &-0.002 &-0.002 & 0.001 &-0.001 &-0.002 & 0.000 &-0.001 &-0.001 & 0.001 &-0.001 &-0.001 & 0.000 &-0.001 & 0.000 \\
 7 &-0.005 & 0.001 & 0.001 &-0.005 & 0.001 & 0.000 &-0.002 & 0.001 & 0.000 &-0.003 & 0.001 & 0.000 &-0.001 & 0.001 & 0.000 \\
 8 & 0.001 &-0.005 &-0.002 & 0.001 &-0.004 &-0.001 & 0.001 &-0.003 &-0.001 & 0.001 &-0.002 &-0.001 & 0.000 &-0.002 &-0.001 \\
 9 &-0.009 & 0.002 & 0.000 &-0.003 & 0.001 & 0.000 &-0.005 & 0.001 & 0.000 & 0.000 & 0.000 & 0.000 &-0.003 & 0.000 & 0.000 \\
10 &-0.006 &-0.008 &-0.003 & 0.000 &-0.004 &-0.001 &-0.001 &-0.003 &-0.001 & 0.000 &-0.002 &-0.001 & 0.000 &-0.001 & 0.000 \\
11 &-0.007 & 0.002 & 0.001 &-0.010 & 0.002 & 0.000 &-0.004 & 0.001 & 0.000 &-0.004 & 0.000 & 0.000 & 0.000 & 0.002 & 0.000 \\
12 & 0.033 & 0.002 &-0.006 &-0.012 &-0.008 &-0.004 & 0.001 &-0.004 &-0.002 &-0.002 &-0.002 &-0.001 &-0.001 &-0.002 &-0.001 \\
13 &-0.364 & 0.017 & 0.000 & 0.068 &-0.002 & 0.001 &-0.036 & 0.002 & 0.000 & 0.009 & 0.001 & 0.000 &-0.010 & 0.001 & 0.000 \\
14 &-0.044 &-0.217 & 0.015 & 0.002 & 0.007 & 0.000 &-0.007 &-0.015 & 0.002 & 0.000 &-0.007 & 0.001 & 0.000 &-0.005 & 0.001 \\
15 &-0.006 &-0.051 &-0.132 &-0.008 & 0.017 &-0.012 &-0.001 &-0.002 &-0.009 & 0.000 & 0.001 &-0.006 & 0.000 & 0.000 &-0.004 \\
16 & \bf{1.000} &-0.025 & 0.003 &-0.431 & 0.017 & 0.000 & 0.114 & 0.000 & 0.001 &-0.057 & 0.002 & 0.000 & 0.017 & 0.003 & 0.000 \\
17 &-0.025 & \bf{1.000} &-0.047 &-0.033 &-0.258 & 0.016 & 0.005 & 0.023 & 0.002 &-0.009 &-0.016 & 0.002 & 0.002 &-0.003 & 0.001 \\
18 & 0.003 &-0.047 & \bf{1.000} & 0.003 &-0.060 &-0.148 &-0.008 & 0.018 &-0.011 &-0.005 &-0.001 &-0.010 & 0.000 & 0.002 &-0.006 \\
19 &-0.431 &-0.033 & 0.003 & \bf{1.000} &-0.012 & 0.001 &-0.507 & 0.011 & 0.000 & 0.185 & 0.002 & 0.001 &-0.099 &-0.004 &-0.001 \\
20 & 0.017 &-0.258 &-0.060 &-0.012 & \bf{1.000} &-0.028 &-0.036 &-0.298 & 0.015 & 0.016 & 0.038 & 0.002 &-0.017 &-0.019 & 0.002 \\
21 & 0.000 & 0.016 &-0.148 & 0.001 &-0.028 & \bf{1.000} & 0.003 &-0.066 &-0.162 & 0.005 & 0.014 &-0.009 &-0.013 & 0.001 &-0.010 \\
22 & 0.114 & 0.005 &-0.008 &-0.507 &-0.036 & 0.003 & \bf{1.000} & 0.012 & 0.002 &-0.591 & 0.001 &-0.002 & 0.282 & 0.026 & 0.002 \\
23 & 0.000 & 0.023 & 0.018 & 0.011 &-0.298 &-0.066 & 0.012 & \bf{1.000} &-0.014 &-0.062 &-0.333 & 0.012 & 0.045 & 0.058 & 0.001 \\
24 & 0.001 & 0.002 &-0.011 & 0.000 & 0.015 &-0.162 & 0.002 &-0.014 & \bf{1.000} &-0.004 &-0.069 &-0.176 & 0.015 & 0.013 &-0.005 \\
25 &-0.057 &-0.009 &-0.005 & 0.185 & 0.016 & 0.005 &-0.591 &-0.062 &-0.004 & \bf{1.000} & 0.031 & 0.007 &-0.689 &-0.056 &-0.005 \\
26 & 0.002 &-0.016 &-0.001 & 0.002 & 0.038 & 0.014 & 0.001 &-0.333 &-0.069 & 0.031 & \bf{1.000} &-0.005 &-0.076 &-0.387 & 0.008 \\
27 & 0.000 & 0.002 &-0.010 & 0.001 & 0.002 &-0.009 &-0.002 & 0.012 &-0.176 & 0.007 &-0.005 & \bf{1.000} &-0.014 &-0.066 &-0.199 \\
28 & 0.017 & 0.002 & 0.000 &-0.099 &-0.017 &-0.013 & 0.282 & 0.045 & 0.015 &-0.689 &-0.076 &-0.014 & \bf{1.000} & 0.145 & 0.010 \\
29 & 0.003 &-0.003 & 0.002 &-0.004 &-0.019 & 0.001 & 0.026 & 0.058 & 0.013 &-0.056 &-0.387 &-0.066 & 0.145 & \bf{1.000} & 0.012 \\
30 & 0.000 & 0.001 &-0.006 &-0.001 & 0.002 &-0.010 & 0.002 & 0.001 &-0.005 &-0.005 & 0.008 &-0.199 & 0.010 & 0.012 & \bf{1.000} \\
31 &-0.012 &-0.003 &-0.003 & 0.042 & 0.008 & 0.004 &-0.148 &-0.038 &-0.016 & 0.353 & 0.071 & 0.026 &-0.714 &-0.226 &-0.023 \\
32 & 0.000 &-0.001 &-0.001 & 0.008 & 0.003 & 0.002 &-0.017 &-0.020 &-0.003 & 0.051 & 0.081 & 0.011 &-0.103 &-0.392 &-0.094 \\
33 & 0.000 & 0.001 &-0.003 & 0.000 & 0.001 &-0.005 &-0.001 & 0.001 &-0.010 & 0.002 & 0.001 & 0.001 &-0.005 & 0.003 &-0.200 \\
34 & 0.003 &-0.001 & 0.001 &-0.025 &-0.010 &-0.004 & 0.069 & 0.017 & 0.006 &-0.175 &-0.052 &-0.021 & 0.360 & 0.160 & 0.038 \\
35 & 0.003 & 0.002 & 0.000 &-0.002 & 0.000 & 0.000 & 0.016 & 0.007 & 0.002 &-0.024 &-0.021 &-0.006 & 0.059 & 0.099 & 0.016 \\
36 & 0.000 & 0.000 &-0.002 & 0.000 & 0.001 &-0.002 & 0.001 & 0.001 &-0.003 &-0.001 & 0.001 &-0.009 & 0.002 & 0.001 & 0.006 \\
37 &-0.007 &-0.003 &-0.002 & 0.010 & 0.003 & 0.000 &-0.042 &-0.013 &-0.005 & 0.079 & 0.020 & 0.008 &-0.175 &-0.094 &-0.027 \\
38 &-0.001 &-0.002 &-0.003 & 0.004 &-0.001 &-0.003 &-0.008 &-0.006 &-0.006 & 0.015 & 0.002 &-0.002 &-0.028 &-0.033 &-0.013 \\
39 & 0.000 &-0.002 &-0.001 & 0.000 &-0.002 &-0.002 &-0.001 &-0.002 &-0.003 & 0.000 &-0.002 &-0.004 &-0.002 &-0.002 &-0.009 \\
40 & 0.004 & 0.006 & 0.004 &-0.010 & 0.003 & 0.006 & 0.024 & 0.014 & 0.010 &-0.041 & 0.000 & 0.002 & 0.077 & 0.048 & 0.018 \\
41 & 0.004 & 0.017 & 0.004 & 0.002 & 0.019 & 0.011 & 0.015 & 0.025 & 0.014 &-0.004 & 0.022 & 0.008 & 0.023 & 0.032 & 0.011 \\
42 & 0.000 & 0.000 & 0.000 & 0.000 & 0.000 &-0.001 & 0.000 & 0.000 &-0.001 & 0.000 & 0.000 &-0.001 & 0.000 & 0.000 &-0.002 \\
43 &-0.006 &-0.026 &-0.008 &-0.004 &-0.030 &-0.021 &-0.023 &-0.038 &-0.025 & 0.005 &-0.036 &-0.018 &-0.034 &-0.046 &-0.020 \\
44 & 0.000 &-0.001 &-0.001 & 0.000 &-0.002 &-0.002 &-0.001 &-0.002 &-0.002 & 0.000 &-0.002 &-0.002 &-0.002 &-0.003 &-0.003 \\
45 & 0.000 & 0.000 & 0.000 & 0.000 & 0.000 & 0.000 & 0.000 & 0.000 & 0.000 & 0.000 & 0.000 & 0.000 & 0.000 & 0.000 & 0.000 \\
\end{tabular}
\end{ruledtabular}
\end{table*}
\endgroup
%
%
%
\addtocounter{table}{-1}
\begingroup
\squeezetable
\begin{table*}[H]
\caption{ -- continued.}
\begin{ruledtabular}
\begin{tabular}{r|rrrrrrrrrrrrrrrrrrr}
 &31 & 32& 33& 34& 35& 36& 37& 38& 39& 40& 41& 42& 43& 44& 45\\
\hline
 1 & 0.000 & 0.001 & 0.000 &-0.002 & 0.000 & 0.000 & 0.001 & 0.001 & 0.000 &-0.002 &-0.001 & 0.000 & 0.001 & 0.000 & 0.000 \\
 2 &-0.001 & 0.000 & 0.000 & 0.000 & 0.001 & 0.000 &-0.001 & 0.000 & 0.000 & 0.000 & 0.001 & 0.000 &-0.002 & 0.000 & 0.000 \\
 3 & 0.001 & 0.000 & 0.000 &-0.002 & 0.000 & 0.000 & 0.002 & 0.001 & 0.000 &-0.002 & 0.001 & 0.000 &-0.002 & 0.000 & 0.000 \\
 4 & 0.000 & 0.000 & 0.000 & 0.000 & 0.000 & 0.000 &-0.002 &-0.001 & 0.000 & 0.002 & 0.002 & 0.000 &-0.003 & 0.000 & 0.000 \\
 5 &-0.001 & 0.000 & 0.000 &-0.001 & 0.001 & 0.000 & 0.000 & 0.001 & 0.000 &-0.002 & 0.001 & 0.000 &-0.001 & 0.000 & 0.000 \\
 6 & 0.001 &-0.001 & 0.000 & 0.000 &-0.001 & 0.000 & 0.001 & 0.000 & 0.000 & 0.000 & 0.000 & 0.000 &-0.001 & 0.000 & 0.000 \\
 7 &-0.001 & 0.001 & 0.000 &-0.001 & 0.001 & 0.000 &-0.001 & 0.001 & 0.000 &-0.002 & 0.000 & 0.000 & 0.000 & 0.000 & 0.000 \\
 8 & 0.001 &-0.001 & 0.000 &-0.001 & 0.000 & 0.000 & 0.000 &-0.001 & 0.000 & 0.001 & 0.003 & 0.000 &-0.005 & 0.000 & 0.000 \\
 9 & 0.001 & 0.001 & 0.000 &-0.002 & 0.000 & 0.000 & 0.000 & 0.001 & 0.000 &-0.003 &-0.002 & 0.000 & 0.002 & 0.000 & 0.000 \\
10 & 0.000 & 0.000 & 0.000 & 0.000 & 0.000 & 0.000 &-0.001 &-0.001 &-0.001 & 0.002 & 0.005 & 0.000 &-0.008 &-0.001 & 0.000 \\
11 &-0.005 &-0.001 & 0.000 & 0.004 & 0.003 & 0.000 &-0.010 &-0.003 & 0.000 & 0.007 & 0.000 & 0.000 & 0.000 & 0.000 & 0.000 \\
12 &-0.001 &-0.001 &-0.001 &-0.001 & 0.000 & 0.000 & 0.001 & 0.000 &-0.001 &-0.001 & 0.003 & 0.000 &-0.006 & 0.000 & 0.000 \\
13 & 0.002 & 0.002 & 0.000 &-0.004 & 0.000 & 0.000 & 0.002 & 0.001 & 0.000 &-0.003 & 0.003 & 0.000 &-0.004 & 0.000 & 0.000 \\
14 & 0.000 &-0.003 & 0.001 &-0.001 &-0.001 & 0.000 & 0.000 &-0.001 & 0.000 & 0.001 & 0.004 & 0.000 &-0.006 & 0.000 & 0.000 \\
15 & 0.000 & 0.000 &-0.002 & 0.000 &-0.001 &-0.001 & 0.000 &-0.001 & 0.000 & 0.001 &-0.001 & 0.000 & 0.000 & 0.000 & 0.000 \\
16 &-0.012 & 0.000 & 0.000 & 0.003 & 0.003 & 0.000 &-0.007 &-0.001 & 0.000 & 0.004 & 0.004 & 0.000 &-0.006 & 0.000 & 0.000 \\
17 &-0.003 &-0.001 & 0.001 &-0.001 & 0.002 & 0.000 &-0.003 &-0.002 &-0.002 & 0.006 & 0.017 & 0.000 &-0.026 &-0.001 & 0.000 \\
18 &-0.003 &-0.001 &-0.003 & 0.001 & 0.000 &-0.002 &-0.002 &-0.003 &-0.001 & 0.004 & 0.004 & 0.000 &-0.008 &-0.001 & 0.000 \\
19 & 0.042 & 0.008 & 0.000 &-0.025 &-0.002 & 0.000 & 0.010 & 0.004 & 0.000 &-0.010 & 0.002 & 0.000 &-0.004 & 0.000 & 0.000 \\
20 & 0.008 & 0.003 & 0.001 &-0.010 & 0.000 & 0.001 & 0.003 &-0.001 &-0.002 & 0.003 & 0.019 & 0.000 &-0.030 &-0.002 & 0.000 \\
21 & 0.004 & 0.002 &-0.005 &-0.004 & 0.000 &-0.002 & 0.000 &-0.003 &-0.002 & 0.006 & 0.011 &-0.001 &-0.021 &-0.002 & 0.000 \\
22 &-0.148 &-0.017 &-0.001 & 0.069 & 0.016 & 0.001 &-0.042 &-0.008 &-0.001 & 0.024 & 0.015 & 0.000 &-0.023 &-0.001 & 0.000 \\
23 &-0.038 &-0.020 & 0.001 & 0.017 & 0.007 & 0.001 &-0.013 &-0.006 &-0.002 & 0.014 & 0.025 & 0.000 &-0.038 &-0.002 & 0.000 \\
24 &-0.016 &-0.003 &-0.010 & 0.006 & 0.002 &-0.003 &-0.005 &-0.006 &-0.003 & 0.010 & 0.014 &-0.001 &-0.025 &-0.002 & 0.000 \\
25 & 0.353 & 0.051 & 0.002 &-0.175 &-0.024 &-0.001 & 0.079 & 0.015 & 0.000 &-0.041 &-0.004 & 0.000 & 0.005 & 0.000 & 0.000 \\
26 & 0.071 & 0.081 & 0.001 &-0.052 &-0.021 & 0.001 & 0.020 & 0.002 &-0.002 & 0.000 & 0.022 & 0.000 &-0.036 &-0.002 & 0.000 \\
27 & 0.026 & 0.011 & 0.001 &-0.021 &-0.006 &-0.009 & 0.008 &-0.002 &-0.004 & 0.002 & 0.008 &-0.001 &-0.018 &-0.002 & 0.000 \\
28 &-0.714 &-0.103 &-0.005 & 0.360 & 0.059 & 0.002 &-0.175 &-0.028 &-0.002 & 0.077 & 0.023 & 0.000 &-0.034 &-0.002 & 0.000 \\
29 &-0.226 &-0.392 & 0.003 & 0.160 & 0.099 & 0.001 &-0.094 &-0.033 &-0.002 & 0.048 & 0.032 & 0.000 &-0.046 &-0.003 & 0.000 \\
30 &-0.023 &-0.094 &-0.200 & 0.038 & 0.016 & 0.006 &-0.027 &-0.013 &-0.009 & 0.018 & 0.011 &-0.002 &-0.020 &-0.003 & 0.000 \\
31 & \bf{1.000} & 0.227 & 0.011 &-0.713 &-0.120 &-0.005 & 0.359 & 0.060 & 0.001 &-0.161 &-0.023 & 0.000 & 0.031 & 0.002 & 0.000 \\
32 & 0.227 & \bf{1.000} & 0.020 &-0.317 &-0.394 & 0.001 & 0.210 & 0.106 &-0.002 &-0.104 &-0.015 & 0.000 &-0.001 & 0.003 & 0.000 \\
33 & 0.011 & 0.020 & \bf{1.000} &-0.026 &-0.111 &-0.202 & 0.042 & 0.022 & 0.009 &-0.025 &-0.008 &-0.006 & 0.000 &-0.003 & 0.000 \\
34 &-0.713 &-0.317 &-0.026 & \bf{1.000} & 0.256 & 0.011 &-0.721 &-0.133 &-0.005 & 0.354 & 0.079 &-0.001 &-0.113 &-0.007 & 0.000 \\
35 &-0.120 &-0.394 &-0.111 & 0.256 & \bf{1.000} & 0.023 &-0.344 &-0.417 &-0.003 & 0.243 & 0.146 &-0.002 &-0.129 &-0.017 & 0.001 \\
36 &-0.005 & 0.001 &-0.202 & 0.011 & 0.023 & \bf{1.000} &-0.026 &-0.118 &-0.211 & 0.046 & 0.037 & 0.016 &-0.032 &-0.015 &-0.003 \\
37 & 0.359 & 0.210 & 0.042 &-0.721 &-0.344 &-0.026 & \bf{1.000} & 0.269 & 0.009 &-0.690 &-0.133 & 0.002 & 0.185 & 0.012 &-0.001 \\
38 & 0.060 & 0.106 & 0.022 &-0.133 &-0.417 &-0.118 & 0.269 & \bf{1.000} & 0.026 &-0.384 &-0.417 & 0.004 & 0.224 & 0.063 &-0.003 \\
39 & 0.001 &-0.002 & 0.009 &-0.005 &-0.003 &-0.211 & 0.009 & 0.026 & \bf{1.000} &-0.039 &-0.128 &-0.216 & 0.066 & 0.028 & 0.014 \\
40 &-0.161 &-0.104 &-0.025 & 0.354 & 0.243 & 0.046 &-0.690 &-0.384 &-0.039 & \bf{1.000} & 0.433 &-0.005 &-0.637 &-0.038 & 0.002 \\
41 &-0.023 &-0.015 &-0.008 & 0.079 & 0.146 & 0.037 &-0.133 &-0.417 &-0.128 & 0.433 & \bf{1.000} & 0.008 &-0.655 &-0.198 & 0.008 \\
42 & 0.000 & 0.000 &-0.006 &-0.001 &-0.002 & 0.016 & 0.002 & 0.004 &-0.216 &-0.005 & 0.008 & \bf{1.000} & 0.007 &-0.226 &-0.132 \\
43 & 0.031 &-0.001 & 0.000 &-0.113 &-0.129 &-0.032 & 0.185 & 0.224 & 0.066 &-0.637 &-0.655 & 0.007 & \bf{1.000} & 0.057 &-0.002 \\
44 & 0.002 & 0.003 &-0.003 &-0.007 &-0.017 &-0.015 & 0.012 & 0.063 & 0.028 &-0.038 &-0.198 &-0.226 & 0.057 & \bf{1.000} &-0.027 \\
45 & 0.000 & 0.000 & 0.000 & 0.000 & 0.001 &-0.003 &-0.001 &-0.003 & 0.014 & 0.002 & 0.008 &-0.132 &-0.002 &-0.027 & \bf{1.000} \\
\end{tabular}
\end{ruledtabular}
\end{table*}
\endgroup
%
%
\begingroup
\squeezetable
\begin{table*}[H]
{\caption{\label{tab:corr-46bins-deu1}{Correlation matrix  for
$A_{||}^d$, $A_1^d$ and $g_1^d$ in 45 $x$-bins for deuteron (left).
 For $\langle x\rangle$ and $\langle Q^2\rangle$ of each bin, see e.g. 
Tabs.~\ref{tab:apard} and \ref{tab:g1_46}.}}}
\begin{ruledtabular}
\begin{tabular}{r|rrrrrrrrrrrrrrrrr}
  & 1 & 2 & 3 & 4 & 5 & 6 & 7 & 8 & 9 & 10 & 11 & 12 & 13 & 14 & 15 \\
\hline
 1 & \bf{1.000} &-0.150 & 0.007 &-0.003 &-0.003 & 0.001 &-0.003 & 0.001 &-0.001 & 0.000 &-0.004 & 0.000 &-0.001 & 0.000 & 0.000 \\
 2 &-0.150 & \bf{1.000} &-0.186 & 0.003 &-0.014 & 0.004 &-0.007 & 0.000 &-0.004 & 0.000 &-0.012 & 0.000 &-0.001 & 0.000 & 0.000 \\
 3 & 0.007 &-0.186 & \bf{1.000} &-0.226 &-0.007 & 0.006 &-0.019 & 0.003 &-0.010 & 0.002 &-0.041 & 0.002 & 0.002 &-0.002 & 0.000 \\
 4 &-0.003 & 0.003 &-0.226 & \bf{1.000} &-0.232 &-0.003 & 0.009 & 0.003 &-0.016 & 0.002 &-0.044 & 0.000 & 0.001 &-0.003 & 0.000 \\
 5 &-0.003 &-0.014 &-0.007 &-0.232 & \bf{1.000} &-0.129 &-0.260 & 0.006 & 0.019 & 0.000 &-0.070 & 0.000 &-0.002 &-0.002 & 0.001 \\
 6 & 0.001 & 0.004 & 0.006 &-0.003 &-0.129 & \bf{1.000} &-0.009 &-0.176 &-0.006 & 0.018 &-0.032 &-0.004 & 0.005 &-0.004 &-0.003 \\
 7 &-0.003 &-0.007 &-0.019 & 0.009 &-0.260 &-0.009 & \bf{1.000} &-0.068 &-0.351 &-0.001 & 0.026 & 0.000 &-0.009 & 0.003 & 0.001 \\
 8 & 0.001 & 0.000 & 0.003 & 0.003 & 0.006 &-0.176 &-0.068 & \bf{1.000} &-0.013 &-0.215 &-0.050 & 0.021 & 0.001 &-0.010 &-0.003 \\
 9 &-0.001 &-0.004 &-0.010 &-0.016 & 0.019 &-0.006 &-0.351 &-0.013 & \bf{1.000} &-0.040 &-0.407 & 0.003 & 0.005 & 0.002 & 0.001 \\
10 & 0.000 & 0.000 & 0.002 & 0.002 & 0.000 & 0.018 &-0.001 &-0.215 &-0.040 & \bf{1.000} &-0.054 &-0.233 & 0.029 & 0.010 & 0.002 \\
11 &-0.004 &-0.012 &-0.041 &-0.044 &-0.070 &-0.032 & 0.026 &-0.050 &-0.407 &-0.054 & \bf{1.000} &-0.065 &-0.052 & 0.010 & 0.002 \\
12 & 0.000 & 0.000 & 0.002 & 0.000 & 0.000 &-0.004 & 0.000 & 0.021 & 0.003 &-0.233 &-0.065 & \bf{1.000} &-0.243 &-0.140 &-0.048 \\
13 &-0.001 &-0.001 & 0.002 & 0.001 &-0.002 & 0.005 &-0.009 & 0.001 & 0.005 & 0.029 &-0.052 &-0.243 & \bf{1.000} &-0.029 & 0.010 \\
14 & 0.000 & 0.000 &-0.002 &-0.003 &-0.002 &-0.004 & 0.003 &-0.010 & 0.002 & 0.010 & 0.010 &-0.140 &-0.029 & \bf{1.000} &-0.075 \\
15 & 0.000 & 0.000 & 0.000 & 0.000 & 0.001 &-0.003 & 0.001 &-0.003 & 0.001 & 0.002 & 0.002 &-0.048 & 0.010 &-0.075 & \bf{1.000} \\
16 &-0.001 &-0.001 &-0.001 & 0.000 &-0.003 & 0.001 &-0.004 & 0.001 &-0.006 &-0.008 &-0.002 & 0.049 &-0.394 &-0.045 &-0.005 \\
17 & 0.000 & 0.000 &-0.002 &-0.002 &-0.002 &-0.002 & 0.001 &-0.003 & 0.001 &-0.007 & 0.001 & 0.009 & 0.018 &-0.229 &-0.055 \\
18 & 0.000 & 0.000 & 0.000 & 0.000 & 0.001 &-0.001 & 0.000 &-0.002 & 0.000 &-0.002 & 0.000 &-0.003 & 0.000 & 0.019 &-0.137 \\
19 &-0.002 &-0.001 & 0.000 &-0.001 & 0.000 & 0.001 &-0.002 & 0.001 &-0.002 & 0.001 &-0.005 &-0.017 & 0.100 & 0.004 &-0.008 \\
20 & 0.000 & 0.000 & 0.000 &-0.002 & 0.000 &-0.001 & 0.001 &-0.003 & 0.001 &-0.002 & 0.001 &-0.007 &-0.003 & 0.020 & 0.019 \\
21 & 0.000 & 0.000 & 0.000 & 0.000 & 0.000 &-0.001 & 0.000 &-0.001 & 0.000 &-0.001 & 0.000 &-0.002 & 0.001 &-0.001 &-0.005 \\
22 & 0.000 & 0.000 &-0.001 & 0.000 &-0.002 & 0.001 &-0.002 & 0.001 &-0.001 & 0.000 &-0.005 & 0.005 &-0.044 &-0.007 & 0.000 \\
23 & 0.000 & 0.000 & 0.000 & 0.000 &-0.002 &-0.001 & 0.001 &-0.001 & 0.001 &-0.002 & 0.001 &-0.002 & 0.001 &-0.013 &-0.003 \\
24 & 0.000 & 0.000 & 0.000 & 0.000 & 0.000 &-0.001 & 0.000 &-0.001 & 0.000 & 0.000 & 0.000 &-0.002 & 0.000 & 0.001 &-0.007 \\
25 &-0.002 &-0.001 & 0.000 & 0.000 & 0.000 & 0.000 &-0.001 & 0.000 &-0.002 &-0.001 & 0.001 &-0.004 & 0.011 & 0.000 & 0.000 \\
26 & 0.000 & 0.000 & 0.000 & 0.000 & 0.000 &-0.001 & 0.000 &-0.001 & 0.000 &-0.002 & 0.001 &-0.002 & 0.000 &-0.003 & 0.001 \\
27 & 0.000 & 0.000 & 0.000 & 0.000 & 0.000 &-0.001 & 0.000 & 0.000 & 0.000 & 0.000 & 0.000 &-0.002 & 0.000 & 0.001 &-0.004 \\
28 & 0.001 & 0.001 &-0.001 &-0.002 & 0.000 & 0.000 & 0.000 & 0.000 & 0.001 & 0.000 &-0.003 & 0.002 &-0.006 &-0.001 & 0.000 \\
29 & 0.000 & 0.000 & 0.000 &-0.001 & 0.000 &-0.001 & 0.001 &-0.001 & 0.001 &-0.001 & 0.000 & 0.000 & 0.001 &-0.004 & 0.000 \\
30 & 0.000 & 0.000 & 0.000 & 0.000 & 0.000 & 0.000 & 0.000 & 0.000 & 0.000 & 0.000 & 0.000 & 0.000 & 0.000 & 0.000 &-0.003 \\
31 &-0.001 &-0.002 & 0.000 & 0.002 &-0.002 & 0.001 &-0.003 & 0.000 &-0.002 & 0.000 & 0.001 &-0.003 &-0.001 & 0.001 & 0.001 \\
32 & 0.000 & 0.000 & 0.000 & 0.001 & 0.000 &-0.001 &-0.001 &-0.001 &-0.001 & 0.000 & 0.001 &-0.002 & 0.000 &-0.001 & 0.000 \\
33 & 0.000 & 0.000 & 0.000 & 0.000 & 0.000 & 0.000 & 0.000 & 0.000 & 0.000 & 0.000 & 0.000 & 0.000 & 0.000 & 0.000 &-0.002 \\
34 & 0.001 & 0.001 & 0.000 &-0.002 & 0.001 & 0.000 & 0.003 & 0.000 & 0.002 &-0.001 &-0.003 & 0.001 & 0.000 &-0.002 & 0.000 \\
35 & 0.000 & 0.001 & 0.000 &-0.001 & 0.001 &-0.001 & 0.001 &-0.001 & 0.001 & 0.000 & 0.000 & 0.000 & 0.002 &-0.001 &-0.001 \\
36 & 0.000 & 0.000 & 0.000 & 0.000 & 0.000 & 0.000 & 0.000 & 0.000 & 0.000 & 0.000 & 0.000 & 0.000 & 0.000 & 0.000 &-0.001 \\
37 & 0.000 &-0.002 & 0.000 & 0.002 &-0.001 & 0.000 &-0.003 & 0.000 &-0.003 & 0.000 & 0.001 & 0.000 &-0.002 & 0.002 & 0.000 \\
38 & 0.000 & 0.000 & 0.000 & 0.000 & 0.000 &-0.001 & 0.000 & 0.000 &-0.001 &-0.001 & 0.001 & 0.000 &-0.001 &-0.001 &-0.001 \\
39 & 0.000 & 0.000 & 0.000 & 0.000 & 0.000 & 0.000 & 0.000 & 0.000 & 0.000 &-0.001 & 0.000 &-0.001 & 0.000 &-0.001 &-0.001 \\
40 &-0.001 & 0.001 & 0.000 & 0.000 &-0.001 & 0.001 & 0.000 & 0.000 & 0.002 & 0.001 &-0.001 &-0.001 & 0.002 & 0.000 & 0.001 \\
41 &-0.001 & 0.001 & 0.001 & 0.002 & 0.001 & 0.001 & 0.000 & 0.001 & 0.001 & 0.003 &-0.002 & 0.003 & 0.003 & 0.003 & 0.000 \\
42 & 0.000 & 0.000 & 0.000 & 0.000 & 0.000 & 0.000 & 0.000 & 0.000 & 0.000 & 0.000 & 0.000 & 0.000 & 0.000 & 0.000 & 0.000 \\
43 & 0.001 &-0.001 &-0.002 &-0.003 &-0.001 &-0.002 & 0.000 &-0.002 &-0.001 &-0.005 & 0.003 &-0.005 &-0.004 &-0.005 & 0.000 \\
44 & 0.000 & 0.000 & 0.000 & 0.000 & 0.000 & 0.000 & 0.000 & 0.000 & 0.000 &-0.001 & 0.000 &-0.001 & 0.000 &-0.001 & 0.000 \\
45 & 0.000 & 0.000 & 0.000 & 0.000 & 0.000 & 0.000 & 0.000 & 0.000 & 0.000 & 0.000 & 0.000 & 0.000 & 0.000 & 0.000 & 0.000 \\
\end{tabular}
\end{ruledtabular}
\end{table*}
\endgroup
\addtocounter{table}{-1}
\begingroup
\squeezetable
\begin{table*}[H]
\caption{ -- continued.
}
\begin{ruledtabular}
\begin{tabular}{r|rrrrrrrrrrrrrrrrrrr}
  &16 & 17 & 18 & 19 & 20 & 21 & 22 & 23 & 24 & 25 & 26 & 27 & 28 & 29 & 30\\
\hline
 1 &-0.001 & 0.000 & 0.000 &-0.002 & 0.000 & 0.000 & 0.000 & 0.000 & 0.000 &-0.002 & 0.000 & 0.000 & 0.001 & 0.000 & 0.000 \\
 2 &-0.001 & 0.000 & 0.000 &-0.001 & 0.000 & 0.000 & 0.000 & 0.000 & 0.000 &-0.001 & 0.000 & 0.000 & 0.001 & 0.000 & 0.000 \\
 3 &-0.001 &-0.002 & 0.000 & 0.000 & 0.000 & 0.000 &-0.001 & 0.000 & 0.000 & 0.000 & 0.000 & 0.000 &-0.001 & 0.000 & 0.000 \\
 4 & 0.000 &-0.002 & 0.000 &-0.001 &-0.002 & 0.000 & 0.000 & 0.000 & 0.000 & 0.000 & 0.000 & 0.000 &-0.002 &-0.001 & 0.000 \\
 5 &-0.003 &-0.002 & 0.001 & 0.000 & 0.000 & 0.000 &-0.002 &-0.002 & 0.000 & 0.000 & 0.000 & 0.000 & 0.000 & 0.000 & 0.000 \\
 6 & 0.001 &-0.002 &-0.001 & 0.001 &-0.001 &-0.001 & 0.001 &-0.001 &-0.001 & 0.000 &-0.001 &-0.001 & 0.000 &-0.001 & 0.000 \\
 7 &-0.004 & 0.001 & 0.000 &-0.002 & 0.001 & 0.000 &-0.002 & 0.001 & 0.000 &-0.001 & 0.000 & 0.000 & 0.000 & 0.001 & 0.000 \\
 8 & 0.001 &-0.003 &-0.002 & 0.001 &-0.003 &-0.001 & 0.001 &-0.001 &-0.001 & 0.000 &-0.001 & 0.000 & 0.000 &-0.001 & 0.000 \\
 9 &-0.006 & 0.001 & 0.000 &-0.002 & 0.001 & 0.000 &-0.001 & 0.001 & 0.000 &-0.002 & 0.000 & 0.000 & 0.001 & 0.001 & 0.000 \\
10 &-0.008 &-0.007 &-0.002 & 0.001 &-0.002 &-0.001 & 0.000 &-0.002 & 0.000 &-0.001 &-0.002 & 0.000 & 0.000 &-0.001 & 0.000 \\
11 &-0.002 & 0.001 & 0.000 &-0.005 & 0.001 & 0.000 &-0.005 & 0.001 & 0.000 & 0.001 & 0.001 & 0.000 &-0.003 & 0.000 & 0.000 \\
12 & 0.049 & 0.009 &-0.003 &-0.017 &-0.007 &-0.002 & 0.005 &-0.002 &-0.002 &-0.004 &-0.002 &-0.002 & 0.002 & 0.000 & 0.000 \\
13 &-0.394 & 0.018 & 0.000 & 0.100 &-0.003 & 0.001 &-0.044 & 0.001 & 0.000 & 0.011 & 0.000 & 0.000 &-0.006 & 0.001 & 0.000 \\
14 &-0.045 &-0.229 & 0.019 & 0.004 & 0.020 &-0.001 &-0.007 &-0.013 & 0.001 & 0.000 &-0.003 & 0.001 &-0.001 &-0.004 & 0.000 \\
15 &-0.005 &-0.055 &-0.137 &-0.008 & 0.019 &-0.005 & 0.000 &-0.003 &-0.007 & 0.000 & 0.001 &-0.004 & 0.000 & 0.000 &-0.003 \\
16 & \bf{1.000} &-0.026 & 0.003 &-0.476 & 0.017 & 0.000 & 0.162 &-0.001 & 0.001 &-0.075 & 0.001 & 0.000 & 0.029 & 0.004 & 0.000 \\
17 &-0.026 & \bf{1.000} &-0.051 &-0.033 &-0.274 & 0.018 & 0.007 & 0.039 & 0.000 &-0.011 &-0.016 & 0.003 & 0.003 & 0.000 & 0.001 \\
18 & 0.003 &-0.051 & \bf{1.000} & 0.005 &-0.065 &-0.154 &-0.010 & 0.022 &-0.003 &-0.003 &-0.003 &-0.008 &-0.001 & 0.002 &-0.004 \\
19 &-0.476 &-0.033 & 0.005 & \bf{1.000} &-0.009 & 0.001 &-0.568 & 0.006 &-0.001 & 0.252 & 0.004 & 0.002 &-0.136 &-0.011 &-0.001 \\
20 & 0.017 &-0.274 &-0.065 &-0.009 & \bf{1.000} &-0.031 &-0.042 &-0.323 & 0.016 & 0.025 & 0.059 & 0.000 &-0.023 &-0.022 & 0.001 \\
21 & 0.000 & 0.018 &-0.154 & 0.001 &-0.031 & \bf{1.000} & 0.003 &-0.071 &-0.170 & 0.006 & 0.019 & 0.001 &-0.013 &-0.002 &-0.008 \\
22 & 0.162 & 0.007 &-0.010 &-0.568 &-0.042 & 0.003 & \bf{1.000} & 0.025 & 0.004 &-0.661 &-0.012 &-0.004 & 0.361 & 0.042 & 0.003 \\
23 &-0.001 & 0.039 & 0.022 & 0.006 &-0.323 &-0.071 & 0.025 & \bf{1.000} &-0.014 &-0.079 &-0.368 & 0.011 & 0.066 & 0.092 & 0.001 \\
24 & 0.001 & 0.000 &-0.003 &-0.001 & 0.016 &-0.170 & 0.004 &-0.014 & \bf{1.000} &-0.007 &-0.074 &-0.188 & 0.020 & 0.021 & 0.006 \\
25 &-0.075 &-0.011 &-0.003 & 0.252 & 0.025 & 0.006 &-0.661 &-0.079 &-0.007 & \bf{1.000} & 0.055 & 0.011 &-0.747 &-0.096 &-0.007 \\
26 & 0.001 &-0.016 &-0.003 & 0.004 & 0.059 & 0.019 &-0.012 &-0.368 &-0.074 & 0.055 & \bf{1.000} &-0.001 &-0.109 &-0.436 & 0.004 \\
27 & 0.000 & 0.003 &-0.008 & 0.002 & 0.000 & 0.001 &-0.004 & 0.011 &-0.188 & 0.011 &-0.001 & \bf{1.000} &-0.021 &-0.074 &-0.213 \\
28 & 0.029 & 0.003 &-0.001 &-0.136 &-0.023 &-0.013 & 0.361 & 0.066 & 0.020 &-0.747 &-0.109 &-0.021 & \bf{1.000} & 0.203 & 0.016 \\
29 & 0.004 & 0.000 & 0.002 &-0.011 &-0.022 &-0.002 & 0.042 & 0.092 & 0.021 &-0.096 &-0.436 &-0.074 & 0.203 & \bf{1.000} & 0.020 \\
30 & 0.000 & 0.001 &-0.004 &-0.001 & 0.001 &-0.008 & 0.003 & 0.001 & 0.006 &-0.007 & 0.004 &-0.213 & 0.016 & 0.020 & \bf{1.000} \\
31 &-0.016 &-0.004 &-0.001 & 0.063 & 0.012 & 0.006 &-0.190 &-0.056 &-0.022 & 0.421 & 0.111 & 0.035 &-0.763 &-0.293 &-0.033 \\
32 & 0.000 &-0.001 & 0.000 & 0.011 & 0.006 & 0.003 &-0.027 &-0.032 &-0.007 & 0.075 & 0.126 & 0.022 &-0.157 &-0.459 &-0.108 \\
33 & 0.000 & 0.000 &-0.002 & 0.000 & 0.001 &-0.003 &-0.001 & 0.000 &-0.009 & 0.004 & 0.002 & 0.012 &-0.009 &-0.003 &-0.217 \\
34 & 0.004 &-0.001 & 0.000 &-0.029 &-0.011 &-0.006 & 0.082 & 0.031 & 0.011 &-0.208 &-0.084 &-0.030 & 0.421 & 0.225 & 0.050 \\
35 & 0.001 & 0.002 & 0.000 &-0.001 &-0.001 & 0.000 & 0.015 & 0.014 & 0.006 &-0.037 &-0.041 &-0.009 & 0.092 & 0.152 & 0.033 \\
36 & 0.000 & 0.001 &-0.001 & 0.000 & 0.001 &-0.001 & 0.000 & 0.002 &-0.001 &-0.001 & 0.001 &-0.008 & 0.004 & 0.003 & 0.017 \\
37 &-0.001 &-0.001 &-0.001 & 0.007 & 0.006 & 0.002 &-0.035 &-0.022 &-0.009 & 0.092 & 0.044 & 0.013 &-0.213 &-0.137 &-0.038 \\
38 & 0.001 &-0.002 &-0.002 & 0.000 &-0.001 &-0.003 &-0.006 &-0.010 &-0.008 & 0.019 & 0.011 & 0.000 &-0.046 &-0.056 &-0.020 \\
39 & 0.000 &-0.002 &-0.001 &-0.001 &-0.002 &-0.002 &-0.001 &-0.003 &-0.003 & 0.000 &-0.002 &-0.003 &-0.002 &-0.003 &-0.009 \\
40 &-0.001 & 0.004 & 0.003 &-0.001 & 0.003 & 0.004 & 0.013 & 0.018 & 0.013 &-0.040 &-0.014 &-0.001 & 0.098 & 0.069 & 0.023 \\
41 & 0.002 & 0.015 & 0.005 & 0.005 & 0.019 & 0.012 & 0.009 & 0.022 & 0.017 &-0.005 & 0.015 & 0.009 & 0.027 & 0.035 & 0.014 \\
42 & 0.000 & 0.000 & 0.000 & 0.000 & 0.000 & 0.000 & 0.000 & 0.000 & 0.000 & 0.000 & 0.000 &-0.001 & 0.000 & 0.000 &-0.001 \\
43 &-0.004 &-0.021 &-0.008 &-0.007 &-0.028 &-0.019 &-0.013 &-0.031 &-0.026 & 0.005 &-0.025 &-0.016 &-0.034 &-0.041 &-0.019 \\
44 & 0.000 &-0.002 &-0.001 &-0.001 &-0.003 &-0.002 &-0.001 &-0.004 &-0.003 & 0.001 &-0.002 &-0.002 &-0.004 &-0.006 &-0.004 \\
45 & 0.000 & 0.000 & 0.000 & 0.000 & 0.000 & 0.000 & 0.000 & 0.000 & 0.000 & 0.000 & 0.000 & 0.000 & 0.000 & 0.000 & 0.000 \\
\end{tabular}
\end{ruledtabular}
\end{table*}
\endgroup
\addtocounter{table}{-1}
\begingroup
\squeezetable
\begin{table*}[H]
\caption{ -- continued.}
\begin{ruledtabular}
\begin{tabular}{r|rrrrrrrrrrrrrrrrrrr}
 &31 & 32& 33& 34& 35& 36& 37& 38& 39& 40& 41& 42& 43& 44& 45\\
\hline
 1 &-0.001 & 0.000 & 0.000 & 0.001 & 0.000 & 0.000 & 0.000 & 0.000 & 0.000 &-0.001 &-0.001 & 0.000 & 0.001 & 0.000 & 0.000 \\
 2 &-0.002 & 0.000 & 0.000 & 0.001 & 0.001 & 0.000 &-0.002 & 0.000 & 0.000 & 0.001 & 0.001 & 0.000 &-0.001 & 0.000 & 0.000 \\
 3 & 0.000 & 0.000 & 0.000 & 0.000 & 0.000 & 0.000 & 0.000 & 0.000 & 0.000 & 0.000 & 0.001 & 0.000 &-0.002 & 0.000 & 0.000 \\
 4 & 0.002 & 0.001 & 0.000 &-0.002 &-0.001 & 0.000 & 0.002 & 0.000 & 0.000 & 0.000 & 0.002 & 0.000 &-0.003 & 0.000 & 0.000 \\
 5 &-0.002 & 0.000 & 0.000 & 0.001 & 0.001 & 0.000 &-0.001 & 0.000 & 0.000 &-0.001 & 0.001 & 0.000 &-0.001 & 0.000 & 0.000 \\
 6 & 0.001 &-0.001 & 0.000 & 0.000 &-0.001 & 0.000 & 0.000 &-0.001 & 0.000 & 0.001 & 0.001 & 0.000 &-0.002 & 0.000 & 0.000 \\
 7 &-0.003 &-0.001 & 0.000 & 0.003 & 0.001 & 0.000 &-0.003 & 0.000 & 0.000 & 0.000 & 0.000 & 0.000 & 0.000 & 0.000 & 0.000 \\
 8 & 0.000 &-0.001 & 0.000 & 0.000 &-0.001 & 0.000 & 0.000 & 0.000 & 0.000 & 0.000 & 0.001 & 0.000 &-0.002 & 0.000 & 0.000 \\
 9 &-0.002 &-0.001 & 0.000 & 0.002 & 0.001 & 0.000 &-0.003 &-0.001 & 0.000 & 0.002 & 0.001 & 0.000 &-0.001 & 0.000 & 0.000 \\
10 & 0.000 & 0.000 & 0.000 &-0.001 & 0.000 & 0.000 & 0.000 &-0.001 &-0.001 & 0.001 & 0.003 & 0.000 &-0.005 &-0.001 & 0.000 \\
11 & 0.001 & 0.001 & 0.000 &-0.003 & 0.000 & 0.000 & 0.001 & 0.001 & 0.000 &-0.001 &-0.002 & 0.000 & 0.003 & 0.000 & 0.000 \\
12 &-0.003 &-0.002 & 0.000 & 0.001 & 0.000 & 0.000 & 0.000 & 0.000 &-0.001 &-0.001 & 0.003 & 0.000 &-0.005 &-0.001 & 0.000 \\
13 &-0.001 & 0.000 & 0.000 & 0.000 & 0.002 & 0.000 &-0.002 &-0.001 & 0.000 & 0.002 & 0.003 & 0.000 &-0.004 & 0.000 & 0.000 \\
14 & 0.001 &-0.001 & 0.000 &-0.002 &-0.001 & 0.000 & 0.002 &-0.001 &-0.001 & 0.000 & 0.003 & 0.000 &-0.005 &-0.001 & 0.000 \\
15 & 0.001 & 0.000 &-0.002 & 0.000 &-0.001 &-0.001 & 0.000 &-0.001 &-0.001 & 0.001 & 0.000 & 0.000 & 0.000 & 0.000 & 0.000 \\
16 &-0.016 & 0.000 & 0.000 & 0.004 & 0.001 & 0.000 &-0.001 & 0.001 & 0.000 &-0.001 & 0.002 & 0.000 &-0.004 & 0.000 & 0.000 \\
17 &-0.004 &-0.001 & 0.000 &-0.001 & 0.002 & 0.001 &-0.001 &-0.002 &-0.002 & 0.004 & 0.015 & 0.000 &-0.021 &-0.002 & 0.000 \\
18 &-0.001 & 0.000 &-0.002 & 0.000 & 0.000 &-0.001 &-0.001 &-0.002 &-0.001 & 0.003 & 0.005 & 0.000 &-0.008 &-0.001 & 0.000 \\
19 & 0.063 & 0.011 & 0.000 &-0.029 &-0.001 & 0.000 & 0.007 & 0.000 &-0.001 &-0.001 & 0.005 & 0.000 &-0.007 &-0.001 & 0.000 \\
20 & 0.012 & 0.006 & 0.001 &-0.011 &-0.001 & 0.001 & 0.006 &-0.001 &-0.002 & 0.003 & 0.019 & 0.000 &-0.028 &-0.003 & 0.000 \\
21 & 0.006 & 0.003 &-0.003 &-0.006 & 0.000 &-0.001 & 0.002 &-0.003 &-0.002 & 0.004 & 0.012 & 0.000 &-0.019 &-0.002 & 0.000 \\
22 &-0.190 &-0.027 &-0.001 & 0.082 & 0.015 & 0.000 &-0.035 &-0.006 &-0.001 & 0.013 & 0.009 & 0.000 &-0.013 &-0.001 & 0.000 \\
23 &-0.056 &-0.032 & 0.000 & 0.031 & 0.014 & 0.002 &-0.022 &-0.010 &-0.003 & 0.018 & 0.022 & 0.000 &-0.031 &-0.004 & 0.000 \\
24 &-0.022 &-0.007 &-0.009 & 0.011 & 0.006 &-0.001 &-0.009 &-0.008 &-0.003 & 0.013 & 0.017 & 0.000 &-0.026 &-0.003 & 0.000 \\
25 & 0.421 & 0.075 & 0.004 &-0.208 &-0.037 &-0.001 & 0.092 & 0.019 & 0.000 &-0.040 &-0.005 & 0.000 & 0.005 & 0.001 & 0.000 \\
26 & 0.111 & 0.126 & 0.002 &-0.084 &-0.041 & 0.001 & 0.044 & 0.011 &-0.002 &-0.014 & 0.015 & 0.000 &-0.025 &-0.002 & 0.000 \\
27 & 0.035 & 0.022 & 0.012 &-0.030 &-0.009 &-0.008 & 0.013 & 0.000 &-0.003 &-0.001 & 0.009 &-0.001 &-0.016 &-0.002 & 0.000 \\
28 &-0.763 &-0.157 &-0.009 & 0.421 & 0.092 & 0.004 &-0.213 &-0.046 &-0.002 & 0.098 & 0.027 & 0.000 &-0.034 &-0.004 & 0.000 \\
29 &-0.293 &-0.459 &-0.003 & 0.225 & 0.152 & 0.003 &-0.137 &-0.056 &-0.003 & 0.069 & 0.035 & 0.000 &-0.041 &-0.006 & 0.000 \\
30 &-0.033 &-0.108 &-0.217 & 0.050 & 0.033 & 0.017 &-0.038 &-0.020 &-0.009 & 0.023 & 0.014 &-0.001 &-0.019 &-0.004 & 0.000 \\
31 & \bf{1.000} & 0.305 & 0.019 &-0.762 &-0.184 &-0.007 & 0.424 & 0.091 & 0.001 &-0.192 &-0.028 & 0.000 & 0.029 & 0.004 & 0.000 \\
32 & 0.305 & \bf{1.000} & 0.033 &-0.398 &-0.466 &-0.004 & 0.287 & 0.160 &-0.001 &-0.148 &-0.034 & 0.000 & 0.011 & 0.007 & 0.000 \\
33 & 0.019 & 0.033 & \bf{1.000} &-0.039 &-0.128 &-0.222 & 0.057 & 0.042 & 0.021 &-0.038 &-0.014 &-0.006 & 0.004 &-0.001 & 0.000 \\
34 &-0.762 &-0.398 &-0.039 & \bf{1.000} & 0.343 & 0.018 &-0.770 &-0.193 &-0.006 & 0.403 & 0.089 &-0.001 &-0.104 &-0.014 & 0.001 \\
35 &-0.184 &-0.466 &-0.128 & 0.343 & \bf{1.000} & 0.039 &-0.433 &-0.493 &-0.009 & 0.323 & 0.195 &-0.002 &-0.153 &-0.035 & 0.002 \\
36 &-0.007 &-0.004 &-0.222 & 0.018 & 0.039 & \bf{1.000} &-0.039 &-0.137 &-0.238 & 0.063 & 0.060 & 0.029 &-0.047 &-0.021 &-0.004 \\
37 & 0.424 & 0.287 & 0.057 &-0.770 &-0.433 &-0.039 & \bf{1.000} & 0.354 & 0.014 &-0.729 &-0.179 & 0.001 & 0.211 & 0.028 &-0.001 \\
38 & 0.091 & 0.160 & 0.042 &-0.193 &-0.493 &-0.137 & 0.354 & \bf{1.000} & 0.044 &-0.479 &-0.497 & 0.000 & 0.301 & 0.103 &-0.004 \\
39 & 0.001 &-0.001 & 0.021 &-0.006 &-0.009 &-0.238 & 0.014 & 0.044 & \bf{1.000} &-0.052 &-0.148 &-0.246 & 0.084 & 0.052 & 0.022 \\
40 &-0.192 &-0.148 &-0.038 & 0.403 & 0.323 & 0.063 &-0.729 &-0.479 &-0.052 & \bf{1.000} & 0.497 &-0.002 &-0.653 &-0.078 & 0.003 \\
41 &-0.028 &-0.034 &-0.014 & 0.089 & 0.195 & 0.060 &-0.179 &-0.497 &-0.148 & 0.497 & \bf{1.000} & 0.020 &-0.713 &-0.257 & 0.010 \\
42 & 0.000 & 0.000 &-0.006 &-0.001 &-0.002 & 0.029 & 0.001 & 0.000 &-0.246 &-0.002 & 0.020 & \bf{1.000} & 0.001 &-0.259 &-0.163 \\
43 & 0.029 & 0.011 & 0.004 &-0.104 &-0.153 &-0.047 & 0.211 & 0.301 & 0.084 &-0.653 &-0.713 & 0.001 & \bf{1.000} & 0.111 &-0.005 \\
44 & 0.004 & 0.007 &-0.001 &-0.014 &-0.035 &-0.021 & 0.028 & 0.103 & 0.052 &-0.078 &-0.257 &-0.259 & 0.111 & \bf{1.000} &-0.023 \\
45 & 0.000 & 0.000 & 0.000 & 0.001 & 0.002 &-0.004 &-0.001 &-0.004 & 0.022 & 0.003 & 0.010 &-0.163 &-0.005 &-0.023 & \bf{1.000} \\
\end{tabular}
\end{ruledtabular}
\end{table*}
\endgroup
%

\begin{table*}[H]
{\renewcommand{\baselinestretch}{1.}\caption{\label{tab:A1_19}Virtual photon asymmetries 
 $A_1^p$ and $A_1^d$
at the average $\langle x \rangle $ and $\langle Q^2 \rangle $ in 19 $x$-bins 
(each $x$-bin is the average over the $Q^2$-bins), including
statistical, systematic and evolution uncertainties.
A normalization uncertainty of 5.2\% for the proton and 5\% for the deuteron has been included in the column `syst'.
}}
\begin{ruledtabular}
\begin{tabular}{rcc|ccccc|ccccc}
$~~$bin$~~$ & $~~\langle x\rangle~~$ & $~\langle Q^2\rangle/$GeV$^2~$ 
& $~A_1^p$  &$~\pm$stat.$~$ & $~\pm$syst.$~$ & $~\pm$par.$~$ & $~\pm$evol.$~$  
& $~A_1^d$  &$~\pm$stat.$~$ & $~\pm$syst.$~$ & $~\pm$par.$~$ & $~\pm$evol.$~$  
\\
\hline
  1 &  0.0058 &  0.26 &  0.0221 &  0.0270 &  0.0028 &  0.0011 &  0.0000 &  0.0092 &  0.0122 &  0.0005 &  0.0005 &  0.0000 \\
  2 &  0.0096 &  0.41 &  0.0158 &  0.0192 &  0.0025 &  0.0007 &  0.0000 & -0.0028 &  0.0090 &  0.0004 &  0.0001 &  0.0000 \\
  3 &  0.0142 &  0.57 &  0.0364 &  0.0190 &  0.0034 &  0.0016 &  0.0000 & -0.0017 &  0.0091 &  0.0004 &  0.0001 &  0.0000 \\
  4 &  0.0190 &  0.73 &  0.0459 &  0.0229 &  0.0042 &  0.0022 &  0.0000 & -0.0033 &  0.0111 &  0.0005 &  0.0002 &  0.0000 \\
  5 &  0.0253 &  0.91 &  0.0866 &  0.0185 &  0.0066 &  0.0044 &  0.0025 &  0.0144 &  0.0091 &  0.0011 &  0.0007 &  0.0010 \\
  6 &  0.0328 &  1.15 &  0.1089 &  0.0243 &  0.0078 &  0.0044 &  0.0025 & -0.0092 &  0.0122 &  0.0006 &  0.0003 &  0.0009 \\
  7 &  0.0403 &  1.33 &  0.0998 &  0.0237 &  0.0077 &  0.0035 &  0.0014 &  0.0048 &  0.0120 &  0.0010 &  0.0002 &  0.0006 \\
  8 &  0.0506 &  1.51 &  0.1117 &  0.0206 &  0.0094 &  0.0037 &  0.0006 &  0.0208 &  0.0106 &  0.0019 &  0.0006 &  0.0003 \\
  9 &  0.0648 &  2.01 &  0.1090 &  0.0202 &  0.0085 &  0.0035 &  0.0016 &  0.0254 &  0.0103 &  0.0019 &  0.0008 &  0.0010 \\
 10 &  0.0829 &  2.45 &  0.1997 &  0.0206 &  0.0129 &  0.0067 &  0.0022 &  0.0645 &  0.0107 &  0.0032 &  0.0019 &  0.0011 \\
 11 &  0.1059 &  2.97 &  0.2011 &  0.0218 &  0.0131 &  0.0069 &  0.0021 &  0.0503 &  0.0115 &  0.0027 &  0.0014 &  0.0008 \\
 12 &  0.1354 &  3.59 &  0.2681 &  0.0236 &  0.0166 &  0.0090 &  0.0022 &  0.0907 &  0.0128 &  0.0042 &  0.0024 &  0.0007 \\
 13 &  0.1730 &  4.31 &  0.3246 &  0.0263 &  0.0200 &  0.0096 &  0.0020 &  0.1318 &  0.0146 &  0.0058 &  0.0031 &  0.0008 \\
 14 &  0.2209 &  5.13 &  0.3361 &  0.0305 &  0.0203 &  0.0079 &  0.0013 &  0.1656 &  0.0173 &  0.0073 &  0.0030 &  0.0006 \\
 15 &  0.2820 &  6.11 &  0.4094 &  0.0370 &  0.0249 &  0.0083 &  0.0007 &  0.1810 &  0.0216 &  0.0080 &  0.0028 &  0.0004 \\
 16 &  0.3598 &  7.24 &  0.5169 &  0.0486 &  0.0316 &  0.0105 &  0.0008 &  0.3255 &  0.0291 &  0.0139 &  0.0062 &  0.0004 \\
 17 &  0.4583 &  8.53 &  0.6573 &  0.0714 &  0.0385 &  0.0158 &  0.0013 &  0.3815 &  0.0440 &  0.0164 &  0.0115 &  0.0007 \\
 18 &  0.5819 & 10.16 &  0.6647 &  0.1302 &  0.0413 &  0.0221 &  0.0018 &  0.4403 &  0.0813 &  0.0191 &  0.0206 &  0.0010 \\
 19 &  0.7248 & 12.21 &  1.1976 &  0.2763 &  0.0698 &  0.0438 &  0.0071 &  0.8641 &  0.1790 &  0.0359 &  0.0667 &  0.0051 \\

\end{tabular}
\end{ruledtabular}
\end{table*}


\begin{table*}[H]
{\renewcommand{\baselinestretch}{1.}\caption{\label{tab:gpd_20}Structure functions 
 $g_1^p$ and $g_1^d$
at the average $\langle x \rangle $ and $\langle Q^2 \rangle $ in 19 $x$-bins 
(each $x$-bin is the average over the $Q^2$-bins), including
statistical, systematic and evolution uncertainties.
A normalization uncertainty of 5.2\% for the proton and 5\% for the deuteron has been included in the column `syst'.
}}
\begin{ruledtabular}
\begin{tabular}{rcc|ccccc|ccccc}
$~~$bin$~~$ & $~~\langle x\rangle~~$ & $~\langle Q^2\rangle/$GeV$^2~$ 
& $~g_1^p$  &$~\pm$stat.$~$ & $~\pm$syst.$~$ & $~\pm$par.$~$ & $~\pm$evol.$~$  
& $~g_1^d$  &$~\pm$stat.$~$ & $~\pm$syst.$~$ & $~\pm$par.$~$ & $~\pm$evol.$~$  
\\
\hline
 1 &  0.0058 &  0.26 &  0.2584 &  0.3138 &  0.0331 &  0.0697 &  0.0000 &  0.1062 &  0.1400 &  0.0055 &  0.0286 &  0.0000 \\
 2 &  0.0096 &  0.41 &  0.1357 &  0.1632 &  0.0213 &  0.0220 &  0.0000 & -0.0224 &  0.0751 &  0.0037 &  0.0036 &  0.0000 \\
 3 &  0.0142 &  0.57 &  0.2361 &  0.1229 &  0.0219 &  0.0231 &  0.0000 & -0.0098 &  0.0575 &  0.0027 &  0.0010 &  0.0000 \\
 4 &  0.0190 &  0.73 &  0.2416 &  0.1200 &  0.0217 &  0.0152 &  0.0000 & -0.0158 &  0.0570 &  0.0027 &  0.0010 &  0.0000 \\
 5 &  0.0253 &  0.91 &  0.3366 &  0.0720 &  0.0253 &  0.0146 &  0.0097 &  0.0552 &  0.0348 &  0.0041 &  0.0025 &  0.0039 \\
 6 &  0.0328 &  1.15 &  0.3559 &  0.0795 &  0.0254 &  0.0116 &  0.0081 & -0.0277 &  0.0390 &  0.0019 &  0.0009 &  0.0028 \\
 7 &  0.0403 &  1.33 &  0.2803 &  0.0663 &  0.0215 &  0.0083 &  0.0039 &  0.0144 &  0.0328 &  0.0028 &  0.0005 &  0.0017 \\
 8 &  0.0506 &  1.51 &  0.2616 &  0.0482 &  0.0219 &  0.0078 &  0.0013 &  0.0479 &  0.0241 &  0.0043 &  0.0013 &  0.0007 \\
 9 &  0.0648 &  2.01 &  0.2155 &  0.0398 &  0.0165 &  0.0061 &  0.0031 &  0.0489 &  0.0198 &  0.0036 &  0.0012 &  0.0019 \\
10 &  0.0829 &  2.45 &  0.3235 &  0.0333 &  0.0205 &  0.0098 &  0.0036 &  0.0998 &  0.0167 &  0.0049 &  0.0024 &  0.0016 \\
11 &  0.1059 &  2.97 &  0.2676 &  0.0289 &  0.0170 &  0.0085 &  0.0027 &  0.0637 &  0.0146 &  0.0033 &  0.0015 &  0.0010 \\
12 &  0.1354 &  3.59 &  0.2887 &  0.0253 &  0.0174 &  0.0091 &  0.0023 &  0.0907 &  0.0128 &  0.0041 &  0.0021 &  0.0007 \\
13 &  0.1730 &  4.31 &  0.2775 &  0.0223 &  0.0166 &  0.0078 &  0.0017 &  0.1017 &  0.0113 &  0.0044 &  0.0022 &  0.0006 \\
14 &  0.2209 &  5.13 &  0.2207 &  0.0197 &  0.0129 &  0.0049 &  0.0008 &  0.0949 &  0.0099 &  0.0041 &  0.0016 &  0.0004 \\
15 &  0.2820 &  6.11 &  0.1935 &  0.0170 &  0.0112 &  0.0037 &  0.0003 &  0.0723 &  0.0085 &  0.0031 &  0.0010 &  0.0002 \\
16 &  0.3598 &  7.24 &  0.1585 &  0.0144 &  0.0090 &  0.0031 &  0.0002 &  0.0805 &  0.0071 &  0.0034 &  0.0014 &  0.0001 \\
17 &  0.4583 &  8.53 &  0.1108 &  0.0116 &  0.0059 &  0.0025 &  0.0002 &  0.0500 &  0.0056 &  0.0021 &  0.0014 &  0.0001 \\
18 &  0.5819 & 10.16 &  0.0471 &  0.0088 &  0.0025 &  0.0015 &  0.0001 &  0.0230 &  0.0041 &  0.0010 &  0.0010 &  0.0001 \\
19 &  0.7248 & 12.21 &  0.0225 &  0.0050 &  0.0011 &  0.0008 &  0.0001 &  0.0116 &  0.0023 &  0.0005 &  0.0009 &  0.0001 \\
\end{tabular}
\end{ruledtabular}
\end{table*}

%
\begingroup
\squeezetable
\begin{table*}[H]
{\renewcommand{\baselinestretch}{1.}\caption{\label{tab:corr-20bins-hyd}{Correlation matrix for
 $g_1^p$ in 19 $x$-bins (averaged over $Q^2$). For
$\langle x\rangle$ and $\langle Q^2\rangle$ of each bin, see
e.g. Tab.~\ref{tab:gpd_20}.}} } 
\begin{ruledtabular}
\begin{tabular}{r|rrrrrrrrrrrrrrrrrrr}
 & 1 & 2 & 3 & 4 & 5 & 6 & 7 & 8 & 9 & 10 & 11 & 12 & 13 & 14 & 15 & 16 & 17 & 18 & 19  \\
\hline

 1  &  \bf{1.000} & -0.159 &  0.005 & -0.005 & -0.002 & -0.001 & -0.001 & -0.001 &  0.000 &  0.000 &  0.000 &  0.000 &  0.000 &  0.000 &  0.000 &  0.000 &  0.000 &  0.000 &  0.000 \\
 2  & -0.159 &  \bf{1.000} & -0.196 &  0.000 & -0.012 & -0.004 & -0.002 & -0.004 & -0.001 & -0.001 &  0.000 &  0.000 &  0.000 &  0.000 &  0.000 &  0.000 &  0.000 &  0.001 &  0.000 \\
 3  &  0.005 & -0.196 &  \bf{1.000} & -0.230 & -0.010 & -0.009 & -0.003 & -0.009 & -0.002 & -0.001 & -0.001 &  0.000 &  0.000 &  0.000 &  0.000 &  0.000 &  0.000 &  0.001 &  0.000 \\
 4  & -0.005 &  0.000 & -0.230 &  \bf{1.000} & -0.206 &  0.003 & -0.006 & -0.011 & -0.002 & -0.002 & -0.001 & -0.001 &  0.000 &  0.000 &  0.000 &  0.000 & -0.001 &  0.001 & -0.001 \\
 5  & -0.002 & -0.012 & -0.010 & -0.206 &  \bf{1.000} & -0.213 &  0.002 & -0.026 & -0.004 & -0.003 & -0.003 & -0.001 & -0.001 &  0.000 &  0.000 &  0.000 &  0.000 &  0.001 &  0.000 \\
 6  & -0.001 & -0.004 & -0.009 &  0.003 & -0.213 &  \bf{1.000} & -0.238 & -0.002 & -0.010 & -0.004 & -0.002 & -0.001 & -0.001 & -0.001 &  0.000 &  0.000 &  0.000 &  0.001 & -0.001 \\
 7  & -0.001 & -0.002 & -0.003 & -0.006 &  0.002 & -0.238 &  \bf{1.000} & -0.255 &  0.011 & -0.009 & -0.003 & -0.002 & -0.001 & -0.001 &  0.000 &  0.000 & -0.001 &  0.002 & -0.001 \\
 8  & -0.001 & -0.004 & -0.009 & -0.011 & -0.026 & -0.002 & -0.255 &  \bf{1.000} & -0.222 &  0.008 & -0.011 & -0.003 & -0.002 & -0.001 & -0.001 &  0.000 & -0.001 &  0.002 & -0.001 \\
 9  &  0.000 & -0.001 & -0.002 & -0.002 & -0.004 & -0.010 &  0.011 & -0.222 &  \bf{1.000} & -0.236 &  0.011 & -0.014 & -0.004 & -0.003 & -0.002 & -0.001 & -0.001 &  0.002 & -0.001 \\
10  &  0.000 & -0.001 & -0.001 & -0.002 & -0.003 & -0.004 & -0.009 &  0.008 & -0.236 &  \bf{1.000} & -0.248 &  0.019 & -0.015 & -0.002 & -0.002 &  0.001 & -0.004 &  0.008 & -0.004 \\
11  &  0.000 &  0.000 & -0.001 & -0.001 & -0.003 & -0.002 & -0.003 & -0.011 &  0.011 & -0.248 &  \bf{1.000} & -0.260 &  0.025 & -0.015 &  0.001 & -0.001 & -0.004 &  0.011 & -0.006 \\
12  &  0.000 &  0.000 &  0.000 & -0.001 & -0.001 & -0.001 & -0.002 & -0.003 & -0.014 &  0.019 & -0.260 &  \bf{1.000} & -0.272 &  0.029 & -0.016 &  0.003 & -0.009 &  0.014 & -0.008 \\
13  &  0.000 &  0.000 &  0.000 &  0.000 & -0.001 & -0.001 & -0.001 & -0.002 & -0.004 & -0.015 &  0.025 & -0.272 &  \bf{1.000} & -0.287 &  0.034 & -0.016 & -0.003 &  0.010 & -0.007 \\
14  &  0.000 &  0.000 &  0.000 &  0.000 &  0.000 & -0.001 & -0.001 & -0.001 & -0.003 & -0.002 & -0.015 &  0.029 & -0.287 &  \bf{1.000} & -0.292 &  0.041 & -0.023 &  0.010 & -0.009 \\
15  &  0.000 &  0.000 &  0.000 &  0.000 &  0.000 &  0.000 &  0.000 & -0.001 & -0.002 & -0.002 &  0.001 & -0.016 &  0.034 & -0.292 &  \bf{1.000} & -0.302 &  0.045 & -0.006 & -0.001 \\
16  &  0.000 &  0.000 &  0.000 &  0.000 &  0.000 &  0.000 &  0.000 &  0.000 & -0.001 &  0.001 & -0.001 &  0.003 & -0.016 &  0.041 & -0.302 &  \bf{1.000} & -0.321 &  0.057 & -0.028 \\
17  &  0.000 &  0.000 &  0.000 & -0.001 &  0.000 &  0.000 & -0.001 & -0.001 & -0.001 & -0.004 & -0.004 & -0.009 & -0.003 & -0.023 &  0.045 & -0.321 &  \bf{1.000} & -0.316 &  0.069 \\
18  &  0.000 &  0.001 &  0.001 &  0.001 &  0.001 &  0.001 &  0.002 &  0.002 &  0.002 &  0.008 &  0.011 &  0.014 &  0.010 &  0.010 & -0.006 &  0.057 & -0.316 &  \bf{1.000} & -0.316 \\
19  &  0.000 &  0.000 &  0.000 & -0.001 &  0.000 & -0.001 & -0.001 & -0.001 & -0.001 & -0.004 & -0.006 & -0.008 & -0.007 & -0.009 & -0.001 & -0.028 &  0.069 & -0.316 &  \bf{1.000} \\
\end{tabular}
\end{ruledtabular}
\end{table*}
\endgroup

%
\begingroup
\squeezetable
\begin{table*}[H]
{\renewcommand{\baselinestretch}{1.}\caption{\label{tab:corr-20bins-deu}{
Correlation matrix for  $g_1^d$ in 19 $x$-bins (averaged over $Q^2$). 
For $\langle x\rangle$ and $\langle Q^2\rangle$ of each bin, see
e.g. Tab.~\ref{tab:gpd_20}.}}} 
\begin{ruledtabular}
\begin{tabular}{r|rrrrrrrrrrrrrrrrrrr}
 & 1 & 2 & 3 & 4 & 5 & 6 & 7 & 8 & 9 & 10 & 11 & 12 & 13 & 14 & 15 & 16 & 17 & 18 & 19  \\
\hline
 1  &  \bf{1.000} & -0.150 &  0.007 & -0.003 & -0.002 & -0.001 &  0.000 & -0.001 &  0.000 &  0.000 &  0.000 &  0.000 &  0.000 &  0.000 &  0.000 &  0.000 &  0.000 &  0.000 &  0.000 \\
 2  & -0.150 &  \bf{1.000} & -0.186 &  0.003 & -0.010 & -0.004 & -0.001 & -0.003 & -0.001 &  0.000 &  0.000 &  0.000 &  0.000 &  0.000 &  0.000 &  0.000 &  0.000 &  0.000 &  0.000 \\
 3  &  0.007 & -0.186 &  \bf{1.000} & -0.226 & -0.002 & -0.008 & -0.002 & -0.008 & -0.001 & -0.002 &  0.000 &  0.000 &  0.000 &  0.000 &  0.000 &  0.000 &  0.000 &  0.001 &  0.000 \\
 4  & -0.003 &  0.003 & -0.226 &  \bf{1.000} & -0.205 &  0.007 & -0.004 & -0.011 & -0.002 & -0.001 & -0.001 &  0.000 &  0.000 & -0.001 &  0.000 &  0.000 &  0.000 &  0.001 &  0.000 \\
 5  & -0.002 & -0.010 & -0.002 & -0.205 &  \bf{1.000} & -0.216 &  0.015 & -0.023 & -0.003 & -0.002 & -0.001 & -0.002 & -0.001 &  0.000 &  0.000 &  0.000 &  0.000 &  0.001 &  0.000 \\
 6  & -0.001 & -0.004 & -0.008 &  0.007 & -0.216 &  \bf{1.000} & -0.249 &  0.011 & -0.008 & -0.003 & -0.002 & -0.001 & -0.001 &  0.000 &  0.000 &  0.000 &  0.000 &  0.000 &  0.000 \\
 7  &  0.000 & -0.001 & -0.002 & -0.004 &  0.015 & -0.249 &  \bf{1.000} & -0.266 &  0.021 & -0.009 & -0.002 & -0.001 & -0.001 & -0.001 &  0.000 &  0.000 & -0.001 &  0.002 & -0.001 \\
 8  & -0.001 & -0.003 & -0.008 & -0.011 & -0.023 &  0.011 & -0.266 &  \bf{1.000} & -0.229 &  0.020 & -0.010 & -0.002 & -0.003 &  0.000 &  0.000 &  0.000 &  0.000 &  0.001 & -0.001 \\
 9  &  0.000 & -0.001 & -0.001 & -0.002 & -0.003 & -0.008 &  0.021 & -0.229 &  \bf{1.000} & -0.246 &  0.023 & -0.013 & -0.002 & -0.003 & -0.001 & -0.001 & -0.001 &  0.002 & -0.001 \\
10  &  0.000 &  0.000 & -0.002 & -0.001 & -0.002 & -0.003 & -0.009 &  0.020 & -0.246 &  \bf{1.000} & -0.259 &  0.032 & -0.013 &  0.000 & -0.001 &  0.001 & -0.003 &  0.007 & -0.003 \\
11  &  0.000 &  0.000 &  0.000 & -0.001 & -0.001 & -0.002 & -0.002 & -0.010 &  0.023 & -0.259 &  \bf{1.000} & -0.274 &  0.038 & -0.013 &  0.001 &  0.001 & -0.005 &  0.011 & -0.005 \\
12  &  0.000 &  0.000 &  0.000 &  0.000 & -0.002 & -0.001 & -0.001 & -0.002 & -0.013 &  0.032 & -0.274 &  \bf{1.000} & -0.287 &  0.043 & -0.015 &  0.005 & -0.008 &  0.012 & -0.007 \\
13  &  0.000 &  0.000 &  0.000 &  0.000 & -0.001 & -0.001 & -0.001 & -0.003 & -0.002 & -0.013 &  0.038 & -0.287 &  \bf{1.000} & -0.303 &  0.049 & -0.015 & -0.003 &  0.010 & -0.004 \\
14  &  0.000 &  0.000 &  0.000 & -0.001 &  0.000 &  0.000 & -0.001 &  0.000 & -0.003 &  0.000 & -0.013 &  0.043 & -0.303 &  \bf{1.000} & -0.310 &  0.055 & -0.023 &  0.008 & -0.007 \\
15  &  0.000 &  0.000 &  0.000 &  0.000 &  0.000 &  0.000 &  0.000 &  0.000 & -0.001 & -0.001 &  0.001 & -0.015 &  0.049 & -0.310 &  \bf{1.000} & -0.320 &  0.060 & -0.005 &  0.002 \\
16  &  0.000 &  0.000 &  0.000 &  0.000 &  0.000 &  0.000 &  0.000 &  0.000 & -0.001 &  0.001 &  0.001 &  0.005 & -0.015 &  0.055 & -0.320 &  \bf{1.000} & -0.346 &  0.072 & -0.034 \\
17  &  0.000 &  0.000 &  0.000 &  0.000 &  0.000 &  0.000 & -0.001 &  0.000 & -0.001 & -0.003 & -0.005 & -0.008 & -0.003 & -0.023 &  0.060 & -0.346 &  \bf{1.000} & -0.347 &  0.092 \\
18  &  0.000 &  0.000 &  0.001 &  0.001 &  0.001 &  0.000 &  0.002 &  0.001 &  0.002 &  0.007 &  0.011 &  0.012 &  0.010 &  0.008 & -0.005 &  0.072 & -0.347 &  \bf{1.000} & -0.355 \\
19  &  0.000 &  0.000 &  0.000 &  0.000 &  0.000 &  0.000 & -0.001 & -0.001 & -0.001 & -0.003 & -0.005 & -0.007 & -0.004 & -0.007 &  0.002 & -0.034 &  0.092 & -0.355 &  \bf{1.000} \\
\end{tabular}
\end{ruledtabular}
\end{table*}
\endgroup

\begin{table*}[H]
{\renewcommand{\baselinestretch}{1.}\caption{\label{tab:gpd_15}Structure
function $g_1^p$ and $g_1^d$  at the average $\langle x \rangle $ and $\langle Q^2 \rangle $ in 15 $x$-bins
(each $x$-bin is the average over the $Q^2$ bins), including
statistical and systematic uncertainties.
A normalization uncertainty of 5.2\% for the proton and 5\% for the deuteron has been included in the column `syst'.
}}
\begin{ruledtabular}
\begin{tabular}{rcc|ccccc|ccccc}
$~~$bin$~~$ & $~~\langle x\rangle~~$ & $~\langle Q^2\rangle/$GeV$^2~$ 
& $~g_1^p$  &$~\pm$stat.$~$ & $~\pm$syst.$~$ & $~\pm$par.$~$  & $~\pm$evol.$~$ 
& $~g_1^d$  &$~\pm$stat.$~$ & $~\pm$syst.$~$ & $~\pm$par.$~$  & $~\pm$evol.$~$ 
\\
\hline
 1 &  0.0264 &  1.12 &  0.4736 &  0.1490 &  0.0375 &  0.0182 &  0.0000 &  0.1140 &  0.0689 &  0.0074 &  0.0045 &  0.0000 \\
 2 &  0.0329 &  1.25 &  0.3459 &  0.0949 &  0.0273 &  0.0109 &  0.0000 & -0.0405 &  0.0458 &  0.0018 &  0.0012 &  0.0000 \\
 3 &  0.0403 &  1.38 &  0.2696 &  0.0706 &  0.0215 &  0.0079 &  0.0000 &  0.0125 &  0.0346 &  0.0030 &  0.0004 &  0.0000 \\
 4 &  0.0506 &  1.54 &  0.2640 &  0.0498 &  0.0220 &  0.0078 &  0.0000 &  0.0510 &  0.0248 &  0.0043 &  0.0014 &  0.0000 \\
 5 &  0.0648 &  2.01 &  0.2155 &  0.0398 &  0.0165 &  0.0061 &  0.0031 &  0.0489 &  0.0198 &  0.0036 &  0.0012 &  0.0019 \\
 6 &  0.0829 &  2.45 &  0.3235 &  0.0333 &  0.0205 &  0.0098 &  0.0036 &  0.0998 &  0.0167 &  0.0049 &  0.0024 &  0.0016 \\
 7 &  0.1059 &  2.97 &  0.2676 &  0.0289 &  0.0170 &  0.0085 &  0.0027 &  0.0637 &  0.0146 &  0.0033 &  0.0015 &  0.0010 \\
 8 &  0.1354 &  3.59 &  0.2887 &  0.0253 &  0.0174 &  0.0091 &  0.0023 &  0.0907 &  0.0128 &  0.0041 &  0.0021 &  0.0007 \\
 9 &  0.1730 &  4.31 &  0.2775 &  0.0223 &  0.0166 &  0.0078 &  0.0017 &  0.1017 &  0.0113 &  0.0044 &  0.0022 &  0.0006 \\
10 &  0.2209 &  5.13 &  0.2207 &  0.0197 &  0.0129 &  0.0049 &  0.0008 &  0.0949 &  0.0099 &  0.0041 &  0.0016 &  0.0004 \\
11 &  0.2820 &  6.11 &  0.1935 &  0.0170 &  0.0112 &  0.0037 &  0.0003 &  0.0723 &  0.0085 &  0.0031 &  0.0010 &  0.0002 \\
12 &  0.3598 &  7.24 &  0.1585 &  0.0144 &  0.0090 &  0.0031 &  0.0002 &  0.0805 &  0.0071 &  0.0034 &  0.0014 &  0.0001 \\
13 &  0.4583 &  8.53 &  0.1108 &  0.0116 &  0.0059 &  0.0025 &  0.0002 &  0.0500 &  0.0056 &  0.0021 &  0.0014 &  0.0001 \\
14 &  0.5819 & 10.16 &  0.0471 &  0.0088 &  0.0025 &  0.0015 &  0.0001 &  0.0230 &  0.0041 &  0.0010 &  0.0010 &  0.0001 \\
15 &  0.7248 & 12.21 &  0.0225 &  0.0050 &  0.0011 &  0.0008 &  0.0001 &  0.0116 &  0.0023 &  0.0005 &  0.0009 &  0.0001 \\
\end{tabular}
\end{ruledtabular}
\end{table*}

\begin{table*}[H]
{\renewcommand{\baselinestretch}{1.}\caption{\label{tab:A1pd_15}Virtual photon asymmetries
 $A_1^p$ and $A_1^d$  at the average $\langle x \rangle $ and $\langle Q^2 \rangle $ in 15 $x$-bins
(each $x$-bin is the average over the $Q^2$ bins), including
statistical and systematic uncertainties.
A normalization uncertainty of 5.2\% for the proton and 5\% for the deuteron has been included in the column `syst'.
}}
\begin{ruledtabular}
\begin{tabular}{rcc|ccccc|ccccc}
$~~$bin$~~$ & $~~\langle x\rangle~~$ & $~\langle Q^2\rangle/$GeV$^2~$ 
& $~A_1^p$  &$~\pm$stat.$~$ & $~\pm$syst.$~$ & $~\pm$par.$~$  & $~\pm$evol.$~$ 
& $~A_1^d$  &$~\pm$stat.$~$ & $~\pm$syst.$~$ & $~\pm$par.$~$  & $~\pm$evol.$~$ 
\\
\hline
  1 &  0.0264 &  1.12 &  0.1113 &  0.0351 &  0.0088 &  0.0036 &  0.0000 &  0.0275 &  0.0167 &  0.0018 &  0.0009 &  0.0000 \\
  2 &  0.0329 &  1.25 &  0.0975 &  0.0268 &  0.0078 &  0.0039 &  0.0000 & -0.0123 &  0.0133 &  0.0005 &  0.0005 &  0.0000 \\
  3 &  0.0403 &  1.38 &  0.0897 &  0.0236 &  0.0072 &  0.0039 &  0.0000 &  0.0039 &  0.0120 &  0.0011 &  0.0002 &  0.0000 \\
  4 &  0.0506 &  1.54 &  0.1061 &  0.0201 &  0.0089 &  0.0046 &  0.0000 &  0.0210 &  0.0104 &  0.0018 &  0.0009 &  0.0000 \\
  5 &  0.0648 &  2.01 &  0.1090 &  0.0202 &  0.0085 &  0.0035 &  0.0016 &  0.0254 &  0.0103 &  0.0019 &  0.0008 &  0.0010 \\
  6 &  0.0829 &  2.45 &  0.1997 &  0.0206 &  0.0129 &  0.0067 &  0.0022 &  0.0645 &  0.0107 &  0.0032 &  0.0019 &  0.0011 \\
  7 &  0.1059 &  2.97 &  0.2011 &  0.0218 &  0.0131 &  0.0069 &  0.0021 &  0.0503 &  0.0115 &  0.0027 &  0.0014 &  0.0008 \\
  8 &  0.1354 &  3.59 &  0.2681 &  0.0236 &  0.0166 &  0.0090 &  0.0022 &  0.0907 &  0.0128 &  0.0042 &  0.0024 &  0.0007 \\
  9 &  0.1730 &  4.31 &  0.3246 &  0.0263 &  0.0200 &  0.0096 &  0.0020 &  0.1318 &  0.0146 &  0.0058 &  0.0031 &  0.0008 \\
 10 &  0.2209 &  5.13 &  0.3361 &  0.0305 &  0.0203 &  0.0079 &  0.0013 &  0.1656 &  0.0173 &  0.0073 &  0.0030 &  0.0006 \\
 11 &  0.2820 &  6.11 &  0.4094 &  0.0370 &  0.0249 &  0.0083 &  0.0007 &  0.1810 &  0.0216 &  0.0080 &  0.0028 &  0.0004 \\
 12 &  0.3598 &  7.24 &  0.5169 &  0.0486 &  0.0316 &  0.0105 &  0.0008 &  0.3255 &  0.0291 &  0.0139 &  0.0062 &  0.0004 \\
 13 &  0.4583 &  8.53 &  0.6573 &  0.0714 &  0.0385 &  0.0158 &  0.0013 &  0.3815 &  0.0440 &  0.0164 &  0.0115 &  0.0007 \\
 14 &  0.5819 & 10.16 &  0.6647 &  0.1302 &  0.0413 &  0.0221 &  0.0018 &  0.4403 &  0.0813 &  0.0191 &  0.0206 &  0.0010 \\
 15 &  0.7248 & 12.21 &  1.1976 &  0.2763 &  0.0698 &  0.0438 &  0.0071 &  0.8641 &  0.1790 &  0.0359 &  0.0667 &  0.0051 \\
\end{tabular}
\end{ruledtabular}
\end{table*}

\begin{table*}[H]
{\renewcommand{\baselinestretch}{1.}\caption{\label{tab:corr-15bins-hyd}{Correlation matrix for $g_1^p$ in 15
$x$-bins ($Q^2>1$GeV$^2$) averaged over $Q^2$. For $\langle
x\rangle$ and $\langle Q^2\rangle$ of each bin, see e.g. Tab.~\ref{tab:gpd_15}.}}}
\begin{ruledtabular}
\begin{tabular}{r|rrrrrrrrrrrrrrr}
   &   1  &   2 & 3 & 4 & 5 & 6 & 7 & 8 & 9 & 10 & 11 & 12& 13 & 14 & 15\\
\hline
 1  &  \bf{1.000} & -0.160 &  0.009 & -0.004 & -0.004 & -0.002 & -0.002 & -0.001 & -0.001 & -0.001 & -0.001 & -0.001 &  0.000 &  0.000 &  0.000 \\
 2  & -0.160 &  \bf{1.000} & -0.201 &  0.011 & -0.010 & -0.004 & -0.003 & -0.002 & -0.001 & -0.001 &  0.000 &  0.000 & -0.001 &  0.002 & -0.001 \\
 3  &  0.009 & -0.201 &  \bf{1.000} & -0.223 &  0.011 & -0.009 & -0.003 & -0.002 & -0.002 & -0.001 &  0.000 &  0.000 & -0.001 &  0.002 & -0.001 \\
 4  & -0.004 &  0.011 & -0.223 &  \bf{1.000} & -0.223 &  0.009 & -0.011 & -0.003 & -0.002 & -0.002 & -0.001 &  0.000 & -0.001 &  0.002 & -0.001 \\
 5  & -0.004 & -0.010 &  0.011 & -0.223 &  \bf{1.000} & -0.236 &  0.011 & -0.014 & -0.004 & -0.003 & -0.002 & -0.001 & -0.001 &  0.002 & -0.001 \\
 6  & -0.002 & -0.004 & -0.009 &  0.009 & -0.236 &  \bf{1.000} & -0.248 &  0.019 & -0.015 & -0.002 & -0.002 &  0.001 & -0.004 &  0.008 & -0.004 \\
 7  & -0.002 & -0.003 & -0.003 & -0.011 &  0.011 & -0.248 &  \bf{1.000} & -0.260 &  0.025 & -0.015 &  0.001 & -0.001 & -0.004 &  0.011 & -0.006 \\
 8  & -0.001 & -0.002 & -0.002 & -0.003 & -0.014 &  0.019 & -0.260 &  \bf{1.000} & -0.272 &  0.029 & -0.016 &  0.003 & -0.009 &  0.014 & -0.008 \\
 9  & -0.001 & -0.001 & -0.002 & -0.002 & -0.004 & -0.015 &  0.025 & -0.272 &  \bf{1.000} & -0.287 &  0.034 & -0.016 & -0.003 &  0.010 & -0.007 \\
10  & -0.001 & -0.001 & -0.001 & -0.002 & -0.003 & -0.002 & -0.015 &  0.029 & -0.287 &  \bf{1.000} & -0.292 &  0.041 & -0.023 &  0.010 & -0.009 \\
11  & -0.001 &  0.000 &  0.000 & -0.001 & -0.002 & -0.002 &  0.001 & -0.016 &  0.034 & -0.292 &  \bf{1.000} & -0.302 &  0.045 & -0.006 & -0.001 \\
12  & -0.001 &  0.000 &  0.000 &  0.000 & -0.001 &  0.001 & -0.001 &  0.003 & -0.016 &  0.041 & -0.302 &  \bf{1.000} & -0.321 &  0.057 & -0.028 \\
13  &  0.000 & -0.001 & -0.001 & -0.001 & -0.001 & -0.004 & -0.004 & -0.009 & -0.003 & -0.023 &  0.045 & -0.321 &  \bf{1.000} & -0.316 &  0.069 \\
14  &  0.000 &  0.002 &  0.002 &  0.002 &  0.002 &  0.008 &  0.011 &  0.014 &  0.010 &  0.010 & -0.006 &  0.057 & -0.316 &  \bf{1.000} & -0.316 \\
15  &  0.000 & -0.001 & -0.001 & -0.001 & -0.001 & -0.004 & -0.006 & -0.008 & -0.007 & -0.009 & -0.001 & -0.028 &  0.069 & -0.316 &  \bf{1.000} \\
\end{tabular}
\end{ruledtabular}
\end{table*}


%
%
\begin{table*}[H]
{\renewcommand{\baselinestretch}{1.}\caption{\label{tab:corr-15bins-deu}{Correlation matrix for
$g_1^d$ in 15 $x$-bins ($Q^2>1$GeV$^2$) averaged
over $Q^2$. For $\langle x\rangle$ and $\langle Q^2\rangle$ of each
bin, see e.g. Tab.~\ref{tab:gpd_15}.}} }
\begin{ruledtabular}
\begin{tabular}{r|rrrrrrrrrrrrrrr}
   &   1  &   2 & 3 & 4 & 5 & 6 & 7 & 8 & 9 & 10 & 11 & 12& 13 & 14 & 15\\
\hline
 1  &  \bf{1.000} & -0.176 &  0.018 & -0.004 & -0.002 & -0.002 & -0.002 & -0.001 & -0.001 &  0.000 &  0.000 &  0.000 &  0.000 &  0.000 &  0.000 \\
 2  & -0.176 &  \bf{1.000} & -0.215 &  0.021 & -0.009 & -0.003 & -0.002 & -0.001 & -0.001 & -0.001 &  0.000 &  0.000 &  0.000 &  0.001 &  0.000 \\
 3  &  0.018 & -0.215 &  \bf{1.000} & -0.233 &  0.020 & -0.009 & -0.002 & -0.001 & -0.001 & -0.001 &  0.000 &  0.000 & -0.001 &  0.002 & -0.001 \\
 4  & -0.004 &  0.021 & -0.233 &  \bf{1.000} & -0.230 &  0.020 & -0.010 & -0.002 & -0.003 &  0.000 & -0.001 &  0.000 & -0.001 &  0.002 & -0.001 \\
 5  & -0.002 & -0.009 &  0.020 & -0.230 &  \bf{1.000} & -0.246 &  0.023 & -0.013 & -0.002 & -0.003 & -0.001 & -0.001 & -0.001 &  0.002 & -0.001 \\
 6  & -0.002 & -0.003 & -0.009 &  0.020 & -0.246 &  \bf{1.000} & -0.259 &  0.032 & -0.013 &  0.000 & -0.001 &  0.001 & -0.003 &  0.007 & -0.003 \\
 7  & -0.002 & -0.002 & -0.002 & -0.010 &  0.023 & -0.259 &  \bf{1.000} & -0.274 &  0.038 & -0.013 &  0.001 &  0.001 & -0.005 &  0.011 & -0.005 \\
 8  & -0.001 & -0.001 & -0.001 & -0.002 & -0.013 &  0.032 & -0.274 &  \bf{1.000} & -0.287 &  0.043 & -0.015 &  0.005 & -0.008 &  0.012 & -0.007 \\
 9  & -0.001 & -0.001 & -0.001 & -0.003 & -0.002 & -0.013 &  0.038 & -0.287 &  \bf{1.000} & -0.303 &  0.049 & -0.015 & -0.003 &  0.010 & -0.004 \\
10  &  0.000 & -0.001 & -0.001 &  0.000 & -0.003 &  0.000 & -0.013 &  0.043 & -0.303 &  \bf{1.000} & -0.310 &  0.055 & -0.023 &  0.008 & -0.007 \\
11  &  0.000 &  0.000 &  0.000 & -0.001 & -0.001 & -0.001 &  0.001 & -0.015 &  0.049 & -0.310 &  \bf{1.000} & -0.320 &  0.060 & -0.005 &  0.002 \\
12  &  0.000 &  0.000 &  0.000 &  0.000 & -0.001 &  0.001 &  0.001 &  0.005 & -0.015 &  0.055 & -0.320 &  \bf{1.000} & -0.346 &  0.072 & -0.034 \\
13  &  0.000 &  0.000 & -0.001 & -0.001 & -0.001 & -0.003 & -0.005 & -0.008 & -0.003 & -0.023 &  0.060 & -0.346 &  \bf{1.000} & -0.347 &  0.092 \\
14  &  0.000 &  0.001 &  0.002 &  0.002 &  0.002 &  0.007 &  0.011 &  0.012 &  0.010 &  0.008 & -0.005 &  0.072 & -0.347 &  \bf{1.000} & -0.355 \\
15  &  0.000 &  0.000 & -0.001 & -0.001 & -0.001 & -0.003 & -0.005 & -0.007 & -0.004 & -0.007 &  0.002 & -0.034 &  0.092 & -0.355 &  \bf{1.000} \\

\end{tabular}
\end{ruledtabular}
\end{table*}
%
%
%
%

\begin{table*}[H]
{\renewcommand{\baselinestretch}{1.}\caption{\label{tab:g1n46}
Structure functions $g_1^n$ and $g_1^{NS}$ at the average $\langle x
\rangle $ and $\langle Q^2 \rangle $  in 45 bins, 
with statistical and systematic uncertainties.}}
\begin{ruledtabular}
\begin{tabular}{rcc|cccc|cccc}
bin &$~\langle x\rangle ~$ & $~\langle Q^2\rangle/$GeV$^2~$ &$~g_1^n~$&$~\pm$stat.$~$ & $~\pm$syst.$~$ &
$~\pm$par.$~$  &$~g_1^{NS}~$&$~\pm$stat.$~$ & $~\pm$syst.$~$ &$~\pm$par.$~$   \\
\hline
 1 &  0.0058 &  0.26 & -0.0288 &  0.4360 &  0.0269 &  0.0068 &  0.2872 &  0.6967 &  0.0591 &  0.0119 \\
 2 &  0.0096 &  0.41 & -0.1841 &  0.2302 &  0.0183 &  0.0011 &  0.3198 &  0.3646 &  0.0388 &  0.0015 \\
 3 &  0.0142 &  0.57 & -0.2572 &  0.1747 &  0.0205 &  0.0008 &  0.4934 &  0.2754 &  0.0420 &  0.0014 \\
 4 &  0.0190 &  0.73 & -0.2757 &  0.1720 &  0.0184 &  0.0008 &  0.5173 &  0.2698 &  0.0398 &  0.0011 \\
 5 &  0.0248 &  0.82 & -0.2154 &  0.1298 &  0.0162 &  0.0012 &  0.4966 &  0.2015 &  0.0366 &  0.0014 \\
 6 &  0.0264 &  1.12 & -0.2270 &  0.2108 &  0.0226 &  0.0041 &  0.7006 &  0.3332 &  0.0598 &  0.0042 \\
 7 &  0.0325 &  0.87 & -0.3594 &  0.2400 &  0.0207 &  0.0006 &  0.7375 &  0.3678 &  0.0413 &  0.0009 \\
 8 &  0.0329 &  1.25 & -0.4335 &  0.1371 &  0.0278 &  0.0016 &  0.7794 &  0.2140 &  0.0550 &  0.0018 \\
 9 &  0.0399 &  0.90 & -0.2867 &  0.3256 &  0.0229 &  0.0013 &  0.6458 &  0.4943 &  0.0444 &  0.0015 \\
10 &  0.0403 &  1.38 & -0.2425 &  0.1029 &  0.0173 &  0.0008 &  0.5121 &  0.1599 &  0.0385 &  0.0011 \\
11 &  0.0498 &  0.93 & -0.2324 &  0.3841 &  0.0114 &  0.0007 &  0.4383 &  0.5833 &  0.0310 &  0.0009 \\
12 &  0.0506 &  1.54 & -0.1536 &  0.0732 &  0.0153 &  0.0019 &  0.4176 &  0.1131 &  0.0367 &  0.0021 \\
13 &  0.0643 &  1.25 & -0.1236 &  0.1531 &  0.0111 &  0.0024 &  0.3873 &  0.2326 &  0.0260 &  0.0025 \\
14 &  0.0645 &  1.85 & -0.0759 &  0.0870 &  0.0110 &  0.0022 &  0.2822 &  0.1344 &  0.0264 &  0.0024 \\
15 &  0.0655 &  2.58 & -0.1421 &  0.1014 &  0.0121 &  0.0012 &  0.3370 &  0.1583 &  0.0301 &  0.0015 \\
16 &  0.0823 &  1.31 & -0.2034 &  0.1623 &  0.0133 &  0.0006 &  0.4210 &  0.2436 &  0.0280 &  0.0009 \\
17 &  0.0824 &  2.06 & -0.0809 &  0.0769 &  0.0117 &  0.0038 &  0.3912 &  0.1184 &  0.0297 &  0.0039 \\
18 &  0.0835 &  3.08 & -0.1121 &  0.0733 &  0.0124 &  0.0040 &  0.4655 &  0.1140 &  0.0360 &  0.0041 \\
19 &  0.1051 &  1.38 & -0.2273 &  0.1804 &  0.0145 &  0.0012 &  0.5215 &  0.2665 &  0.0324 &  0.0014 \\
20 &  0.1054 &  2.29 & -0.1089 &  0.0703 &  0.0123 &  0.0018 &  0.3194 &  0.1079 &  0.0255 &  0.0020 \\
21 &  0.1064 &  3.65 & -0.1333 &  0.0578 &  0.0115 &  0.0028 &  0.4322 &  0.0897 &  0.0313 &  0.0030 \\
22 &  0.1344 &  1.46 &  0.0484 &  0.2049 &  0.0104 &  0.0046 &  0.1799 &  0.2968 &  0.0229 &  0.0046 \\
23 &  0.1347 &  2.56 & -0.0555 &  0.0647 &  0.0110 &  0.0036 &  0.3279 &  0.0987 &  0.0257 &  0.0037 \\
24 &  0.1358 &  4.30 & -0.1213 &  0.0475 &  0.0109 &  0.0030 &  0.4190 &  0.0737 &  0.0297 &  0.0031 \\
25 &  0.1719 &  1.56 & -0.1105 &  0.2376 &  0.0158 &  0.0025 &  0.3728 &  0.3390 &  0.0313 &  0.0027 \\
26 &  0.1722 &  2.87 & -0.0203 &  0.0602 &  0.0107 &  0.0042 &  0.2943 &  0.0915 &  0.0250 &  0.0043 \\
27 &  0.1734 &  5.05 & -0.0732 &  0.0400 &  0.0083 &  0.0034 &  0.3500 &  0.0621 &  0.0254 &  0.0035 \\
28 &  0.2191 &  1.67 & -0.1138 &  0.2663 &  0.0138 &  0.0016 &  0.3212 &  0.3776 &  0.0277 &  0.0018 \\
29 &  0.2200 &  3.18 & -0.0682 &  0.0561 &  0.0104 &  0.0024 &  0.2758 &  0.0848 &  0.0219 &  0.0025 \\
30 &  0.2213 &  5.87 &  0.0014 &  0.0336 &  0.0042 &  0.0038 &  0.2232 &  0.0524 &  0.0172 &  0.0039 \\
31 &  0.2786 &  1.98 &  0.0376 &  0.2296 &  0.0142 &  0.0048 &  0.2172 &  0.3251 &  0.0262 &  0.0049 \\
32 &  0.2810 &  3.77 & -0.0526 &  0.0487 &  0.0086 &  0.0024 &  0.2450 &  0.0735 &  0.0186 &  0.0025 \\
33 &  0.2824 &  6.94 & -0.0311 &  0.0288 &  0.0048 &  0.0027 &  0.2248 &  0.0449 &  0.0163 &  0.0028 \\
34 &  0.3550 &  2.46 & -0.0528 &  0.1771 &  0.0146 &  0.0020 &  0.2212 &  0.2537 &  0.0268 &  0.0021 \\
35 &  0.3585 &  4.62 & -0.0070 &  0.0394 &  0.0064 &  0.0025 &  0.1646 &  0.0597 &  0.0156 &  0.0026 \\
36 &  0.3603 &  8.25 &  0.0228 &  0.0245 &  0.0017 &  0.0030 &  0.1367 &  0.0384 &  0.0102 &  0.0031 \\
37 &  0.4520 &  3.08 &  0.0783 &  0.1331 &  0.0120 &  0.0031 &  0.0288 &  0.1931 &  0.0183 &  0.0031 \\
38 &  0.4567 &  5.61 &  0.0051 &  0.0305 &  0.0034 &  0.0022 &  0.1215 &  0.0466 &  0.0096 &  0.0023 \\
39 &  0.4589 &  9.72 & -0.0037 &  0.0197 &  0.0016 &  0.0017 &  0.1080 &  0.0311 &  0.0069 &  0.0018 \\
40 &  0.5629 &  3.90 & -0.0722 &  0.0910 &  0.0092 &  0.0003 &  0.1482 &  0.1353 &  0.0146 &  0.0005 \\
41 &  0.5798 &  6.77 & -0.0011 &  0.0244 &  0.0015 &  0.0007 &  0.0400 &  0.0375 &  0.0037 &  0.0007 \\
42 &  0.5823 & 11.36 &  0.0015 &  0.0146 &  0.0008 &  0.0009 &  0.0502 &  0.0233 &  0.0034 &  0.0009 \\
43 &  0.6921 &  6.32 &  0.0246 &  0.0448 &  0.0030 &  0.0008 &  0.0014 &  0.0686 &  0.0042 &  0.0009 \\
44 &  0.7173 &  9.56 & -0.0057 &  0.0112 &  0.0006 &  0.0003 &  0.0314 &  0.0177 &  0.0020 &  0.0004 \\
45 &  0.7311 & 14.29 &  0.0079 &  0.0095 &  0.0003 &  0.0005 &  0.0125 &  0.0151 &  0.0009 &  0.0005 \\
\end{tabular}
\end{ruledtabular}
\end{table*}
\begin{table*}[H]
{\renewcommand{\baselinestretch}{1.}\caption{\label{tab:g1n20} Structure functions $g_1^n$ and $g_1^{NS}$ 
at the average $\langle x \rangle $ and $\langle Q^2 \rangle $ in 19
 $x$-bins, averaged over all {\Qs},
with statistical and systematic uncertainties.}}
\begin{ruledtabular}
\begin{tabular}{rcc|ccccc|ccccc}
bin &$~\langle x\rangle~$ & $~\langle Q^2\rangle/$GeV$^2~$ &$~g_1^n~$&$~\pm$stat.$~$ & $~\pm$syst.$~$ &
$~\pm$par.$~$ &$~\pm$evol.$~$ &$~g_1^{NS}~$&$~\pm$stat.$~$ & $~\pm$syst.$~$ &$~\pm$par.$~$ &$~\pm$evol.$~$ \\
\hline
 1 &  0.0058 &  0.26 & -0.0288 &  0.4360 &  0.0269 &  0.0068 &  0.0000 &  0.2872 &  0.6967 &  0.0591 &  0.0119 &  0.0000 \\
 2 &  0.0096 &  0.41 & -0.1841 &  0.2302 &  0.0183 &  0.0011 &  0.0000 &  0.3198 &  0.3646 &  0.0388 &  0.0015 &  0.0000 \\
 3 &  0.0142 &  0.57 & -0.2572 &  0.1747 &  0.0205 &  0.0008 &  0.0000 &  0.4934 &  0.2754 &  0.0420 &  0.0014 &  0.0000 \\
 4 &  0.0190 &  0.73 & -0.2757 &  0.1720 &  0.0184 &  0.0008 &  0.0000 &  0.5173 &  0.2698 &  0.0398 &  0.0011 &  0.0000 \\
 5 &  0.0253 &  0.91 & -0.2172 &  0.1042 &  0.0179 &  0.0021 &  0.0023 &  0.5537 &  0.1625 &  0.0430 &  0.0022 &  0.0112 \\
 6 &  0.0328 &  1.15 & -0.4158 &  0.1159 &  0.0258 &  0.0012 &  0.0027 &  0.7717 &  0.1800 &  0.0510 &  0.0014 &  0.0104 \\
 7 &  0.0403 &  1.33 & -0.2492 &  0.0970 &  0.0179 &  0.0008 &  0.0014 &  0.5296 &  0.1503 &  0.0391 &  0.0011 &  0.0046 \\
 8 &  0.0506 &  1.51 & -0.1581 &  0.0711 &  0.0149 &  0.0018 &  0.0006 &  0.4197 &  0.1097 &  0.0362 &  0.0020 &  0.0013 \\
 9 &  0.0648 &  2.01 & -0.1098 &  0.0584 &  0.0109 &  0.0019 &  0.0019 &  0.3252 &  0.0904 &  0.0269 &  0.0020 &  0.0030 \\
10 &  0.0829 &  2.45 & -0.1078 &  0.0491 &  0.0113 &  0.0036 &  0.0017 &  0.4313 &  0.0758 &  0.0314 &  0.0037 &  0.0044 \\
11 &  0.1059 &  2.97 & -0.1298 &  0.0427 &  0.0109 &  0.0024 &  0.0012 &  0.3975 &  0.0658 &  0.0277 &  0.0025 &  0.0037 \\
12 &  0.1354 &  3.59 & -0.0927 &  0.0375 &  0.0092 &  0.0033 &  0.0008 &  0.3814 &  0.0576 &  0.0264 &  0.0034 &  0.0031 \\
13 &  0.1730 &  4.31 & -0.0576 &  0.0330 &  0.0076 &  0.0037 &  0.0006 &  0.3351 &  0.0509 &  0.0240 &  0.0038 &  0.0022 \\
14 &  0.2209 &  5.13 & -0.0156 &  0.0290 &  0.0047 &  0.0034 &  0.0004 &  0.2362 &  0.0448 &  0.0173 &  0.0035 &  0.0010 \\
15 &  0.2820 &  6.11 & -0.0371 &  0.0250 &  0.0048 &  0.0026 &  0.0002 &  0.2306 &  0.0387 &  0.0158 &  0.0027 &  0.0004 \\
16 &  0.3598 &  7.24 &  0.0157 &  0.0210 &  0.0023 &  0.0029 &  0.0001 &  0.1428 &  0.0326 &  0.0110 &  0.0030 &  0.0003 \\
17 &  0.4583 &  8.53 & -0.0026 &  0.0168 &  0.0017 &  0.0018 &  0.0001 &  0.1134 &  0.0262 &  0.0074 &  0.0019 &  0.0003 \\
18 &  0.5819 & 10.16 &  0.0026 &  0.0125 &  0.0007 &  0.0008 &  0.0000 &  0.0445 &  0.0197 &  0.0030 &  0.0009 &  0.0001 \\
19 &  0.7248 & 12.21 &  0.0025 &  0.0071 &  0.0003 &  0.0004 &  0.0000 &  0.0199 &  0.0113 &  0.0013 &  0.0004 &  0.0001 \\
\end{tabular}
\end{ruledtabular}
\end{table*}
\begin{table*}[H]
{\renewcommand{\baselinestretch}{1.}\caption{ \label{tab:g1n15} Structure functions $g_1^n$ and $g_1^{NS}$ 
at the average $\langle x \rangle $ and $\langle Q^2 \rangle $  in 15
 $x$-bins, averaged over $Q^2>$1~GeV$^2$,
with statistical and systematic uncertainties.}}
\begin{ruledtabular}
\begin{tabular}{rcc|ccccc|ccccc}
bin &$~\langle x\rangle ~$ & $~\langle Q^2\rangle /$GeV$^2~$ &$~g_1^n~$&$~\pm$stat.$~$ & $~\pm$syst.$~$ &
$~\pm$par.$~$ &$~\pm$evol.$~$ &$~g_1^{NS}~$&$~\pm$stat.$~$ & $~\pm$syst.$~$ &$~\pm$par.$~$ &$~\pm$evol.$~$  \\
\hline
 1 &  0.0264 &  1.12 & -0.2270 &  0.2108 &  0.0226 &  0.0041 &  0.0000 &  0.7006 &  0.3332 &  0.0598 &  0.0042 &  0.0000 \\
 2 &  0.0329 &  1.25 & -0.4335 &  0.1371 &  0.0278 &  0.0016 &  0.0000 &  0.7794 &  0.2140 &  0.0550 &  0.0018 &  0.0000 \\
 3 &  0.0403 &  1.38 & -0.2425 &  0.1029 &  0.0173 &  0.0008 &  0.0000 &  0.5121 &  0.1599 &  0.0385 &  0.0011 &  0.0000 \\
 4 &  0.0506 &  1.54 & -0.1536 &  0.0732 &  0.0153 &  0.0019 &  0.0000 &  0.4176 &  0.1131 &  0.0367 &  0.0021 &  0.0000 \\
 5 &  0.0648 &  2.01 & -0.1098 &  0.0584 &  0.0109 &  0.0019 &  0.0019 &  0.3252 &  0.0904 &  0.0269 &  0.0020 &  0.0030 \\
 6 &  0.0829 &  2.45 & -0.1078 &  0.0491 &  0.0113 &  0.0036 &  0.0017 &  0.4313 &  0.0758 &  0.0314 &  0.0037 &  0.0044 \\
 7 &  0.1059 &  2.97 & -0.1298 &  0.0427 &  0.0109 &  0.0024 &  0.0012 &  0.3975 &  0.0658 &  0.0277 &  0.0025 &  0.0037 \\
 8 &  0.1354 &  3.59 & -0.0927 &  0.0375 &  0.0092 &  0.0033 &  0.0008 &  0.3814 &  0.0576 &  0.0264 &  0.0034 &  0.0031 \\
 9 &  0.1730 &  4.31 & -0.0576 &  0.0330 &  0.0076 &  0.0037 &  0.0006 &  0.3351 &  0.0509 &  0.0240 &  0.0038 &  0.0022 \\
10 &  0.2209 &  5.13 & -0.0156 &  0.0290 &  0.0047 &  0.0034 &  0.0004 &  0.2362 &  0.0448 &  0.0173 &  0.0035 &  0.0010 \\
11 &  0.2820 &  6.11 & -0.0371 &  0.0250 &  0.0048 &  0.0026 &  0.0002 &  0.2306 &  0.0387 &  0.0158 &  0.0027 &  0.0004 \\
12 &  0.3598 &  7.24 &  0.0157 &  0.0210 &  0.0023 &  0.0029 &  0.0001 &  0.1428 &  0.0326 &  0.0110 &  0.0030 &  0.0003 \\
13 &  0.4583 &  8.53 & -0.0026 &  0.0168 &  0.0017 &  0.0018 &  0.0001 &  0.1134 &  0.0262 &  0.0074 &  0.0019 &  0.0003 \\
14 &  0.5819 & 10.16 &  0.0026 &  0.0125 &  0.0007 &  0.0008 &  0.0000 &  0.0445 &  0.0197 &  0.0030 &  0.0009 &  0.0001 \\
15 &  0.7248 & 12.21 &  0.0025 &  0.0071 &  0.0003 &  0.0004 &  0.0000 &  0.0199 &  0.0113 &  0.0013 &  0.0004 &  0.0001 \\
\end{tabular}
\end{ruledtabular}
\end{table*}

\begin{table*}[H]
{\renewcommand{\baselinestretch}{1.}\caption{ \label{tab:deltasigma}
Recent published values of $\Delta \Sigma$, separated into experimental evaluations based on the $g_1$ integral on a given target, 
and evaluations from QCD fits, published from 2000 on, based on all $g_1$ data available at the times of publication. 
All evaluations are in the $\overline{MS}$ scheme.
The HERMES results refer to the present analysis.
The results for E154 and SMC were calculated from values of $\Gamma_1$ given by
those collaborations, as values of $\Delta \Sigma$ were not provided. 
Some analyses used $\Delta C_{NS}$ in order $\alpha_s^3$ (NNNLO) 
and $\Delta C_{S}$ in order $\alpha_s^2$ (NNLO).
Thus the order in the table  was labelled as (N)NNLO, being a mixture 
of NNNLO and NNLO.
The last column refers to the method used by the experimental groups to 
calculate the low-$x$ extrapolation needed to compute the integral over the full $x$ range. 
In the case of E143, the low-$x$ extrapolation ($0\leq x \leq 0.03$) is calculated as an average of four fits, details of which 
can be found in Ref.~\cite{Abe:1998wq}. 
}}
\begin{ruledtabular}
\begin{tabular}{l|c|c|l|c|c|l}
Analysis                  &  year & $Q^2$  & $\Delta \Sigma $       &     target            & order       & low-$x$ extr.\\        
                          &       &(GeV$^2$)&                       &                       &             & \\
\hline
\hline
\multicolumn{4}{c}{~~~~~~~~~~~~~~~~~~~~~~~~~~Experimental evaluations}\\
\hline
E142~\cite{E142n}         & 1996 & 2  & $0.43\pm 0.12$(total)                                                      & n($^3$He) &    (N)NNLO & Regge      \\ 
E154~\cite{e154qcd}       & 1997 & 5  & $0.191 \pm 0.011\mathrm{(theor.)}\pm 0.080\mathrm{(exp.)}\pm 0.070$(evol.) & n($^3$He) &     ``     & E154 QCD fit~\cite{e154qcd}\\ 
SMC~\cite{Adeva:1998vv}   & 1998 & 10 & $0.116 \pm 0.011\mathrm{(theor.)}\pm 0.079\mathrm{(exp.)}\pm 0.138$(evol.) &    p      &     ``     & SMC QCD fit~\cite{smcqcd}\\ 
SMC~\cite{Adeva:1998vv}   & 1998 & 10 & $0.060 \pm 0.008\mathrm{(theor.)}\pm 0.075\mathrm{(exp.)}\pm 0.139$(evol.) &    d      &     ``     & ``  \\ 
E143~\cite{Abe:1998wq}    & 1998 & 3  & $0.32 \pm 0.10 $(total)                                                    &    p      &     ``     & see caption\\ 
``                        &  ``  & `` & $0.37 \pm 0.08 $(total)                                                    &    d      &     ``     &  ``        \\ 
E155~\cite{Anthony:1999rm}& 1999 & 5  & ${0.15}\pm 0.03\mathrm{(stat.)} \pm 0.08$(syst.)                           &    d      &     ``     &E154 QCD fit~\cite{e154qcd}\\ %
``                        &''    & `` & ${0.18}\pm 0.03\mathrm{(stat.)} \pm 0.08$(syst.)                           &    ``     &     ``     &SMC QCD fit~\cite{smcqcd}\\ %
HERMES                    & 2006 & 5  & $0.321 \pm 0.011\mathrm{(theor.)}\pm 0.024\mathrm{(exp.)}\pm 0.028$(evol.) &    d      &     NLO    &none        \\ 
``                        & ``   & `` & $0.330 \pm 0.011\mathrm{(theor.)}\pm 0.025\mathrm{(exp.)}\pm 0.028$(evol.) &    ``     &     NNLO   &  ``        \\ 
``                        & ``   & `` & $0.333 \pm 0.011\mathrm{(theor.)}\pm 0.025\mathrm{(exp.)}\pm 0.028$(evol.) &    ``     &    (N)NNLO &  ``        \\ 
\hline
\multicolumn{3}{c}{~~~~~~~~~~~~~~~~~~~~~~~~~~QCD fits}\\
\hline
E155~\cite{e155p}         & 2000 & 5  & $0.23\pm0.04$(stat.) $\pm0.06$(syst.)                                      &    -      &     NLO    &            \\ 
GRSV\cite{grsv}           & 2001 & 1  & 0.204 (standard scenario)                                                  &    -      &     ``     &            \\ 
``                        &  ``  & `` & 0.282 (valence scenario)                                                   &    -      &     ``     &            \\
``                        &  ``  & 5  & 0.197 (standard scenario)                                                  &    -      &     ``     &            \\
``                        &  ``  & `` & 0.273 (valence scenario)                                                   &    -      &     ``     &            \\
``                        &  ``  & 10 & 0.197 (standard scenario)                                                  &    -      &     ``     &            \\
``                        &  ``  & `` & 0.272 (valence scenario)                                                   &    -      &     ``     &            \\
BB~\cite{bb}              & 2002 & 4  & $0.14\pm0.08$(stat.)                                                       &    -      &     ``     &            \\ 
dFNS~\cite{dFS:2005}      & 2005 &10  & 0.284 (KRE)                                                                &    -      &     ``     &            \\ 
 ``                       & ``   & `` & 0.311 (KKP)                                                                &    -      &     ``     &            \\
AAC~\cite{AAC06}          & 2006 & 1  & $0.25 \pm 0.10$                                                            &    -      &     ``     &            \\
LSS\cite{lss06}           & 2006 & 1  & $0.219 \pm 0.042$                                                          &    -      &     ``     &            \\
\end{tabular}
\end{ruledtabular}
\end{table*}
\end{appendix}
\end{document}

%% file: rec-g1long.tex










\def\groupargonne{\affiliation{Physics Division, Argonne National Laboratory, Argonne, Illinois 60439-4843, USA}}
\def\groupbari{\affiliation{Istituto Nazionale di Fisica Nucleare, Sezione di Bari, 70124 Bari, Italy}}
\def\groupbeijing{\affiliation{School of Physics, Peking University, Beijing 100871, China}}
\def\groupchina{\affiliation{Department of Modern Physics, University of Science and Technology of China, Hefei, Anhui 230026, China}}
\def\groupcolorado{\affiliation{Nuclear Physics Laboratory, University of Colorado, Boulder, Colorado 80309-0390, USA}}
\def\groupdesy{\affiliation{DESY, 22603 Hamburg, Germany}}
\def\groupzeuthen{\affiliation{DESY, 15738 Zeuthen, Germany}}
\def\groupdubna{\affiliation{Joint Institute for Nuclear Research, 141980 Dubna, Russia}}
\def\grouperlangen{\affiliation{Physikalisches Institut, Universit\"at Erlangen-N\"urnberg, 91058 Erlangen, Germany}}
\def\groupferrara{\affiliation{Istituto Nazionale di Fisica Nucleare, Sezione di Ferrara and Dipartimento di Fisica, Universit\`a di Ferrara, 44100 Ferrara, Italy}}
\def\groupfrascati{\affiliation{Istituto Nazionale di Fisica Nucleare, Laboratori Nazionali di Frascati, 00044 Frascati, Italy}}
\def\groupgent{\affiliation{Department of Subatomic and Radiation Physics, University of Gent, 9000 Gent, Belgium}}
\def\groupgiessen{\affiliation{Physikalisches Institut, Universit\"at Gie{\ss}en, 35392 Gie{\ss}en, Germany}}
\def\groupglasgow{\affiliation{Department of Physics and Astronomy, University of Glasgow, Glasgow G12 8QQ, United Kingdom}}
\def\groupillinois{\affiliation{Department of Physics, University of Illinois, Urbana, Illinois 61801-3080, USA}}
\def\groupmichigan{\affiliation{Randall Laboratory of Physics, University of Michigan, Ann Arbor, Michigan 48109-1040, USA }}
\def\groupmoscow{\affiliation{Lebedev Physical Institute, 117924 Moscow, Russia}}
\def\groupnikhef{\affiliation{Nationaal Instituut voor Kernfysica en Hoge-Energiefysica (NIKHEF), 1009 DB Amsterdam, The Netherlands}}
\def\groupstpetersburg{\affiliation{Petersburg Nuclear Physics Institute, St. Petersburg, Gatchina, 188350 Russia}}
\def\groupprotvino{\affiliation{Institute for High Energy Physics, Protvino, Moscow region, 142281 Russia}}
\def\groupregensburg{\affiliation{Institut f\"ur Theoretische Physik, Universit\"at Regensburg, 93040 Regensburg, Germany}}
\def\grouprome{\affiliation{Istituto Nazionale di Fisica Nucleare, Sezione Roma 1, Gruppo Sanit\`a and Physics Laboratory, Istituto Superiore di Sanit\`a, 00161 Roma, Italy}}
\def\grouptriumf{\affiliation{TRIUMF, Vancouver, British Columbia V6T 2A3, Canada}}
\def\grouptokyo{\affiliation{Department of Physics, Tokyo Institute of Technology, Tokyo 152, Japan}}
\def\groupamsterdam{\affiliation{Department of Physics and Astronomy, Vrije Universiteit, 1081 HV Amsterdam, The Netherlands}}
\def\groupwarsaw{\affiliation{Andrzej Soltan Institute for Nuclear Studies, 00-689 Warsaw, Poland}}
\def\groupyerevan{\affiliation{Yerevan Physics Institute, 375036 Yerevan, Armenia}}
\def\groupnone{\noaffiliation}


\groupargonne
\groupbari
\groupbeijing
\groupchina
\groupcolorado
\groupdesy
\groupzeuthen
\groupdubna
\grouperlangen
\groupferrara
\groupfrascati
\groupgent
\groupgiessen
\groupglasgow
\groupillinois
\groupmichigan
\groupmoscow
\groupnikhef
\groupstpetersburg
\groupprotvino
\groupregensburg
\grouprome
\grouptriumf
\grouptokyo
\groupamsterdam
\groupwarsaw
\groupyerevan


\author{A.~Airapetian}  \groupmichigan
\author{N.~Akopov}  \groupyerevan
\author{Z.~Akopov}  \groupyerevan
\author{A.~Andrus}  \groupillinois
\author{E.C.~Aschenauer}  \groupzeuthen
\author{W.~Augustyniak}  \groupwarsaw
\author{R.~Avakian}  \groupyerevan
\author{A.~Avetissian}  \groupyerevan
\author{E.~Avetissian}  \groupfrascati
\author{S.~Belostotski}  \groupstpetersburg
% ...

\collaboration{The HERMES Collaboration} \noaffiliation




